\documentclass[aps,prd,twocolumn,10pt,superscriptaddress,nofootinbib,longbibliography,floatfix]{revtex4-2}

\usepackage{amsmath,amssymb,amsfonts}
\usepackage{bm}
\usepackage{graphicx}
\usepackage{booktabs}
\usepackage{xcolor}
\usepackage{hyperref}
\usepackage{cleveref}
\graphicspath{{figures/}}
\hypersetup{colorlinks=true,linkcolor=blue,citecolor=blue,urlcolor=blue}

\begin{document}

\title{Numerical-Relativity Particle-in-Cell Simulations:\\  Self-Gravitating Collisionless Matter in Dynamical Spacetimes}

\author{Jiaxi Wu}
\affiliation{TAPIR, Mailcode 350-17, California Institute of Technology, Pasadena, CA 91125, USA}
\author{Elias R. Most}
\affiliation{TAPIR, Mailcode 350-17, California Institute of Technology, Pasadena, CA 91125, USA}
\affiliation{Walter Burke Institute for Theoretical Physics, California Institute of Technology, Pasadena, CA 91125, USA}

\begin{abstract}
Gravitational collapse and instability, black-hole dynamics in systems with large scale separations, and questions surrounding extremal black hole formation require full nonlinear studies using accurate and adapted numerical tools. To this end, we present a new algorithm to study
collisionless relativistic systems using a fully kinetic description of matter coupled to the evolving spacetime. In detail, we present a numerical-relativity particle-in-cell (NRPIC) framework for solving
the Einstein--Vlasov system in three spatial dimensions. Implemented on top of the
performance-portable \texttt{AthenaK} code, the method couples particle dynamics
to Z4c spacetime evolution in moving puncture gauge through conservative stress-energy deposition on
adaptively refined meshes. Our implementation makes use of advances in the computational modeling
of relativistic plasmas using modern PIC models, now made available to general-relativistic gravity
via an equivalent electrodynamic formulation. We introduce a 
Boris-type particle pusher that exploits the electrodynamic (Lorentz force)
form of the geodesic force.
We validate the coupled evolution against spherical dust collapse, follow
prolate collapse through collisionless caustics into black-hole formation, and
investigate the stability and nonlinear migration of relativistic clusters.
Highlighting its potential to study gravitational instability in NR, 
we simulate the collapse of two overdensities within a
common collisionless envelope, following black-hole formation, binary inspiral
and merger, and subsequent mass accretion onto the remnant. 
\end{abstract}

\maketitle

\section{Introduction}
\label{sec:intro}
Gravity plays a central role in a wide range of astrophysical phenomena across many scales, including stellar collapse \cite{Janka:2012wk,Burrows:2020qrp} and compact-object formation \cite{Burrows:2024pur,Burrows:2024wqv}, the inspiral and merger of compact binaries \cite{LIGOScientific:2025slb,LIGOScientific:2026wfs,Baiotti:2016qnr,Kyutoku:2021icp}, planet formation in protoplanetary disks \cite{Armitage:2010vb,Chiang:2009xs}, star formation in molecular clouds \cite{McKee:2007ai}, the dynamical evolution of stellar clusters and galaxies \cite{2019ARA&A..57..227K,2008gady.book.....B}, and the growth of large-scale cosmic structure \cite{Springel:2006vs,Angulo:2021kes}. Over the past decades, the diversity of these systems has motivated the development of a broad range of theoretical and numerical approaches, each adapted to the relevant matter properties and gravitational regime.

Gravitational instability is common to many of these processes. It arises when the self-gravity of a matter overdensity is strong enough to overcome stabilizing effects such as thermal and radiation pressure, differential orbital rotation, and velocity dispersion \cite{McKee:2007ai,1965MNRAS.130...97G,2016ARA&A..54..271K,Metzger:2024ujc}. Initially small perturbations can then grow into bound structures or undergo gravitational collapse. In Newtonian gravity, this competition is described by the Jeans instability in approximately homogeneous systems and the Toomre instability in rotating stellar disks \cite{Toomre:1964zx,2008gady.book.....B}. The nonlinear consequences have been studied extensively in different setups using various numerical approaches \cite{Federrath:2012ni,Hopkins:2013vha,Chen:2025uwd}. In general relativity, the stability and subsequent evolution of self-gravitating matter are further coupled to the nonlinear spacetime evolution. Early numerical studies of the Einstein-Vlasov system in axisymmetry demonstrated that sufficiently compact spherical and rotating collisionless clusters can become dynamically unstable and collapse to black holes \cite{1985ApJ...298...34S,1985ApJ...298...58S}. Subsequent studies have investigated nonspherical gravitational collapse, relativistic structure formation, and the stability of different equilibrium distributions \cite{Shibata:1999va,East:2019bwu,East:2019chx,Gunther:2021pbf}, also in the context of critical collapse \cite{Rein:1998uf,Rein:2002fx,Olabarrieta:2001wy,Akbarian:2014gna}.
In other scenarios, e.g., for supermassive Pop-III stars, different studies have investigated black hole formation using fully general-relativistic hydrodynamics simulations in reduced dimensions \cite{Shibata:2024xsl,Shibata:2025lde,Lam:2026asp}.
Depending on the scenario in question, different methods or algorithms can be more appropriate. In particular, particle-based approaches \cite{Diener:2022hui} and moving meshes \cite{Lioutas:2022ghb} offer advantages in systems with widely separated scales, such as protoplanet formation by collapse \cite{Mayer:2002ylv,2025ApJ...995...96N,2024A&A...682L...6F}. Collisionless collapse is also directly relevant to fundamental questions of cosmic censorship and extremal black-hole formation \cite{Andreasson:2011ng,Ames:2023akp,Kehle:2024vyt}.

Motivated by these questions and recent advances in particle-based numerical relativity \cite{Diener:2022hui,Rosswog:2023nnl}, we present a new algorithm that draws on insights gained over the past decades from studies of collisionless relativistic plasmas \cite{fonseca2002osiris,Spitkovsky:2005vsy,Cerutti:2013mma,EntityDevelopmentTeam:2025tlp}.

Numerical investigations of relativistic collisionless systems have a long history. Refs. \cite{1985ApJ...298...34S,1985ApJ...298...58S} developed particle-based solutions of the Einstein–Vlasov equations to study the equilibrium, stability, and collapse of relativistic stellar clusters, including the formation of black holes from unstable configurations. Ref.~\cite{Ames:2020ssl} used a simple particle-in-cell scheme coupled to a partially constrained $(2+1)+1$ formulation of the Einstein equations to study axisymmetric critical collapse. Collisionless systems have also provided a setting for investigating critical phenomena such as extremal black hole formation and cosmic censorship, where the possible formation of naked singularities is explored through simulations of the collapse of highly prolate particle distributions \cite{Shapiro:1991zza,Yoo:2016kzu,East:2019bwu}. Recent interest in this field comes from mathematical advances showing that collisionless charged matter can lead to the formation of an extremally charged black hole in finite time \cite{Kehle:2024vyt}. Several studies have investigated aspects of this numerically \cite{Gelles:2025gxi,Gelles:2026bix,East:2025nfb}.

In this paper, we present the Numerical Relativity Particle-In-Cell (NRPIC) method for three-dimensional simulations of self-gravitating collisionless matter, which builds on previous Einstein–Vlasov calculations \cite{1985ApJ...298...34S,1985ApJ...298...58S,Shibata:1999va,Shibata:1999wi,Pretorius:2018lfb,Ames:2020ssl} and the particle-in-cell approach \cite{1991ppcs.book.....B} in the form widely used in relativistic plasma physics \cite{fonseca2002osiris,Spitkovsky:2005vsy,Cerutti:2013mma,Fedeli:2022oxl}.
NRPIC solves the general-relativistic Klimontovich equation, representing the distribution function as a finite collection of macroparticles following geodesics. The spacetime metric is evolved using Einstein's equations, with a stress-energy tensor reconstructed from the particles. Therefore, the self-gravitating matter is evolved consistently with the spacetime at each timestep. In addition to the GR-Boris pusher widely used in GRPIC codes, we introduce a new DGREM--Boris pusher that uses the DGREM formalism \cite{Olivares:2021cfa,Boyeneni:2025tsx} to recast the Einstein equations into Maxwell-like equations. In this formalism, the geodesic force can be written in a Lorentz-like form and treated together with electrodynamics. We implement this algorithm within the \texttt{AthenaK} framework and couple it to the spacetime-evolution and adaptive mesh-refinement modules.

This work is structured as follows.
Section~\ref{sec:algorithm} presents the NRPIC algorithm, including Z4c spacetime evolution in \texttt{AthenaK} (Section~\ref{sec:z4c}), particle pushers (Section~\ref{sec:pusher}), stress-energy deposition (Section~\ref{sec:stree-energy-deposition}), and mass-conserving deposition on adaptively refined meshes (Section~\ref{sec:amr}). The benchmarks in Section~\ref{sec:tests} test geodesic accuracy (Section~\ref{sec:kerr-tests}), black-hole formation through gravitational collapse (Sections~\ref{sec:os-collapse} and \ref{sec:censorship}), and the long-term evolution and branch migration of relativistic clusters (Section~\ref{sec:st-cluster}). Section~\ref{sec:Jeans} applies the method to a toy model of Jeans collapse in a collisionless cluster, following black-hole formation, binary evolution, and interactions with the surrounding collisionless envelope. The Appendix examines convergence and errors from finite particle sampling, metric interpolation, and stress-energy deposition.

Throughout the paper, we use geometrized units $G=c=1$ and metric signature $(-,+,+,+)$.  Greek indices denote spacetime components, Latin indices denote spatial coordinate components, and hatted indices denote tetrad components.

\section{Method}
\label{sec:algorithm}

\begin{figure*}[t]
 \centering
 \includegraphics[width=0.88\textwidth]{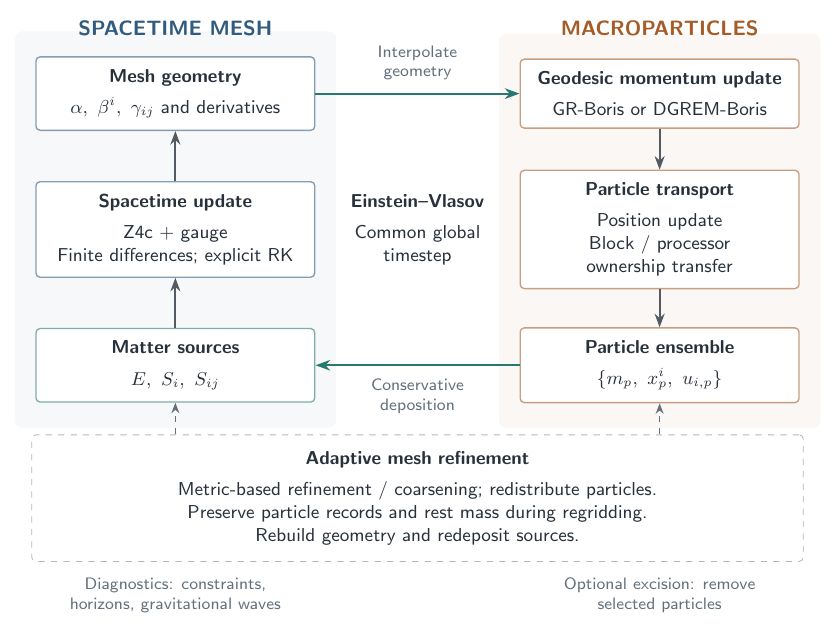}
 \caption{Overview of the NRPIC particle--mesh cycle. Interpolated geometry advances the particle momenta and positions, conservative first-order deposition reconstructs the matter sources ${E}$, $S_i$, and $S_{ij}$ for Z4c spacetime evolution. Particles and fields share a common global time step. Adaptive mesh refinement preserves the particle records during redistribution and rebuilds the geometry and sources on the new hierarchy (dashed arrows).}
 \label{fig:nrpic-cycle}
\end{figure*}

NRPIC evolves the coupled Einstein--Vlasov system for collisionless massive matter distributions. The spacetime is a four-dimensional Lorentzian manifold with metric $g_{\mu\nu}$, foliated by spacelike hypersurfaces of constant coordinate time $t$. In the $3+1$ decomposition \cite{Arnowitt:1962hi}, the metric line element can be written as
\begin{equation}
 ds^2=-\alpha^2dt^2
 +\gamma_{ij}(dx^i+\beta^i dt)(dx^j+\beta^j dt),
 \label{eq:adm-metric}
\end{equation}
where $\alpha>0$ is the lapse, $\beta^i$ is the shift, and $\gamma_{ij}$ is the induced spatial metric. The future-directed unit normal has covariant components $n_\mu=(-\alpha,0,0,0)$ and contravariant components $n^\mu=\alpha^{-1}(1,-\beta^i)$, and the spatial projector is $\gamma_{\mu\nu}=g_{\mu\nu}+n_\mu n_\nu$. Writing $g=\det g_{\mu\nu}$ and $\gamma=\det\gamma_{ij}$, the invariant spacetime and spatial volume elements are $dV_4=\sqrt{-g}\,d^4x=\alpha\sqrt{\gamma}\,dt\,d^3x$ and $d\Sigma=\sqrt{\gamma}\,d^3x$.

The geometry satisfies Einstein's equation,
\begin{equation}
 G_{\mu\nu}\equiv R_{\mu\nu}-\frac{1}{2}g_{\mu\nu}R
 =8\pi T_{\mu\nu},
 \label{eq:einstein}
\end{equation}
where $R_{\mu\nu}$ is the Ricci tensor and $R=g^{\mu\nu}R_{\mu\nu}$ is the Ricci scalar, and $T_{\mu\nu}$ is the matter energy-momentum tensor. The Levi-Civita derivative $\nabla$ is determined by the Christoffel symbols
\begin{equation}
 \Gamma^\mu{}_{\alpha\beta}
 =\frac{1}{2}g^{\mu\nu}
 \left(\partial_\alpha g_{\nu\beta}
      +\partial_\beta g_{\nu\alpha}
      -\partial_\nu g_{\alpha\beta}\right).
 \label{eq:christoffel}
\end{equation}

The matter is described by a scalar distribution function $f(x,u)$, weighted by particle rest mass, on the future unit mass shell $\mathcal P_x=\{u\in T_x\mathcal M:g_{\mu\nu}u^\mu u^\nu=-1,\ u^0>0\}$, where $T_x\mathcal{M}$ is the tangent space at the spacetime point $x$. Expressed in terms of all four contravariant components of the four-velocity, the invariant momentum-space measure is
\begin{equation}
 \begin{aligned}
 d\Pi_x&=2\theta(u^0)\,
 \delta\!\left(1+g_{\mu\nu}u^\mu u^\nu\right)
 \sqrt{-g}\,d^4u,\\
 &=\frac{d^3u}{\sqrt{-g}\,u^0}
  =\frac{d^3u}{\sqrt{\gamma}\,W},
 \end{aligned}
 \label{eq:mass-shell-measure}
\end{equation}
where $\theta$ is the Heaviside function and $d^4u=du^0\,du^1\,du^2\,du^3$. The second line follows by changing to the covariant components $u_\mu=g_{\mu\nu}u^\nu$, using
\begin{equation}
    \sqrt{-g}\,d^4u=\frac{du_0\,du_1\,du_2\,du_3}{\sqrt{-g}},
\end{equation}
and integrating the delta function over $u_0$ on the future-directed branch. Here $d^3u\equiv du_1\,du_2\,du_3$, and $W=-n_\mu u^\mu=\alpha u^0$ is the Lorentz factor measured by observers normal to the spatial slices. We normalize $f$ so that the rest mass carried by particles crossing a slice element $d\Sigma$, with four-velocities in the momentum-space element $d\Pi_x$, is $dM_0=fW\,d\Sigma\,d\Pi_x=f\,d^3x\,d^3u$ \cite{Sarbach:2013fya,Andersson:2019yxx}. In the absence of nongravitational forces, $f$ satisfies the relativistic Vlasov equation,
\begin{equation}
 u^\mu\frac{\partial f}{\partial x^\mu}
 -\Gamma^\mu{}_{\alpha\beta}u^\alpha u^\beta
       \frac{\partial f}{\partial u^\mu}=0.
 \label{eq:vlasov}
\end{equation}

Given a distribution function, $f$, we can introduce a natural expansion of momentum space using the method of moments \cite{Sarbach:2013fya,Acuna-Cardenas:2021nkj}. These moments define a hierarchy of symmetric spacetime tensors,
\begin{equation}
 \mathcal M^{\mu_1\cdots\mu_\ell}(x)
 =\int_{\mathcal P_x} f(x,u)\,
 u^{\mu_1}\cdots u^{\mu_\ell}\,d\Pi_x,
 \label{eq:kinetic-moments}
\end{equation}
where $\ell\geq0$ is an integer and the empty product for $\ell=0$ is defined as unity. With the rest-mass normalization adopted above, the first moment is the rest-mass current,
\begin{equation}
 J^\mu=\mathcal M^\mu
 =\int_{\mathcal P_x}f u^\mu\,d\Pi_x,
 \qquad \nabla_\mu J^\mu=0,
 \label{eq:mass-current}
\end{equation}
where current conservation follows by integrating Eq.~\eqref{eq:vlasov} with vanishing momentum-space boundary flux. The rest-mass density measured by normal observers is $\rho_0=-n_\mu J^\mu$, and the total rest mass on a slice is $M_0=-\int n_\mu J^\mu\,d\Sigma$. The second moment supplies the stress-energy tensor in Eq.~\eqref{eq:einstein},
\begin{equation}
 \begin{aligned}
 T^{\mu\nu}(x)&=\mathcal M^{\mu\nu}
 =\int_{\mathcal P_x}f u^\mu u^\nu\,d\Pi_x\\
 &=\frac{1}{\sqrt{-g}}
 \int f(x,u_i)\frac{u^\mu u^\nu}{u^0}\,d^3u.
 \end{aligned}
 \label{eq:vlasov-stress-energy}
\end{equation}
This allows us to connect the stress-energy of the massive particle distribution function to the matter source in the Einstein equations \eqref{eq:einstein}.

In the particle-based approach presented here, we do not solve for a continuum distribution function (see, e.g., Refs. \cite{2018JCoPh.353..110J,Juno:2026qms} for direct Vlasov approaches used in plasma physics). As such, we need to convert the Vlasov equation \eqref{eq:vlasov} into a discrete form.
To construct a particle representation covariantly, we define the scalar distribution on the full tangent bundle by $\mathcal F=2\theta(u^0)\delta(1+g_{\mu\nu}u^\mu u^\nu)f$, with invariant volume $d\Omega=(-g)\,d^4x\,d^4u$. Its Klimontovich representation \cite{Gratus:2008dd} is the worldline distribution
\begin{equation}
 \begin{aligned}
 \mathcal F_{\rm K}(x,u)
 &=\sum_{p=1}^{N}m_p\int d\tau\\
 &\quad\times\frac{\delta^{(4)}\!\left(x-z_p(\tau)\right)
       \delta^{(4)}\!\left(u-u_p(\tau)\right)}{-g(x)},
 \end{aligned}
 \label{eq:klimontovich-distribution}
\end{equation}
where $z_p^\mu(\tau)$ and $u_p^\mu(\tau)$ are the particle worldline and four-velocity, and $m_p$ is its fixed rest-mass weight. {$N$ is the particle count and $\tau$ is proper time.} The spacetime and tangent-fiber delta functions each carry a normalization factor $1/\sqrt{-g}$, yielding the combined factor $1/(-g)$ in Eq.~\eqref{eq:klimontovich-distribution}. Equivalently, for a smooth test scalar $\Phi$, $\int\mathcal F_{\rm K}\Phi\,d\Omega=\sum_p m_p\int\Phi(z_p,u_p)\,d\tau$. Geodesic flow preserves $d\Omega$. When the particle worldlines follow this flow and their weights remain constant, $\mathcal F_{\rm K}$ satisfies the general-relativistic Klimontovich equation
\begin{equation}
 u^\mu\frac{\partial\mathcal F_{\rm K}}{\partial x^\mu}
 -\Gamma^\mu{}_{\alpha\beta}u^\alpha u^\beta
       \frac{\partial\mathcal F_{\rm K}}{\partial u^\mu}=0,
 \label{eq:klimontovich}
\end{equation}
in the distributional sense. We solve this equation by evolving each particle contributing to Eq.~\eqref{eq:klimontovich-distribution} along a geodesic,
\begin{equation}
 \begin{aligned}
 \frac{dz_p^\mu}{d\tau}&=u_p^\mu,\\
 \frac{du_p^\mu}{d\tau}
 &=-\Gamma^\mu{}_{\alpha\beta}(z_p)\,
 u_p^\alpha u_p^\beta.
 \end{aligned}
 \label{eq:kinetic-particle-geodesics}
\end{equation}
Thus the support of each phase-space delta moves with the geodesic flow, while its rest-mass weight remains constant. These trajectories determine the time-dependent Klimontovich distribution, which we next express on constant-$t$ slices.

Resolving the mass-shell delta function and parametrizing the worldlines by coordinate time gives $\mathcal F_{\rm K}=2\theta(u^0)\delta(1+g_{\mu\nu}u^\mu u^\nu)\bar f_{\rm K}$, where
\begin{equation}
 \begin{aligned}
 \bar f_{\rm K}(t,\boldsymbol{x},u_i)
 &=\sum_{p=1}^{N}m_p
   \delta^{(3)}\!\left(\boldsymbol{x}-\boldsymbol{x}_p\right)\,\times\prod_{i=1}^{3}\delta\!\left(u_i-u_{i,p}\right).
 \end{aligned}
 \label{eq:klimontovich-canonical}
\end{equation}
The positions and momenta appearing in Eq.~\eqref{eq:klimontovich-canonical} are evolved using Eq.~\eqref{eq:kinetic-particle-geodesics}. Setting $t=z_p^0(\tau)$, with inverse $\tau_p(t)$, and defining $x_p^i(t)=z_p^i(\tau_p(t))$ and $u_{i,p}=g_{i\mu}(z_p)u_p^\mu$, we obtain the coordinate-time particle equations
\begin{equation}
 \begin{aligned}
 \frac{dx_p^i}{dt}
 &=\left.\frac{\partial H}{\partial u_i}\right|_p
 =\frac{u_p^i}{u_p^0},\\
 \frac{du_{i,p}}{dt}
 &=-\left.\frac{\partial H}{\partial x^i}\right|_p
 =\frac{u_p^\alpha u_p^\beta}{2u_p^0}
       \left.\partial_i g_{\alpha\beta}\right|_{z_p},
 \end{aligned}
 \label{eq:kinetic-characteristics}
\end{equation}
Here $H(t,\boldsymbol{x},u_i)=-u_0$ is the reduced geodesic Hamiltonian per unit rest mass for evolution in coordinate time, with $u_0$ determined by the mass-shell condition on the future-directed branch. The notation $|_p$ denotes evaluation at $(t,\boldsymbol{x}_p,u_{i,p})$.

The stress-energy source follows by inserting the Klimontovich distribution into the second moment, Eq.~\eqref{eq:vlasov-stress-energy}. Using the full four-velocity representation, the velocity-space delta function evaluates the moment at each particle's four-velocity,
\begin{equation}
 \begin{aligned}
 T^{\mu\nu}(x)
 &=\sqrt{-g}\int\mathcal F_{\rm K}(x,u)\,u^\mu u^\nu\,d^4u\\
 &=\sum_{p=1}^{N}m_p\int
 u_p^\mu u_p^\nu
 \frac{\delta^{(4)}\!\left(x-z_p(\tau)\right)}{\sqrt{-g}}
 \,d\tau.
 \end{aligned}
 \label{eq:klimontovich-stress-worldline}
\end{equation}
Since $dz_p^0/d\tau=u_p^0>0$, the remaining temporal delta function obeys $\delta(t-z_p^0(\tau))=\delta(\tau-\tau_p(t))/u_p^0(\tau_p(t))$, where $z_p^0(\tau_p(t))=t$. Integrating over proper time therefore gives
\begin{equation}
 \begin{aligned}
 T^{\mu\nu}(t,\boldsymbol{x})
 &=\frac{1}{\sqrt{-g}}
 \sum_{p=1}^{N}\frac{m_p u_p^\mu u_p^\nu}{u_p^0}
 \delta^{(3)}\!\left(\boldsymbol{x}-\boldsymbol{x}_p\right)\\
 &=\frac{1}{\sqrt{\gamma}}
 \sum_{p=1}^{N}\frac{m_p u_p^\mu u_p^\nu}{W_p}
 \delta^{(3)}\!\left(\boldsymbol{x}-\boldsymbol{x}_p\right),
 \end{aligned}
 \label{eq:klimontovich-stress-slice}
\end{equation}
where the particle quantities are evaluated at $\tau_p(t)$ and $W_p=\alpha u_p^0$. The same expression follows by inserting the canonical particle sum, Eq.~\eqref{eq:klimontovich-canonical}, into Eq.~\eqref{eq:vlasov-stress-energy}. Equation~\eqref{eq:klimontovich-stress-slice} supplies the source for Einstein's equation~\eqref{eq:einstein}.

We have now assembled the main mathematical formalism to describe a collisionless self-gravitating system in full general relativity. In order to solve it numerically, we will use the NRPIC method, summarized in Figure~\ref{fig:nrpic-cycle}. The mesh geometry of Eq.~\eqref{eq:adm-metric} and its derivatives are interpolated to the particles for the geodesic momentum and position updates of Eq.~\eqref{eq:kinetic-characteristics}, as described in Section~\ref{sec:pusher}. These updates advance the particle ensemble in Eq.~\eqref{eq:klimontovich-canonical} and thereby solve the Klimontovich equation~\eqref{eq:klimontovich}. Conservative stress-energy deposition then evaluates the discrete stress-energy moment in Eq.~\eqref{eq:klimontovich-stress-slice} on the mesh, supplying the matter sources (Section~\ref{sec:stree-energy-deposition}). The spacetime update uses these sources to solve Einstein's equation~\eqref{eq:einstein} in the Z4c formulation \cite{Hilditch:2012fp}(Sec. \ref{sec:z4c}) in moving-puncture gauge \cite{Campanelli:2005dd,Baker:2005vv}. The AMR stage in Fig.~\ref{fig:nrpic-cycle} refines or coarsens the computational mesh according to the geometry, redistributes particles without changing their rest-mass weights, rebuilds the geometry, and reconstructs the matter sources with conservative deposition across refinement boundaries (Section~\ref{sec:amr}). Constraint, horizon, and gravitational-wave diagnostics monitor the evolution, while optional excision removes selected particles and their contributions to Eq.~\eqref{eq:klimontovich-stress-slice} (Section~\ref{sec:z4c}). In the following, we will describe each of these components individually.

\subsection{Spacetime evolution}
\label{sec:z4c}

We solve the Einstein equations \eqref{eq:einstein} using the Z4c formulation \cite{Hilditch:2012fp}. The underlying Z4 formulation introduces an additional constraint four-vector $Z_\mu$ \cite{Bona:2003fj}, together with damping terms that control violations of the constraints \cite{Gundlach:2005eh},
\begin{align}
 &G_{\mu\nu}+2\nabla_{(\mu}Z_{\nu)}
       -g_{\mu\nu}\nabla_\rho Z^\rho
 \notag\\
 &-\kappa_1\left[2n_{(\mu}Z_{\nu)}
       +\kappa_2g_{\mu\nu}n_\rho Z^\rho\right]
 =8\pi T_{\mu\nu}.
 \label{eq:z4-einstein}
\end{align}
{The parameters $\kappa_1$ and $\kappa_2$ control constraint damping.} Here $\nabla_\mu$ is the Levi-Civita derivative associated with $g_{\mu\nu}$, and $n^\mu$ is the future-directed unit normal to the spatial slices. The physical constraint surface is $Z_\mu=0$, on which Eq.~\eqref{eq:z4-einstein} reduces to Eq.~\eqref{eq:einstein}. The Z4c formulation uses a conformal $3+1$ decomposition with constraint additions chosen to retain the propagating Z4 constraints while keeping the evolution system close to BSSNOK~\cite{Hilditch:2012fp}.
These properties are particularly useful in particle simulations, where finite sampling and particle-to-mesh deposition can introduce localized constraint violations that need cleaning. In a regular PIC algorithm, this step is equivalent to a hyperbolic relaxation of the Gauss constraint \cite{2000JCoPh.161..484M}. In addition, we also find it advantageous to add a parabolic constraint relaxation term \cite{Duez:2002bn,Etienne:2024ncu}, an approach also used in fully relativistic smoothed particle hydrodynamics \cite{Biswas:2026wph}. The equations below describe the baseline Z4c system.

Starting from the spatial metric $\gamma_{ij}$ and extrinsic curvature $K_{ij}$, the conformal variables are
\begin{align}
  \chi &= \gamma^{-1/3},
  &\widetilde{\gamma}_{ij}&=\chi\gamma_{ij},
  \\
  \widetilde{A}_{ij}
  &=\chi\left(K_{ij}-\frac{1}{3}\gamma_{ij}K\right),
  &\widehat{K}&= K-2\Theta,
  \label{eq:z4c-variables}
\end{align}
where $\gamma=\det\gamma_{ij}$, $K=\gamma^{ij}K_{ij}$, and $\Theta=-n^\mu Z_\mu$ is the normal projection of the Z4 constraint vector. These quantities are evolved together with the conformal connection functions $\widetilde{\Gamma}^{i}$. The Z4c system includes constraint-damping terms controlled by $\kappa_1$ and $\kappa_2$, while the algebraic conditions $\det\widetilde{\gamma}_{ij}=1$ and $\widetilde{\gamma}^{ij}\widetilde{A}_{ij}=0$ are enforced during evolution. Collisionless matter enters through the energy density, momentum density, and spatial stress measured by normal observers,
\begin{align}
    {E}&=n^\mu n^\nu T_{\mu\nu},\notag\\
    S_i&=-\gamma_i{}^\mu n^\nu T_{\mu\nu},\notag\\
    S_{ij}&=\gamma_i{}^\mu\gamma_j{}^\nu T_{\mu\nu},
\end{align}
where $n^\mu$ is the future-directed unit normal to the spatial slices and $\gamma_i{}^\mu$ is the spatial projector. These sources are reconstructed from the particles through the deposition procedure described in Section \ref{sec:stree-energy-deposition}.

{Here $D_i$ is the covariant derivative of $\gamma_{ij}$, and $R_{ij}$ and $R$ are its Ricci tensor and scalar.} With $S=\gamma^{ij}S_{ij}$, the conformal metric variables evolve according to~\cite{Hilditch:2012fp}
\begin{align}
 \partial_t\chi
 &=\frac{2}{3}\chi
 \left[\alpha(\widehat K+2\Theta)-D_i\beta^i\right],
 \label{eq:z4c-chi}\\
 \partial_t\widetilde{\gamma}_{ij}
 &=-2\alpha\widetilde A_{ij}
 +\beta^k\partial_k\widetilde\gamma_{ij}
 +2\widetilde\gamma_{k(i}\partial_{j)}\beta^k
 -\frac{2}{3}\widetilde\gamma_{ij}\partial_k\beta^k.
 \label{eq:z4c-gammatilde}
\end{align}
The extrinsic-curvature variables satisfy
\begin{align}
 \partial_t\widehat K
 &=-D^iD_i\alpha
 +\alpha\left[\widetilde A_{ij}\widetilde A^{ij}
 +\frac{1}{3}(\widehat K+2\Theta)^2\right]
 \nonumber\\
 &\quad+4\pi\alpha(S+{E})
 +\alpha\kappa_1(1-\kappa_2)\Theta
 +\beta^i\partial_i\widehat K,
 \label{eq:z4c-khat}\\
 \partial_t\widetilde A_{ij}
 &=\chi\left[-D_iD_j\alpha
 +\alpha(R_{ij}-8\pi S_{ij})\right]^{\rm TF}
 \nonumber\\
 &\quad+\alpha\left[(\widehat K+2\Theta)\widetilde A_{ij}
 -2\widetilde A^k{}_i\widetilde A_{kj}\right]
 \nonumber\\
 &\quad+\beta^k\partial_k\widetilde A_{ij}
 +2\widetilde A_{k(i}\partial_{j)}\beta^k
 -\frac{2}{3}\widetilde A_{ij}\partial_k\beta^k.
 \label{eq:z4c-atilde}
\end{align}
{The superscript ${\rm TF}$ denotes the trace-free part with respect to $\gamma_{ij}$.}

The remaining Z4c fields obey
\begin{widetext}
\begin{align}
 \partial_t\widetilde\Gamma^i
 &=-2\widetilde A^{ij}\partial_j\alpha
 +2\alpha\left[
 \widetilde\Gamma^i{}_{jk}\widetilde A^{jk}
 -\frac{3}{2}\widetilde A^{ij}\partial_j\ln\chi
 -\frac{1}{3}\widetilde\gamma^{ij}\partial_j(2\widehat K+\Theta)
 -8\pi\widetilde\gamma^{ij}S_j\right]
 \nonumber\\
 &\quad+\widetilde\gamma^{jk}\partial_j\partial_k\beta^i
 +\frac{1}{3}\widetilde\gamma^{ij}\partial_j\partial_k\beta^k
 +\beta^j\partial_j\widetilde\Gamma^i
 -(\widetilde\Gamma_{\rm d})^j\partial_j\beta^i
 +\frac{2}{3}(\widetilde\Gamma_{\rm d})^i\partial_j\beta^j
 -2\alpha\kappa_1
 \left[\widetilde\Gamma^i-(\widetilde\Gamma_{\rm d})^i\right],
 \label{eq:z4c-gamma}\\
 \partial_t\Theta
 &=\frac{\alpha}{2}\left[
 R-\widetilde A_{ij}\widetilde A^{ij}
 +\frac{2}{3}(\widehat K+2\Theta)^2\right]
 -\alpha\left[8\pi{E}
 +\kappa_1(2+\kappa_2)\Theta\right]
 +\beta^i\partial_i\Theta,
 \label{eq:z4c-theta}
\end{align}
\end{widetext}
where
$(\widetilde\Gamma_{\rm d})^i
\equiv\widetilde\gamma^{jk}\widetilde\Gamma^i{}_{jk}$ is constructed from the conformal metric, {with $\widetilde\Gamma^i{}_{jk}$ its Christoffel symbols.} The damping parameters $\kappa_1$ and $\kappa_2$ suppress violations of the Z4 constraints.

The spatial Ricci terms in Eqs.~\eqref{eq:z4c-atilde} and \eqref{eq:z4c-theta} are evaluated using the decomposition
\begin{equation}
    R_{ij}=R^\chi{}_{ij}+\widetilde R_{ij},
 \label{eq:z4c-ricci-split}
\end{equation}
where
\begin{align}
 R^\chi{}_{ij}
 &=\frac{1}{2\chi}\widetilde D_i\widetilde D_j\chi
 +\frac{1}{2\chi}\widetilde\gamma_{ij}
   \widetilde D^k\widetilde D_k\chi
 \nonumber\\
 &\quad-\frac{1}{4\chi^2}\widetilde D_i\chi\widetilde D_j\chi
 -\frac{3}{4\chi^2}\widetilde\gamma_{ij}
   \widetilde D^k\chi\widetilde D_k\chi,
 \label{eq:z4c-rchi}\\
 \widetilde R_{ij}
 &=-\frac{1}{2}\widetilde\gamma^{lm}
   \partial_l\partial_m\widetilde\gamma_{ij}
 +\widetilde\gamma_{k(i}\partial_{j)}\widetilde\Gamma^k
 +(\widetilde\Gamma_{\rm d})^k\widetilde\Gamma_{(ij)k}
 \nonumber\\
 &\quad+\widetilde\gamma^{lm}\left(
 2\widetilde\Gamma^k{}_{l(i}\widetilde\Gamma_{j)km}
 +\widetilde\Gamma^k{}_{im}\widetilde\Gamma_{klj}\right),
 \label{eq:z4c-rtilde}
\end{align}
where $\widetilde D_i$ is compatible with $\widetilde\gamma_{ij}$ and $\widetilde\Gamma_{ijk}=\widetilde\gamma_{i\ell}\widetilde\Gamma^\ell{}_{jk}$. Indices on conformal quantities are raised and lowered with $\widetilde\gamma_{ij}$.
The expression for $R_{ij}$ uses the independently evolved conformal connection functions, with $R=\gamma^{ij}R_{ij}$. These expressions coincide with the Ricci tensor and scalar of the physical spatial metric when the connection constraint is satisfied. The continuum constraint surface is defined by $\Theta=0$ and $2\widetilde Z^i\equiv\widetilde\Gamma^i-(\widetilde\Gamma_{\rm d})^i=0$, together with the Hamiltonian and momentum constraints.

We close the system with moving-puncture gauge conditions, combining Bona–Massó slicing \cite{Bona:1994dr} with a $\Gamma$-driver shift. In the Z4c implementation, these can be written as \cite{Hilditch:2012fp}
\begin{align}
 \partial_t\alpha
 &=-\alpha^2\mu_L\widehat K+\beta^i\partial_i\alpha,
 \label{eq:z4c-lapse}\\
 \partial_t\beta^i
 &=\alpha^2\mu_S\widetilde\Gamma^i-\eta\beta^i
 +\beta^j\partial_j\beta^i,
 \label{eq:z4c-shift}
\end{align}
where the choice $\mu_L=2/\alpha$ gives $1+\log$ slicing \cite{Alcubierre:2002kk} and $\mu_S$ controls the
shift-driver strength. We also implement a $\Gamma$-driver that explicitly
evolves an auxiliary field $B^i$ \cite{vanMeter:2006vi},
\begin{align}
(\partial_t-\beta^j\partial_j)\beta^i&=\frac{3}{4}B^i,\notag\\
(\partial_t-\beta^j\partial_j)B^i&=(\partial_t-\beta^j\partial_j)\widetilde{\Gamma}^i-\eta(\boldsymbol{x},t)B^i.
\end{align}

For some gravitational-collapse problems, we find it advantageous to use a shift-damping coefficient that depends on the tracked black-hole positions and masses, adapting the prescription in Ref. \cite{Muller:2010zze}.
This gauge allows the coordinates to respond to gravitational collapse and subsequent motion of black holes with a mass hierarchy. For spatially dependent damping, we use
\begin{equation}
    \eta(\boldsymbol{x},t)=\eta_0\left[1+\sum_A\frac{M_{\rm ref}/m_A -1}{1+|\boldsymbol{x}-\boldsymbol{x}_A(t)|^2/(4m_{\rm min}^2)}\right],
\end{equation}
where $\boldsymbol{x}_A(t)$ is the position of black-hole tracker $A$, $m_A$ is the mass of each black hole, $m_{\rm min}$ is the minimum of $m_A$, and $M_{\rm ref}$ is the fixed reference mass defining the simulation units. The profile follows the tracked objects and approaches $\eta_0$ far from them. When multiple black holes form, we enable this spatial dependence and apply the same dimensionless profile to the Z4c damping coefficient
\begin{equation}
    \kappa_1(\boldsymbol{x},t)=\kappa_{1,0}\,\frac{\eta(\boldsymbol{x},t)}{\eta_0},
\end{equation}
where $\kappa_{1,0}$ is the baseline constraint-damping coefficient.

\subsection{Particle pusher}
\label{sec:pusher}
The particle push advances the six canonical variables $(x^i,u_i)$ through the interpolated spacetime. We use two implicit updates: a midpoint discretization of the coordinate equations \cite{Bacchini:2020cze} and a spacetime Boris scheme based on the DGREM force decomposition \cite{Olivares:2021cfa}. Both evolve the characteristics of Eq.~\eqref{eq:kinetic-characteristics}.

The covariant equations of motion are
\begin{align}
    \frac{\text{d}x^\mu}{\text{d}\tau}&=u^\mu,\label{eq:4vel}\\
    u^\nu \nabla_\nu u_\mu&=f_\mu,\label{eq:geodesic}
\end{align}
where $\tau$ is proper time and $f_\mu$ is the nongravitational force per unit rest mass. The tests below have $f_\mu=0$. In the $3+1$ metric of Eq.~\eqref{eq:adm-metric}, the position evolves according to
\begin{equation}
    \frac{\mathrm{d}x^i}{\mathrm{d}t}=\alpha\frac{\gamma^{ij}u_j}{W}-\beta^i,\label{eq:position_eq}
\end{equation}
with Eulerian Lorentz factor $W=-n_\mu u^\mu=\alpha u^0=\sqrt{1+\gamma^{ij}u_i u_j}$. The covariant spatial momentum obeys
\begin{equation}
    \frac{\mathrm{d} u_i}{\mathrm{d} t} = -W\partial_i\alpha
     + u_j\partial_i\beta^j
     -\frac{\alpha}{2W}u_j u_k\partial_i\gamma^{jk}.\label{eq:momentum_eq}
\end{equation}
For a charged particle, the electromagnetic force is
\begin{equation}
    f_\mu=\frac{q}{m}F_{\mu\nu}u^\nu,
\end{equation}
{where $q$ is charge, $m$ is rest mass, and $F_{\mu\nu}$ is the electromagnetic tensor. It contributes} the coordinate-time acceleration
\begin{equation}
 \begin{aligned}
 \mathfrak f_i^{\rm EM}\equiv\frac{f_i}{u^0}
 &=\alpha\frac{q}{m}\gamma_{ij}E^j\\
 &\quad+\alpha\frac{q}{m}e_{ijk}\frac{\gamma^{jl}u_l}{W}B^k.
 \end{aligned}
 \label{eq:lorentz-force}
\end{equation}
Here $E^i$ and $B^i$ are Eulerian fields and $e_{ijk}=\sqrt{\gamma}[ijk]$. The conversion from proper to coordinate time supplies $1/u^0=\alpha/W$.

GR-Boris schemes commonly separate the electromagnetic and geometric forces. In Ref.~\cite{Bacchini:2020cze}, the update is a Strang composition,
\begin{equation}
 \mathcal P_{\Delta t}
 =\mathcal B_{\Delta t/2}\circ\mathcal G_{\Delta t}
  \circ\mathcal B_{\Delta t/2},
 \label{eq:gr-boris-split}
\end{equation}
where $\mathcal B_h$ applies the electromagnetic Boris push in a local orthonormal frame and $\mathcal G_h$ advances the geodesic equations by implicit midpoint. This construction is second order and stores position and momentum at the same time level. The Lorentz--geodesic split is separate from the electric kicks and magnetic rotation within $\mathcal B_h$.

For the uncharged matter considered here, the update reduces to geodesic motion. We integrate this motion directly, either by midpoint or by incorporating the geometric force into the Boris rotation.

\subsubsection{Implicit midpoint pusher}
The midpoint pusher stores $(x^i,u_i)$ at integer times. We define
\begin{equation}
    \overline{x}^i=\frac{x^{i,n+1}+x^{i,n}}{2},\qquad \overline{u}_i=\frac{u_i^{n+1}+u_i^n}{2}.
\end{equation}
The update reads
\begin{equation}
    \frac{x^{i,n+1}-x^{i,n}}{\Delta t}=\overline{\alpha}\frac{\overline{\gamma}^{ij}\overline{u}_j}{\overline{W}}-\overline{\beta}^i,
\end{equation}
\begin{equation}
    \frac{u_i^{n+1}-u_i^n}{\Delta t}=-\overline{W}\partial_i\overline{\alpha}+\overline{u}_j\partial_i\overline{\beta}^j-\frac{\overline{\alpha}}{2\overline{W}}\overline{u}_j\overline{u}_k\partial_i\overline{\gamma}^{jk}
\end{equation}
where $\overline W=\sqrt{1+\overline\gamma^{ij}\overline u_i\overline u_j}$. The ADM fields are interpolated to $\overline x^i$ at $t^n$ and $t^{n+1}$ and averaged in time. We obtain $\overline\gamma^{ij}$ by inverting the averaged spatial metric. The field values and their spatial derivatives use the same Hermite interpolation, specified in Section~\ref{sec:pusher-reconstruction}.

The six endpoint components enter both right-hand sides and are solved together. {We use fixed-point iteration, stopping when the maximum absolute change in any endpoint component falls below $10^{-7}$ in code units.} For smooth geometry and converged iterations, the update is second order in coordinate time. Implicit midpoint is also used for the gravitational substeps of GRPIC schemes~\cite{Zou:2024prf}.

\subsubsection{Spacetime Boris pusher}
To formulate a Boris-like pusher for the geodesic equation, we first briefly sketch the DGREM formalism \cite{Olivares:2021cfa}. DGREM recasts general relativity into a local tetrad observer form, where curved spacetime is recast into deformations of a reference observer akin to elastodynamics.

First, we define a local tetrad basis, $A^{\hat{\alpha}}{}_\mu$, such that
\begin{align}
    A^\mu{}_{\hat{\alpha}} A^\nu{}_{\hat{\beta}} g_{\mu\nu} = \eta_{\hat{\alpha}\hat{\beta}}\,,\label{eq:def-tetrad}
\end{align}
where $\eta_{\hat{\alpha}\hat{\beta}}$ is the Minkowski metric {and $A^\mu{}_{\hat\alpha}$ is the inverse coframe.} Here, the hatted index $\hat{\alpha}$ denotes the locally flat tetradic frame. In order to ensure proper transformation between different local tetrad bases, the covariant derivatives need to be augmented by a spin connection $\omega^{\hat{\beta}}{}_{\hat{\alpha}\mu}$, which plays a role analogous to that of the Christoffel symbol, so that
\begin{align}
    D_\mu V^\nu{}_{\hat{\alpha}} = \partial_\mu V^\nu{}_{\hat{\alpha}} + \Gamma^\nu{}_{\mu\kappa} V^\kappa{}_{\hat{\alpha}} - \omega^{\hat{\beta}}{}_{\hat{\alpha}\mu} V^\nu{}_{\hat{\beta}}\,,
\end{align}
is fully covariant. Next, we treat the tetrad basis as a set of vector potentials and introduce the field strength tensor
\begin{equation}
    F^{\hat{\alpha}}{}_{\mu\nu} = \partial_\mu A^{\hat{\alpha}}{}_\nu - \partial_\nu A^{\hat{\alpha}}{}_\mu\,
\end{equation}
where $F^{\hat{\alpha}}{}_{\mu\nu}$ plays the role of the Faraday tensor in electromagnetism and can be projected to define gravitational electric and magnetic fields with respect to the normal observer
\begin{align}
    B^{\hat{\alpha}\mu} &= -n_\lambda {}^{\ast}\!F^{\hat{\alpha}\mu\lambda}\label{eq:B_tensor},\\
E^{\hat{\alpha}\mu} &= n_\lambda F^{\hat{\alpha}\mu\lambda}\,\label{eq:E_tensor},
\end{align}
where ${}^{\ast}F^{\hat{a}\mu\nu}$ is the dual tensor. Similarly, we can project the Nester-Witten form \cite{Frauendiener:2006dx} constructed from the spin connection as
\begin{equation}
    {}^\ast\!\mathcal{U}^{\hat{\beta}\hat{\gamma}}{}_{\hat{\alpha}} = \omega^{[\hat{\beta}\hat{\gamma}]}{}_{\hat{\alpha}} + \delta^{\hat{\beta}}{}_{\hat{\alpha}} \omega^{[\hat{\gamma}\hat{\delta}]}{}_{\hat{\delta}} - \delta^{\hat{\gamma}}{}_{\hat{\alpha}} \omega^{[\hat{\beta}\hat{\delta}]}{}_{\hat{\delta}}\,,
\end{equation}
onto auxiliary electric and magnetic fields
\begin{align}
    H^{\hat{\alpha}\mu} \left[E,B\right] &= n_\lambda \mathcal{U}^{\mu\lambda\hat{\alpha}}\,,\\
D^{\hat{\alpha}\mu} \left[E,B\right]&= n_\lambda {}^{\ast}\!\mathcal{U}^{\mu\lambda\hat{\alpha}}\,,
\end{align}
From these fields, we can show that the Einstein equations can be written in a form resembling Maxwell's equations \cite{Boyeneni:2025tsx}
\begin{align}
\partial_t {\sqrt{\gamma} {\bf B}}^{\hat{\alpha}} + \nabla\times \left[ {\bf {E}}^{\hat{\alpha}}-{\boldsymbol \beta}\times{\bf B}^{\hat{\alpha}}\right] &=0\,\\ \label{eq:maxwell_equation_vector_form}
\partial_t {\sqrt{\gamma}{\bf D}}_{\hat{\alpha}} - \nabla\times \left[{\bf {H}}_{\hat{\alpha}}-{\boldsymbol \beta}\times{\bf D}_{\hat{\alpha}} \right] &= - \alpha \left[{\bf j}_{\hat{\alpha}}+ 8\pi{\bf J}_{\hat{\alpha}}\right] + \boldsymbol{\beta}q_{\hat{\alpha}}\,,\\
\nabla\cdot {\bf {B}}^{\hat{\alpha}} &=0\,,\\
\nabla\cdot {\bf {D}}_{\hat{\alpha}} &= q_{\hat{\alpha}} + 8\pi Q_{\hat{\alpha}}\,, \label{eqn:poisson}
\end{align}
where $\sqrt{\gamma} = \sqrt{-g}/\alpha$, and $\alpha$ and $\boldsymbol{\beta}$ are the gravitational lapse and shift, respectively. In doing so, we have also introduced spacetime ($q_{\hat{\alpha}}/{\bf j}_{\hat{\alpha}}$) and matter ($Q_{\hat{\alpha}}/{\bf J}_{\hat{\alpha}}$) charge and current densities, which follow directly from the stress-energy tensor.

Returning to the geodesic equation, the spin connection is related to the field strength tensor by
\begin{equation}
    \omega_{\hat{a}\hat{b}\hat{c}}=-\frac{1}{2}(F_{\hat{b}\hat{a}\hat{c}}+F_{\hat{c}\hat{a}\hat{b}}-F_{\hat{a}\hat{b}\hat{c}})
\end{equation}
and the geodesic Eq.~\eqref{eq:geodesic} (assuming no external nongravitational forces) in the local tetrad frame gives
\begin{equation}
    G_{\hat{i}}\equiv\frac{1}{W}\omega_{\hat{a}\hat{i}\hat{b}}u^{\hat{a}}u^{\hat{b}}=\frac{1}{W}F_{\hat{a}\hat{i}\hat{b}}u^{\hat{a}}u^{\hat{b}}=\frac{1}{W}e^{\hat{a}}{}_{\hat{i}}u_{\hat{a}}
\end{equation}
where the final equality uses a projection into the comoving frame. The comoving electric field $e$ is related to the normal-frame electric and magnetic fields by
\begin{equation}
    e^{\hat{a}}{}_{\hat{i}}=n_{\hat{i}}(E^{\hat{a}\hat{b}}u_{\hat{b}})-(n_{\hat{b}}u^{\hat{b}})E^{\hat{a}}{}_{\hat{i}}+\varepsilon_{\hat{i}\hat{b}\hat{c}\hat{d}}n^{\hat{b}}u^{\hat{c}}B^{\hat{a}\hat{d}}
\end{equation}
Therefore, the geodesic term becomes (noting that $n^{\hat{i}}=0$ and $n_{\hat{a}}=-\delta_{\hat{a}}^{\hat{t}}$)
\begin{equation}
    G_{\hat{i}}=E^{\hat{a}}{}_{\hat{i}}u_{\hat{a}}+\varepsilon_{\hat{i}\hat{j}\hat{k}}\frac{u^{\hat{j}}}{W}B^{\hat{a}\hat{k}}u_{\hat{a}}
\end{equation}
with the same electric-plus-magnetic structure as the Lorentz force.
Here $W$ is the Eulerian Lorentz factor. The centered spacetime Boris update stores position and covariant momentum at the same integer times $(x^{i,n},u_i^n)$, where $n$ labels the time level $t^n$. Therefore, the coupled spacetime Boris pusher equations are
\begin{align}
    \frac{\mathrm{d}x^i}{\mathrm{d}t}&=\alpha\frac{\gamma^{ij}u_j}{W}-\beta^i,\label{eq:spacetime-boris-position}\\
    \frac{\mathrm{d}u_{\hat{i}}}{\mathrm{d}\hat{t}}&=\widetilde{E}_{\hat{i}} +\varepsilon_{\hat{i}\hat{j}\hat{k}}\frac{u^{\hat{j}}}{W}\widetilde{B}^{\hat{k}}.\label{eq:spacetime-boris-mom}
\end{align}
Here, the effective electric and magnetic fields include the contribution from Eq.~\eqref{eq:lorentz-force}
\begin{align}
 \widetilde E_{\hat i}&=\frac qm E_{\hat i}
   +E^{\hat a}{}_{\hat i}u_{\hat a},\nonumber\\
 \widetilde B^{\hat k}&=\frac qm B^{\hat k}
   +B^{\hat a\hat k} u_{\hat a},\label{eq:effective-EM-field}
\end{align}
and the right-hand sides are reconstructed at the midpoint $t^{n+1/2}$. Although we do not evolve Maxwell fields here, we include them to highlight the usefulness of our pusher algorithm for GRPIC applications as well. The position update Eq.~\eqref{eq:spacetime-boris-position} is solved in the global coordinate frame, while the momentum update Eq.~\eqref{eq:spacetime-boris-mom} is solved in the local orthonormal frame. The momentum components in the two frames are related by the tetrad defined in Eq.~\eqref{eq:def-tetrad}, evaluated at the corresponding time and position
\begin{equation}
    u_{i,m}=A^{\hat{a}}{}_i(t^m,x^{i,m})u_{\hat{a},m},\label{eq:tetrad_transformation}
\end{equation}
where $m\in\{n,n+1/2,n+1\}$.

Explicitly, the discretized version of Eqs.~\eqref{eq:spacetime-boris-position} and \eqref{eq:spacetime-boris-mom} can be written as
\begin{align}
 \frac{x^{i,n+1}-x^{i,n}}{\Delta t}
 &=\alpha^{n+1/2}
       \frac{u^{i,n+1/2}}{W^{n+1/2}}-\beta^{i,n+1/2},
 \label{eq:disc_pos_eq}\\
 \frac{u_{\hat i}^{n+1}-u_{\hat i}^n}{\Delta\hat t}
 &=\widetilde E_{\hat i}^{n+1/2}
  +\varepsilon_{\hat i\hat j\hat k}
     \frac{u^{\hat{j},n+1/2}}{W^{n+1/2}}\widetilde B^{\hat{k},n+1/2}.
 \label{eq:disc_mom_eq}
\end{align}
Here, $u^{i,n+1/2}=A^{i}{}_{\hat j}{}(t^{n+1/2},x^{i,n+1/2})u^{\hat{j},n+1/2}$ is the inverse transformation of Eq.~\eqref{eq:tetrad_transformation}, and the interval $\Delta\hat t=\alpha^{n+1/2} \Delta t$ is the time increment measured by Eulerian observers.

We describe the steps of the spacetime Boris pusher from time $t^n$ to $t^{n+1}$ as follows:
\begin{enumerate}
 \item At the end of cycle $n$, store the Maxwell (if applicable) and ADM variables from time $t^n$, and proceed with the Z4c spacetime evolution as described in Section \ref{sec:z4c}.
 \item At the end of cycle $n+1$, take the old particle state
       $x^{i,n},u_{i}^n$ and the stored field variables to calculate the initial tetrad $A^{\hat{a}}{}_i(t^n,x^{i,n})$ and convert
       $u_{i}^n$ to $u_{\hat i}^{n}$.
 \item Initialize guesses for $x^{i,n+1},u_{\hat i}^{n+1}$, for example
       $x^{i,n+1}=x^{i,n}+\Delta t\,\mathcal{V}^{i,n}$ and $u_{\hat i}^{n+1}=u_{\hat i}^n$, {where $\mathcal V^{i,n}=(dx^i/dt)^n$,} and form the midpoint state by averaging the old state and the current endpoint guess, with the midpoint Lorentz factor defined as
       \begin{equation}
           W^{n+1/2}=\sqrt{1+\delta^{\hat{i}\hat{j}}u_{\hat{i}}^{n+1/2}u_{\hat{j}}^{n+1/2}}.
       \end{equation}
 \item Interpolate the metric and electromagnetic fields (if applicable) at $(t^{n+1/2},x^{i,n+1/2})$. Construct the tetrad, gravitational electric and magnetic fields, and their contractions with $u_{\hat a}^{n+1/2}$ to obtain $\widetilde E^{n+1/2},\widetilde B^{n+1/2}$, according to Eq.~\eqref{eq:effective-EM-field}.
 \item Apply the Boris electric kick, Cayley rotation, and electric kick,
       \begin{align*}
        \boldsymbol u^-&=\boldsymbol u^n
                  +\tfrac12 \Delta \hat{t}\,\widetilde{\boldsymbol E}^{n+1/2},\\
        \boldsymbol u'&=\boldsymbol u^-
                            +\boldsymbol u^-\times\boldsymbol t,\\
        \boldsymbol u^+&=\boldsymbol u^-
                            +\boldsymbol u'\times\boldsymbol s,\\
        \boldsymbol u_{\mathrm{next}}^{n+1}&=\boldsymbol u^+
                  +\tfrac12 \Delta \hat{t}\,\widetilde{\boldsymbol E}^{n+1/2}.
       \end{align*}
       Here, $\boldsymbol t=\frac{\Delta\hat t}{2W^{n+1/2}}
                             \widetilde{\boldsymbol B}^{n+1/2},$ and $\boldsymbol s=\frac{2\boldsymbol t}{1+|\boldsymbol t|^2}$.
 \item Update $x_{\mathrm{next}}^{i,n+1}$ using
       Eq.~\eqref{eq:disc_pos_eq} and the same midpoint guesses.
 \item Check the position and tetrad-momentum residuals. If the iteration has not converged, retain the new endpoint guesses, recompute the midpoint state, and repeat Steps 4–7.
 \item At convergence, calculate the endpoint coframe and store
       $u_{i}^{n+1}=A^{\hat a}{}_i(t^{n+1},x^{i,n+1})u_{\hat{a}}^{n+1}$.
\end{enumerate}

Once the coupled position and momentum iteration converges, the Boris kick–rotate–kick sequence satisfies the discrete midpoint momentum equation above. The two electric half-kicks together provide the full electric impulse, while the magnetic rotation uses the average of the endpoint tetrad momenta. The resulting update is therefore an implicit midpoint method applied to the coordinate position $x^i$ and the tetrad momentum $u_{\hat i}$. Because the tetrad varies with position and time, averaging tetrad momentum components generally differs from averaging the coordinate components $u_i$. The two formulations consequently generally have different finite-timestep errors, while both retain second-order accuracy for smooth fields and sufficiently converged iterations.

\subsubsection{Metric reconstruction}
\label{sec:pusher-reconstruction}
The metric and its derivatives are reconstructed using Hermite interpolation, which preserves both the input values and their finite-difference slopes. In one dimension, for a particle position between cell centers $x_i\leq\overline x<x_{i+1}$, cell spacing $h_x$, and dimensionless offset $s_x=(\overline x-x_i)/h_x$, the four one-dimensional cubic Hermite weights are
\begin{align}
 w_{-1}(s)&=\tfrac12(-s+2s^2-s^3),\nonumber\\
 w_0(s)&=\tfrac12(2-5s^2+3s^3),\nonumber\\
 w_1(s)&=\tfrac12(s+4s^2-3s^3),\nonumber\\
 w_2(s)&=\tfrac12(-s^2+s^3).
 \label{eq:midpoint-hermite-weights}
\end{align}
In three dimensions, the reconstruction is a tensor product. That is, a metric component $q$ is reconstructed as
\begin{equation}
 q(\overline{\boldsymbol x})=
 \sum_{a,b,c=-1}^{2}w_a(s_x)w_b(s_y)w_c(s_z)q_{i+a,j+b,k+c}.
\end{equation}
Its spatial derivative follows from the same polynomial,
\begin{equation}
 (\partial_xq)(\overline{\boldsymbol x})=\frac1{h_x}
 \sum_{a,b,c=-1}^{2}w'_a(s_x)w_b(s_y)w_c(s_z)q_{i+a,j+b,k+c}.
\end{equation}
Analogous expressions hold in the other directions.


\subsection{Stress-energy deposition}
\label{sec:stree-energy-deposition}
After the particle push, the particle stress-energy is deposited onto the cell-centered spacetime grid. Because momentum is stored at integer times, the deposited $T^{\mu\nu}$ is naturally synchronized with the right-hand side of Einstein's equations.

Projecting the Klimontovich stress-energy tensor in Eq.~\eqref{eq:klimontovich-stress-slice} onto the spatial slices gives the ADM matter sources,
\begin{align}
E &\equiv n_\mu n_\nu T^{\mu\nu}=\sum_p\frac{m_p W_p}{\sqrt{\gamma}}\delta^{(3)}\left(\boldsymbol{x}-\boldsymbol{x}_p\right),\\
S_i &\equiv -\gamma_{i\mu}n_\nu T^{\mu\nu}=\sum_p\frac{m_p u_{i,p}}{\sqrt{\gamma}}\delta^{(3)}\left(\boldsymbol{x}-\boldsymbol{x}_p\right),\\
S_{ij} &\equiv \gamma_{i\mu}\gamma_{j\nu}T^{\mu\nu}=\sum_p\frac{m_p u_{i,p}u_{j,p}}{W_p\sqrt{\gamma}}\delta^{(3)}\left(\boldsymbol{x}-\boldsymbol{x}_p\right).
\end{align}
The coordinate-space delta function is approximated on the mesh by a compact particle shape function that distributes each particle’s contribution among neighboring grid points \cite{1988csup.book.....H}. A common choice is the family of B-splines, which are localized, piecewise-polynomial basis functions. On a one-dimensional uniform grid, the normalized, centered kernels can be constructed from the zeroth-degree box function,
\begin{equation}
    B_0(\xi)=
\begin{cases}
1, & -\dfrac12\leq \xi<\dfrac12,\\[4pt]
0, & \text{otherwise},
\end{cases}
\end{equation}
through repeated convolution,
\begin{equation}
    B_n(\xi)=(B_{n-1}*B_0)(\xi)=\int_{-1/2}^{1/2} B_{n-1}(\xi-\eta)\,d\eta.
\end{equation}
Here $\xi\equiv\frac{x_p-x_i}{\Delta x}$ is a dimensionless coordinate offset measured in units of the grid spacing, and $n$ denotes the polynomial degree of the grid-assignment kernel, thus the order of the shape function. Each convolution raises the degree by one and broadens the support by one grid spacing.

First-order particle shapes are employed in standard PIC implementations \cite{Buneman1993,Spitkovsky:2005vsy}, including the \texttt{TRISTAN-MP} code \cite{Parfrey:2018dnc,Sironi:2023epb} and \texttt{Entity} \cite{EntityDevelopmentTeam:2025tlp}, including its GRPIC formulation \cite{Galishnikova:2025pfi}. Higher-degree kernels are available in \texttt{WarpX} \cite{Myers:2021gfn}, which supports linear through quartic splines, and in recent extensions of \texttt{Entity}, which support degrees up to eleven \cite{Boss:2026fyj}. Increasing the degree makes the particle–mesh coupling smoother and suppresses short-wavelength numerical noise more strongly, at the expense of a wider deposition stencil and greater computational cost \cite{Boss:2026fyj}.

In this work, unless otherwise stated, we adopt a first-order shape function for deposition onto the cell-centered spacetime grid. The corresponding one-dimensional kernel is
\begin{equation}
    B_1(\xi)=
\begin{cases}
1-|\xi|, & |\xi|<1,\\[3pt]
0, & |\xi|\geq1.
\end{cases}\label{eq:first-order-shape}
\end{equation}
The three-dimensional assignment weight is constructed as a tensor product,
\begin{equation}
    S_{ijk}(\boldsymbol{x}_p)=
B_1\!\left(\frac{x_i-x_p}{\Delta x}\right)
B_1\!\left(\frac{y_j-y_p}{\Delta y}\right)
B_1\!\left(\frac{z_k-z_p}{\Delta z}\right),
\end{equation}
where $\boldsymbol{x}_p=(x_p,y_p,z_p)$ is the particle position, $(x_i,y_j,z_k)$ denotes the receiving cell center, and $\Delta x,\Delta y,\Delta z$ are the coordinate grid spacings. These weights are nonnegative, satisfy the partition-of-unity condition
\begin{equation}
    \sum_{ijk}S_{ijk}(\boldsymbol{x}_p)=1,
\end{equation}
and distribute each particle’s contribution among at most eight surrounding cell centers. The coordinate-space delta function is therefore represented discretely as
\begin{equation}
    \delta^{(3)}(\boldsymbol{x}-\boldsymbol{x}_p)\;\longrightarrow\;\frac{S_{ijk}(\boldsymbol{x}_p)}{\Delta V}, \qquad \Delta V=\Delta x\,\Delta y\,\Delta z.
\end{equation}
Substituting this representation into the ADM matter sources, with the metric volume factor evaluated at the receiving cell center, gives the deposition updates
\begin{align}
E_{ijk}&=\sum_p\frac{m_p W_p S_{ijk}(\boldsymbol{x}_p)}{\sqrt{\gamma}_{ijk}\Delta V},\\
S_{a,ijk}&=\sum_p\frac{m_p u_{a,p} S_{ijk}(\boldsymbol{x}_p)}{\sqrt{\gamma}_{ijk}\Delta V},\\
S_{ab,ijk}&=\sum_p\frac{m_p u_{a,p}u_{b,p}S_{ijk}(\boldsymbol{x}_p)}{W_p\sqrt{\gamma}_{ijk}\Delta V}.
\end{align}
Here $\sqrt{\gamma}_{ijk}$ is evaluated at the cell center where the source is deposited. When AMR is activated, we implement a mass-conserving deposition scheme described in detail in Section \ref{sec:mass-conserved-amr}.

\subsection{Numerical Implementation}
\label{sec:numerical_implementation}
We implement the NRPIC algorithm on top of the \texttt{AthenaK} framework \cite{2026ApJS..283...27S}, which provides a block-based octree mesh, a general adaptive mesh refinement framework \cite{2020ApJS..249....4S}, and infrastructure for particle storage and communication. Particles are stored on the same rank as the meshblock to which they belong. \texttt{AthenaK} uses the Kokkos library \cite{2022ITPDS..33..805T}, which makes the code portable and executable on both CPUs and GPUs, allowing the particle and spacetime calculations to use the existing parallel infrastructure. \texttt{AthenaK} also provides a cell-centered Z4c implementation \cite{Zhu:2024utz}, using up to fourth-order finite differences and fourth-order explicit Runge–Kutta time integration. Apparent horizons are located using the currently available \texttt{Fastflow} implementation in the code \cite{Gundlach:1997us}, while gravitational radiation is extracted using the outgoing Weyl scalar $\Psi_4$, its interpolation onto extraction spheres, and its decomposition into spin-weighted spherical harmonics (see, e.g., Ref. \cite{Bishop:2016lgv} for a discussion).

During each timestep, the spacetime solver first advances the geometry from $t^n$ to $t^{n+1}$ using the matter sources deposited at $t^n$, which are held fixed throughout its Runge–Kutta stages. The particle pusher then advances the ensemble over the same interval, with access to both the stored geometry at $t^n$ and the updated geometry at $t^{n+1}$. After the push, particles satisfying an enabled excision criterion are marked for removal, as discussed below, and particle communication assigns the remaining particles to their destination meshblocks and ranks. Finally, the updated particle ensemble is deposited onto the spacetime grid to construct the matter sources for the next timestep, using the deposition procedure described in Sections \ref{sec:stree-energy-deposition} and \ref{sec:amr} when AMR is activated.

To improve numerical robustness and reduce the cost of evolving particles deep inside black holes, we employ particle excision. As the appearance of an apparent horizon is slice-dependent and can happen after the event horizon has already formed \cite{Wald:1991zz}, we use a lapse-based prescription \cite{Rosswog:2023nnl} following empirically determined horizon lapse thresholds for moving puncture gauge \cite{Bernuzzi:2020txg}. Particles satisfying $\alpha(t^{n},\mathbf{x}_p^{n})<\alpha_{\mathrm{exc}}$ are marked after the push and removed before redistribution and source deposition. Apart from lapse excision, the code implementation also supports apparent-horizon-based excision. We find no significant difference between these two schemes in our benchmark tests and therefore choose lapse excision as the default.

\subsubsection{Filtering}

To reduce fluctuations in the fields, we adopt the approach common to PIC codes of filtering the deposited source terms \cite{2014CRMec.342..610V}. After deposition, we apply the binomial filter independently to each of the ten ADM source components $Q\in\{E,S_i,S_{ij}\}$. For a grid cell labeled by $(i,j,k)$, one filter pass gives
\begin{equation}
    Q^{\mathrm{filt}}_{ijk}=\sum_{a,b,c=-1}^{1} w_a w_b w_c\,Q_{i+a,j+b,k+c}
\end{equation}
where $w_0=\tfrac12,\quad w_{\pm1}=\tfrac14$. This stencil assigns weight $1/8$ to the central cell and weights $1/16$, $1/32$, and $1/64$ to each face, edge, and corner neighbor, respectively. We have also implemented higher-degree deposition kernels, such as those used in smoothed particle hydrodynamics codes \cite{Diener:2022hui}, but have found no advantage for the problems considered here.

In addition, in very challenging situations, we found that hyperbolic constraint cleaning in the Z4c system can be insufficient. We therefore optionally add a parabolic relaxation term to clean the Hamiltonian constraint following \cite{Duez:2002bn,Raithel:2022san}, by supplementing the conformal-factor evolution Eq.~\eqref{eq:z4c-chi} with a diffusion term
\begin{equation}
\partial_t\chi=\left(\partial_t\chi\right)_{\mathrm{Z4c}}+\frac{c_H}{2}\chi^{-1/4}\mathcal H ,
\end{equation}
where the Hamiltonian constraint is
\begin{equation}
    \mathcal H\equiv R+\frac{2}{3}K^2-\widetilde A_{ij}\widetilde A^{ij}-16\pi E,
\end{equation}
where $c_H\geq0$ is the effective diffusion coefficient and has dimensions of length in geometrized units. This implementation introduces an additional parabolic timestep restriction depending on $c_H$ \cite{Etienne:2024ncu}.

\subsection{Adaptive Mesh Refinement}
\label{sec:amr}
Many scenarios, including black hole formation in gravitational collapse or the interaction of multiple black holes, involve a hierarchy of scales, all of which have to be resolved down to the horizon scale in our approach.
As such, NRPIC uses adaptive mesh refinement (AMR) to resolve the contracting matter distribution and follow compact-object migration while resolving the horizons.
This functionality is already provided by the \texttt{AthenaK} framework, which uses an octree-based AMR mesh \cite{2026ApJS..283...27S,Zhu:2024utz}.

\subsubsection{Refinement criterion}
Apart from existing mesh refinement criteria, we introduce a dimensionless L\"ohner-type indicator \cite{1987CMAME..61..323L} to identify regions requiring additional spatial resolution (see also Ref. \cite{Porth:2016rfi}). The indicator measures local curvature (i.e., the second derivative) relative to the variation and amplitude of the monitored field. Empirically, we found the following combination to work well for our applications: $q=\alpha\psi^7=\alpha\chi^{-7/4}$, {where $\psi=\chi^{-1/4}$.} Let $\mathbf{i}$ denote a cell index including all three dimensions and $\boldsymbol{e}_a$ a displacement of one cell in spatial direction $a$. We define the centered difference
\begin{equation}
D_a q_{\mathbf i}=q_{\mathbf i+\mathbf e_a}-q_{\mathbf i-\mathbf e_a},
\end{equation}
and the local amplitude scale
\begin{equation}
S_{ab,\mathbf i}
=\sum_{\sigma,\tau=\pm1}
\left|q_{\mathbf i+\sigma\mathbf e_a+\tau\mathbf e_b}\right|.
\end{equation}
The refinement indicator at each cell is then \cite{Porth:2016rfi}
\begin{equation}
    E_{\mathbf i}=\sqrt{
\frac{\displaystyle\sum_{a,b=1}^{3} \left(D_a q_{\mathbf i+\mathbf e_b} -D_a q_{\mathbf i-\mathbf e_b}\right)^2}
{\displaystyle\sum_{a,b=1}^{3} \left( |D_a q_{\mathbf i+\mathbf e_b}| +|D_a q_{\mathbf i-\mathbf e_b}| +\epsilon S_{ab,\mathbf i} +\delta \right)^2}}
\end{equation}
The sums include all pairs of spatial directions, incorporating both pure and mixed second differences. The numerator measures local curvature, while the denominator normalizes this measure using neighboring first differences and field amplitudes. We set $\epsilon=0.01$ to reduce sensitivity to small fluctuations in nearly uniform regions; $\delta=10^{-99}$ is a positive regularizer that prevents a vanishing denominator.

This refinement criterion is particularly efficient and sensitive for dynamical spacetime-matter evolution and the formation of black holes during the process. For each mesh block B, we take the maximum indicator over its interior cells
$E_B=\max_{\mathbf i\in B}E_{\mathbf i}$. A block is flagged for refinement when $E_B$ exceeds the refinement threshold $0.2$ and for derefinement when it falls below the derefinement threshold $0.02$.

\subsubsection{Mass-conservative particle deposition}
\label{sec:mass-conserved-amr}
Particle tasks should be carefully adapted to be compatible with AMR to ensure mass conservation. First, when a block is refined, its particles are assigned to the children containing their positions; when children are derefined, their particles are assigned to the parent. Changes of processor ownership transfer the complete particle records. These operations leave $m_p$, $\boldsymbol{x}_p$, and $u_{i,p}$ unchanged and introduce no particle splitting or merging. Particle communication and removal for crossing block boundaries and simulation outer boundaries are described in Section \ref{sec:numerical_implementation}.

\begin{figure}[t]
    \centering
    \includegraphics[width=\linewidth]{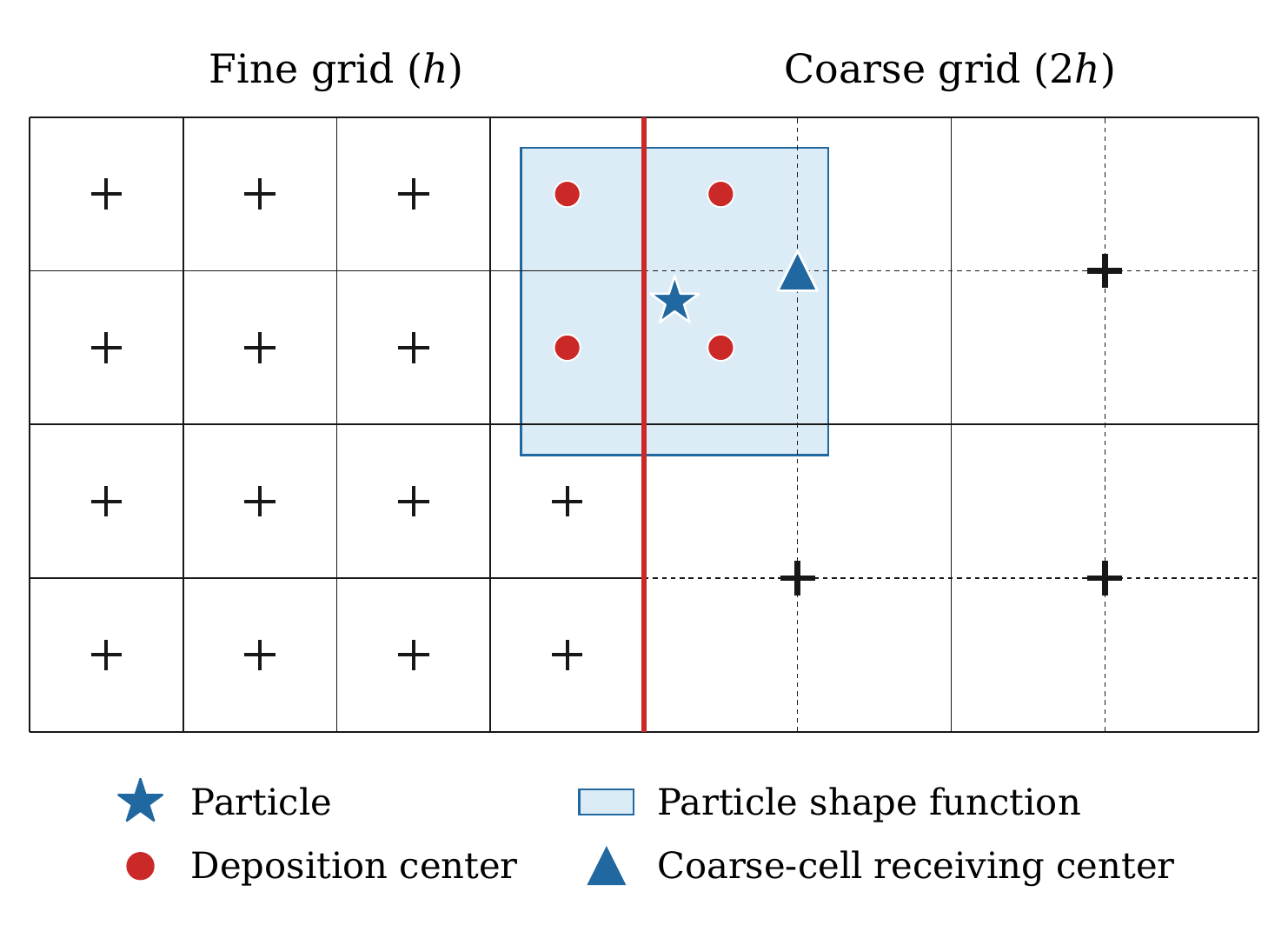}
    \caption{Schematic of conservative particle deposition across a $2{:}1$ mesh-refinement interface in 2D for visualization. The fine mesh (left) and coarse mesh (right) have spacings $h$ and $2h$, respectively, and are separated by the thick red line. Thin solid black lines denote physical cell boundaries, dashed lines indicate virtual subdivisions of the coarse cells, and black plus signs mark physical cell centers. The particle (blue star) lies on the coarse side, with its shape function evaluated at the fine-grid spacing throughout the blue shaded shape function region. Red circles mark the receiving real and virtual fine-cell centers. The two virtual-cell contributions are subsequently accumulated at their common parent coarse-cell center (blue triangle) through volume-preserving restriction, preserving the integral of each deposited source component over proper volume.}
    \label{fig:amr_particle}
\end{figure}

When AMR is activated, the particle size is changed either when the particle crosses an existing mesh refinement boundary or when its meshblock is refined or derefined. Independently evaluating a deposition weight using Eq.~\eqref{eq:first-order-shape} at each receiving block's native resolution generally fails to give weights that sum to unity on the combined mesh, violating conservation of mass. We instead represent the entire particle shape on a single refinement level, chosen as the finest level among the nearest eight grid points. In particular, a particle whose owner meshblock is coarser but encounters a finer neighbor is represented at that finer spacing, including the part of its cloud that remains within coarse blocks. This use of fine-level deposition follows the general AMR-PIC strategy used in \texttt{WarpX} \cite{Myers:2021gfn}, although here the transfer is formulated for relativistic sources on a mesh of nonoverlapping domains.

The schematic of AMR particle deposition is shown in Fig.~\ref{fig:amr_particle}. The plot shows a two-dimensional case for visualization, which can be easily generalized to three dimensions. Let $\ell_p$ denote the finest refinement level among the neighboring cell centers of particle $p$, and let $h_{\ell_p}^{a}$ denote the corresponding grid spacing in direction $a$. We evaluate the entire particle shape on a uniform lattice at this level, extending the lattice virtually into neighboring coarser blocks. The weight assigned to a real or virtual cell $f$, centered at $\mathbf{x}_f$, is
\begin{equation}
    \omega_{fp}^{(\ell_p)}=\prod_{a=1}^{3}B_1\!\left(\frac{x_f^a-x_p^a}{h_{\ell_p}^{a}}\right),
\end{equation}
where $B_1$ is the first-order shape function defined in Eq.~\eqref{eq:first-order-shape} and the weights sum to unity, $\sum_f \omega_{fp}^{(\ell_p)}=1$. All contributions from a given particle therefore use the same shape function and grid spacing, regardless of the refinement level of the receiving physical cell.

After deposition at the finest level, the effective assignment weight on a physical leaf cell $C$ is then obtained by summing the weights of the deposition-level cells contained within it,
\begin{equation}
    \omega_{Cp}^{\mathrm{AMR}}=\sum_{f\subseteq C}\omega_{fp}^{(\ell_p)}.
\end{equation}
For a receiving cell already at level $\ell_p$, this sum contains only $f=C$. For a coarser cell, it combines the contributions from its virtual fine cells, as marked by the blue triangle in Fig.~\ref{fig:amr_particle}. Because the physical leaf cells form a nonoverlapping partition of the domain, an untruncated particle deposition weight satisfies
\begin{equation}
    \sum_{C\in\mathcal L}\omega_{Cp}^{\mathrm{AMR}}=\sum_f\omega_{fp}^{(\ell_p)}=1,\label{eq:weight-unity}
\end{equation}
where $\mathcal L$ denotes the set of physical leaf cells. Consequently, the deposited sources can be expressed as
\begin{equation}
    \begin{pmatrix}E_C\\[2pt]S_{i,C}\\[2pt]S_{ij,C}\end{pmatrix}=\frac{1}{\sqrt{\gamma_C}\,\Delta V_C}\sum_p m_p\,\omega_{Cp}^{\mathrm{AMR}}
    \begin{pmatrix}W_p\\[2pt]u_{i,p}\\[2pt]u_{i,p}u_{j,p}/W_p\end{pmatrix}.\label{eq:tmunu-source-cell}
\end{equation}
Therefore, combining Eqs.~\eqref{eq:weight-unity} and \eqref{eq:tmunu-source-cell}, the stress-energy is fully conserved such that
\begin{equation}
    \sum_{C\in\mathcal L}\begin{pmatrix}E_C\\[2pt]S_{i,C}\\[2pt]S_{ij,C}\end{pmatrix}\sqrt{\gamma_C}\Delta V_C=\sum_p m_p \begin{pmatrix}W_p\\[2pt]u_{i,p}\\[2pt]u_{i,p}u_{j,p}/W_p\end{pmatrix}.
\end{equation}
These identities hold algebraically, up to floating-point accumulation, including for clouds crossing a coarse--fine interface. However, AMR needs to proceed carefully since, at fixed local particle number density, each factor-of-two refinement in three dimensions reduces the mean number of particles per cell by a factor of eight, increasing the relative density fluctuations, which scale approximately as $N_{\mathrm{ppc}}^{-1/2}$ for independent Monte Carlo sampling, where $N_{\mathrm{ppc}}$ is the mean number of particles per cell. Future work should be devoted to potentially performing adaptive particle resampling in those regions.

\section{Numerical tests}
\label{sec:tests}
\subsection{Kerr geodesics}
\label{sec:kerr-tests}

We begin by testing both the spacetime (DGREM) Boris pusher and the implicit midpoint algorithm.
We test the centered DGREM--Boris prescription in
Eqs.~\eqref{eq:disc_pos_eq} and \eqref{eq:disc_mom_eq} in a fixed
Kerr spacetime with $a=0.9M$. Both particle methods use 
identical physical initial data, global timesteps, and
output times. For these tests, we enforce a convergence threshold of $2\times10^{-14}$ in code units on the nonlinear iterations.

\begin{figure*}[t]
  \centering
  \includegraphics[width=\linewidth]{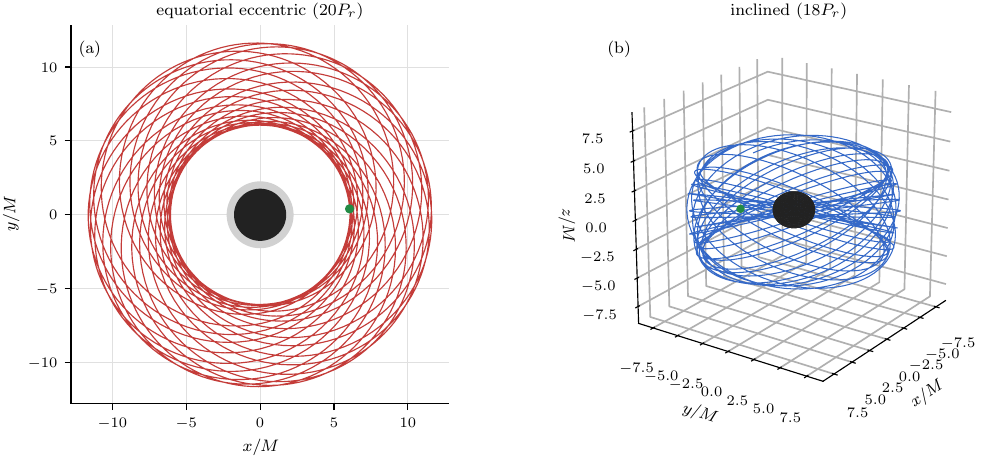}
  \caption{Boyer--Lindquist trajectories for the
  equatorial-eccentric and inclined geodesics evolved using the spacetime Boris pusher. The event horizon is
  black, the equatorial ergosphere gray, and the initial points green.}
  \label{fig:bl-trajectories}
\end{figure*}

{We first consider two bound geodesics initialized in
Boyer--Lindquist coordinates. The equatorial-eccentric orbit starts
at $(r_0,\theta_0,\phi_0)=(6M,\pi/2,0)$ with covariant spatial
four-velocity components
$(u_{r,0},u_{\theta,0},u_{\phi,0})=(0,0,3.144275M)$.
The inclined orbit starts at
$(r_0,\theta_0,\phi_0)=(8M,\pi/2-0.35,0)$ with
$(u_{r,0},u_{\theta,0},u_{\phi,0})=(0,0.9M,2.9M)$.
In both cases, $u^t_0>0$ is fixed by $u^\mu u_\mu=-1$.}
The resulting trajectories are shown in Fig.~\ref{fig:bl-trajectories}. The equatorial-eccentric orbit
spans $6.00M\leq r\leq11.61M$ and exhibits periapsis precession over
20 radial periods; the inclined orbit samples both hemispheres over
the 18 periods shown.

\begin{figure*}[t]
  \centering
  \includegraphics[width=\linewidth]{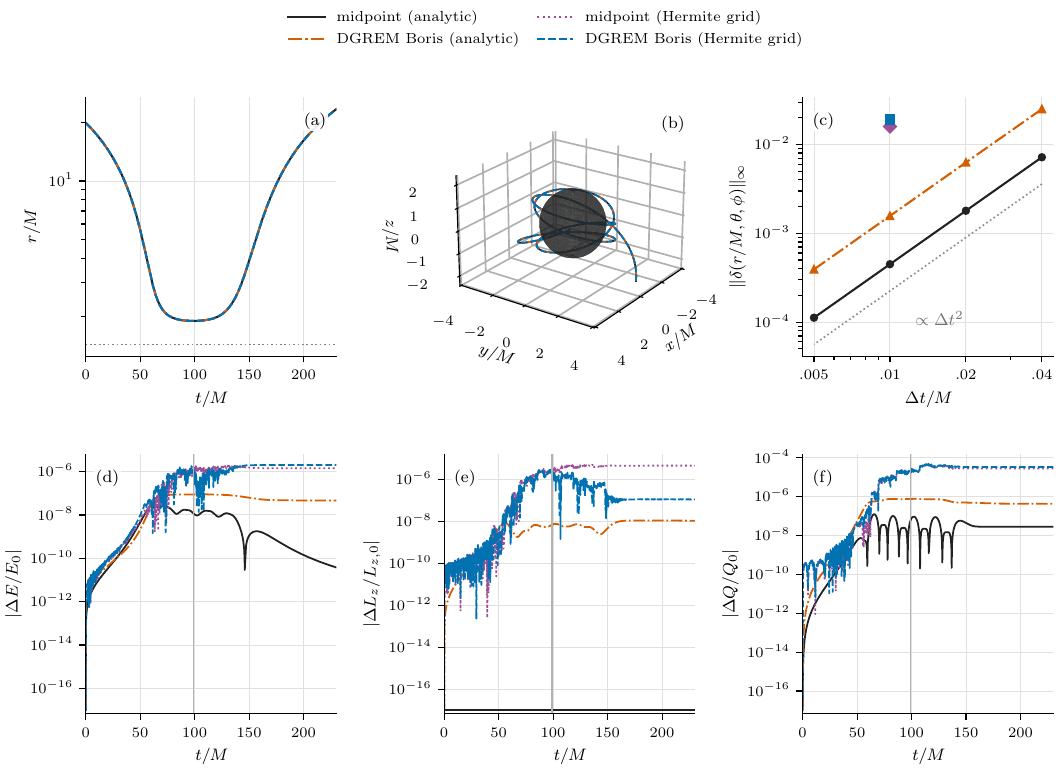}
  \caption{Off-equatorial $e=0.99$ Kerr--Schild passage integrated with different pushers, also comparing the impact of numerically interpolating the metric on the massive particle.
  Analytic coordinate midpoint (black), centered spacetime (DGREM) Boris (orange),
  Hermite-grid coordinate midpoint (purple), and Hermite-grid
  spacetime (DGREM) Boris (blue) use identical physical initial data and
  $\Delta t=0.01M$. The grid spacing is $0.025M$.
  Panels show the radial trajectory (a), the $r<4M$ segment (b),
  final position-error convergence for analytic input and numerical
  results at $\Delta t=0.01M$ (c), conserved geodesic invariants, including energy, $E$, angular momentum $L_z$, and Carter constant, $Q$, relative to their initial values
  (d)--(f). Vertical lines mark pericenter passage.}
  \label{fig:ks-inclined-passage}
\end{figure*}

Figure~\ref{fig:ks-inclined-passage} compares the centered DGREM--Boris
and coordinate implicit-midpoint updates for an off-equatorial
$e=0.99$ passage. {At $t=0$, the particle is on the
inbound branch at $(r_0,\theta_0,\phi_0)=(20M,\pi/2,0)$ in
spherical Kerr--Schild coordinates, with
\begin{align}
 u_{r,0}&=-0.20131797516752664,\nonumber\\
 u_{\theta,0}/M&=1.4049911031746785,\nonumber\\
 u_{\phi,0}/M&=2.4328.\label{eq:ks-passage-ic}
\end{align}
The corresponding specific geodesic invariants are
{energy} $E=-u_t=0.99738$, {axial angular momentum} $L_z/M=2.4328$, and {Carter constant} $Q/M^2=1.974$,
and the pericenter is $r_{\rm p}\simeq1.8983M$. The Cartesian-grid
runs use the same physical position and four-velocity transformed
to Cartesian Kerr--Schild coordinates.}

{We compare both pushers with analytic metric input in
spherical Kerr--Schild coordinates and with metric values sampled
on a Cartesian Kerr--Schild grid. The grid runs use
$\Delta x=\Delta y=\Delta z=0.025M$ and the}
same cubic Hermite reconstruction specified in
Section~\ref{sec:pusher-reconstruction}.
All four histories use $\Delta t=0.01M$ through $t=230M$, with one
particle advance per global step.

For analytic metric input, the maximum relative errors in $E,Q$ are
$9.0\times10^{-8},7.8\times10^{-7}$ for DGREM--Boris and
$2.8\times10^{-8},1.3\times10^{-7}$ for coordinate midpoint.
The DGREM--Boris relative $L_z$ error is $1.1\times10^{-8}$,
whereas coordinate midpoint retains the cyclic momentum to the
stored precision. On the Hermite grid, the maximum relative errors
in $E,Q$ are $2.0\times10^{-6},5.3\times10^{-5}$ for DGREM--Boris
and $2.2\times10^{-6},5.1\times10^{-5}$ for coordinate midpoint,
and their $L_z$ errors are $3.8\times10^{-6}$ and $5.3\times10^{-6}$.

Panel (c) measures the final maximum component error in
$(r/M,\theta,\phi)$ against an independent adaptive integration of
the continuous Kerr geodesic equations. Both methods are second-order accurate.
In practice, the accuracy of both methods is limited by the interpolation error 
of the spacetime, as can clearly be seen from this comparison.
As a result, in all tests presented here, we choose the algorithmically simpler implicit midpoint algorithm and leave a more detailed comparison of pushers in numerical spacetimes to future work.

\begin{figure*}[t]
 \centering
 \includegraphics[width=\linewidth]{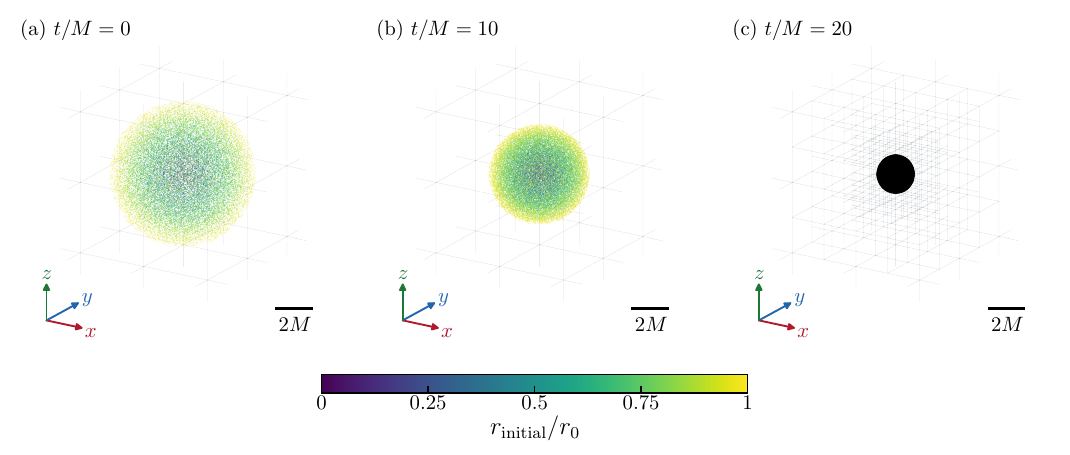}
 \caption{Three-dimensional visualization of Oppenheimer–Snyder dust collapse at $t/M=0$, $10$, and $20$, showing the initial configuration, an intermediate stage of collapse, and the resulting black hole. Particles are colored by their initial isotropic radius, normalized by the initial surface radius. Thin gray lines mark block boundaries of the adaptive mesh illustrating dynamical refinement during collapse.}
 \label{fig:os_particles_3d_amr}
\end{figure*}

\subsection{Oppenheimer--Snyder spherical dust collapse}
\label{sec:os-collapse}

The Oppenheimer–Snyder (OS) solution describes a homogeneous, spherically symmetric ball of pressureless dust that is initially at rest and subsequently collapses under its own gravity \cite{Oppenheimer:1939ue}. Its analytic description in comoving coordinates makes it a particularly useful benchmark for the coupled evolution of matter and spacetime. The interior consists of a portion of a closed Friedmann–Lemaître–Robertson–Walker (FLRW) spacetime matched at the stellar surface to a Schwarzschild exterior. For ADM mass $M$ and initial surface areal radius $R_0>2M$, the initial rest-mass density and subsequent evolution are given by the cycloidal solution \cite{Staley:2011ss}
\begin{align}
 \rho_0 &= \frac{3M}{4\pi R_0^3},
 &\frac{R(\eta)}{R_0}&=\frac{1+\cos\eta}{2},
 \label{eq:os-cycloid-radius}\\
 \tau(\eta)&=\sqrt{\frac{R_0^3}{8M}}(\eta+\sin\eta),
 &\frac{\rho(\eta)}{\rho_0}
 &=\left(\frac{2}{1+\cos\eta}\right)^3.
 \label{eq:os-cycloid-time-density}
\end{align}
Here, $\tau$ is the proper time along the dust worldlines, and $0\leq\eta<\pi$ is the dimensionless conformal time of the interior FLRW geometry, related to proper time by $d\tau=a\,d\eta$, where $a$ is the FLRW scale factor. The formal singularity occurs at $\eta=\pi$, corresponding to $\tau_{\rm coll}=\pi\sqrt{R_0^3/(8M)}$. For comparison with the numerical central-density evolution, we use the central proper time $\tau_c(t)=\int_0^t\alpha_c(t')\,dt'$, which removes the coordinate-time dependence of the slicing from the cycloid test and expresses the numerical history in the same proper time as the analytic solution.

Several studies have simulated OS collapse with different spacetime slicings \cite{Petrich:1985jko,1986PhRvD..33.2100P}. In particular, Staley et al. studied this solution with the moving-puncture gauge \cite{Staley:2011ss}. For $1+\log$ slicing with an initially unit lapse, and assuming that the lapse remains spatially uniform near the center or has a central minimum throughout the evolution, they derived the central lower limit (LL)
\begin{equation}
 \alpha_c \geq \alpha_{\rm LL}
 =1+6\ln\!\left(\frac{a(\tau_c)}{a_m}\right)
 =1+6\ln\!\left(\frac{R(\tau_c)}{R_0}\right),
 \label{eq:os-staley-lapse}
\end{equation}
where $a_m=a(0)=\sqrt{R_0^3/(2M)}$, and $R(\tau_c)$ denotes the analytic surface radius evaluated at the central proper time. In numerical simulations, the bound is saturated until an inward-propagating gauge mode from the stellar surface reaches the center. Thereafter, the central lapse remains above the bound. With the standard moving-puncture shift condition, the resolved geometry approaches a vacuum Schwarzschild trumpet at late times. Together, these analytic relations and moving-puncture benchmarks provide complementary tests of particle sampling, geodesic evolution, stress-energy deposition, Z4c evolution, adaptive mesh refinement, and particle excision.

Following the isotropic construction in Ref.~\cite{Staley:2011ss}, we initialize the time-symmetric OS slice in isotropic Cartesian coordinates. In terms of the surface areal radius, the isotropic coordinate radius of the surface is
\begin{equation}
 \bar r_0=\frac{R_0}{2} \left(1-\frac{M}{R_0}+\sqrt{1-2M/R_0}\right).
 \label{eq:os-isotropic-radius}
\end{equation}
The initial spatial metric and extrinsic curvature are $\gamma_{ij}=\psi^4\delta_{ij}$ and $K_{ij}=0$, with
\begin{align}
 \psi^2(r)&=
 \frac{(1+\sqrt{1-2M/R_0})\bar r_0R_0^2}{2\bar r_0^3+Mr^2},
 &&r\leq\bar r_0,
 \label{eq:os-psi-interior}\\
 \psi^2(r)&=\left(1+\frac{M}{2r}\right)^2,
 &&r>\bar r_0.
 \label{eq:os-psi-exterior}
\end{align}
The lapse is initially unity and the shift vanishes. The interior conformal factor satisfies the Hamiltonian constraint for the uniform rest-mass density $\rho_0$ in Eq.~\eqref{eq:os-cycloid-radius} and matches continuously to the exterior Schwarzschild solution.

We compare two particle-initialization strategies: Monte Carlo sampling and a mass-weighted lattice. Both methods initialize the dust at rest, $u_{i,p}=0$, and represent the same proper rest-mass distribution inside $r\leq\bar r_0$,
\begin{equation}
    dM_{\rm rest}=4\pi\rho_0\psi^6(r)\,r^2dr.
\end{equation}
where $\sqrt{\gamma}=\psi^6$ converts coordinate volume into proper volume. For Monte Carlo sampling, we draw trial positions uniformly in coordinate volume inside the sphere and accept each with probability
\begin{equation}
    P_{\rm acc}(r)=\frac{\psi^6(r)}{\psi^6(0)}.
\end{equation}
The process is repeated until $N_p$ particles are accepted, and we assign each the same rest mass $m_p=M_{\rm rest}/N_p$. For the mass-weighted lattice method, we place particles at the centers of a Cartesian lattice with spacing $h=2\overline{r}_0/n_{\rm lat}$ within radius $\overline{r}_0$, and assign each the local proper-volume weight
\begin{equation}
    m_p=\rho_0\psi^6(r_p)h^3.
\end{equation}
{Here $n_{\rm lat}$ counts lattice cells across the initial diameter.} The choice of particle initialization affects both the initial constraint violations and the subsequent evolution errors. For the OS configuration considered here, the mass-weighted lattice produces less sampling noise than Monte Carlo initialization, as demonstrated in Appendix \ref{sec:os-convergence}. The fiducial evolution presented below uses the lattice method with $n_{\rm lat}=124$, equivalent to $N_p=998,592$ particles, and a total particle rest mass $\sum_p m_p\simeq1.155M$.

To facilitate comparison with the moving-puncture results in Ref.~\cite{Staley:2011ss}, we adopt their fiducial initial configuration, setting the ADM mass to $M=1$ and the surface areal radius to $R_0=5M$. This is equivalent to the isotropic radius $\bar r_0\approx3.936M$. The particles are deposited with the first-order shape function described in Section~\ref{sec:stree-energy-deposition}.

\begin{figure}[t]
 \centering
 \includegraphics[width=\linewidth]{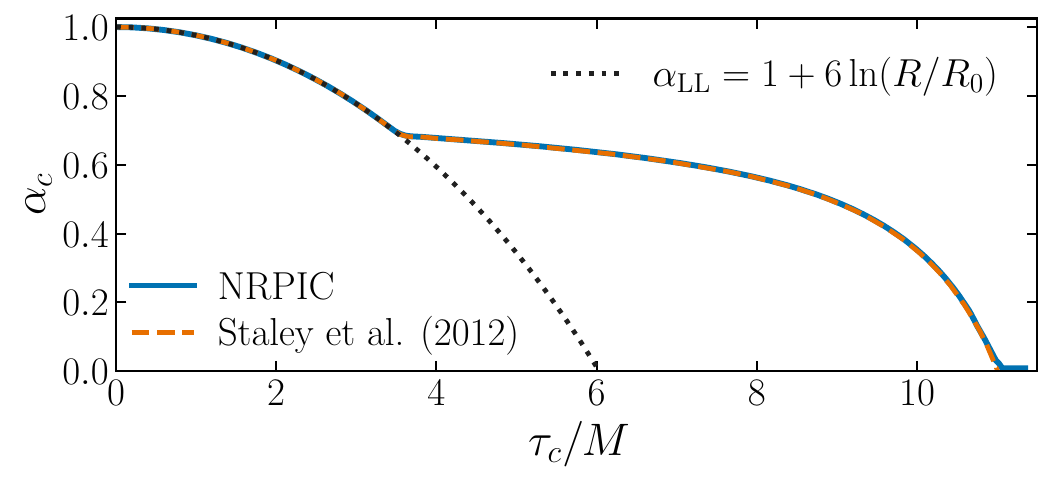}
 \caption{Central lapse $\alpha_c$ as a function of central proper time $\tau_c$. The blue solid curve shows the NRPIC result, and the orange dashed curve shows the reference result of Staley et al. \cite{Staley:2011ss}. The black dotted curve denotes the analytic lower bound $\alpha_{\rm LL}$ given in Eq. \eqref{eq:os-staley-lapse}. The numerical lapse follows this bound at early times and departs from it near $\tau_c\simeq3.5M$, when an inward-propagating gauge mode from the stellar surface reaches the center and the region of spatially uniform lapse disappears, as shown in Fig.~\ref{fig:os_lapse_energy_profiles}. The NRPIC evolution closely reproduces the reference result.}
 \label{fig:os_central_lapse}
\end{figure}

\begin{figure*}[t]
 \centering
 \includegraphics[width=\linewidth]{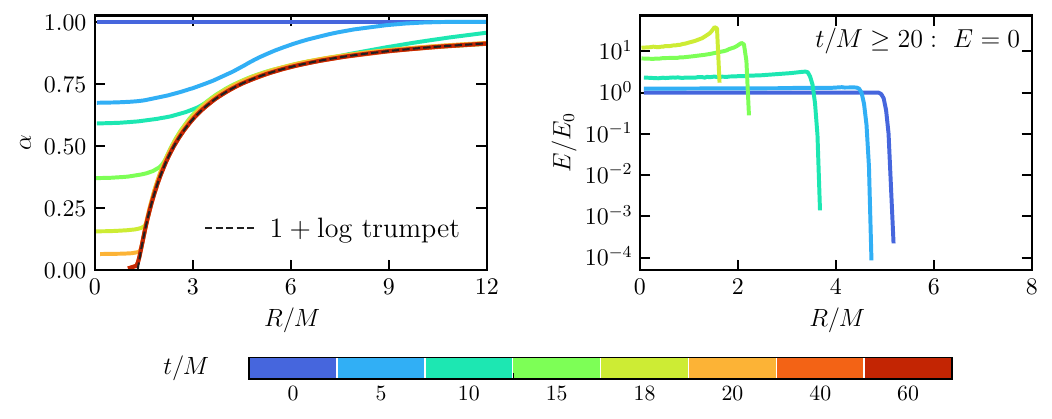}
 \caption{Radial profiles during Oppenheimer–Snyder collapse, with colors indicating the coordinate times shown in the shared color bar. \textbf{Left}: Lapse $\alpha$ as a function of areal radius $R$. The black dashed curve shows the analytic stationary $1+\log$ trumpet solution for a Schwarzschild black hole \cite{Hannam:2008sg}, which the numerical profiles approach at late times. \textbf{Right}: Eulerian energy density $E=T_{\mu\nu}n^\mu n^\nu$, normalized by its initial interior value $E_0=\rho_0=3M/(4\pi R_0^3)$. The outer edge of the density profile follows the contracting matter distribution. No profiles exist for $t\geq20M$ due to particle excision.}
 \label{fig:os_lapse_energy_profiles}
\end{figure*}

The fiducial evolution uses a full simulation domain of $[-128M, 128M]^3$ and $64^3$ root cells, with a maximum of seven AMR levels, corresponding to a finest grid spacing of $\Delta=M/32$. The inner region $[-8M,8M]^3$ is initially refined to level five, corresponding to $\Delta x=M/8$, with further refinement allowed as the collapse proceeds, as shown in Fig.~\ref{fig:os_particles_3d_amr}. We enable lapse-based particle excision, removing particles when their interpolated lapse falls below $\alpha_{\rm exc}=0.1$. After the deposited matter source has vanished, we continue the spacetime evolution to $t=60M$ to assess the stability of the vacuum black-hole remnant in moving-puncture coordinates.

We also perform a full suite of convergence tests by varying particle number on a fixed mesh, varying grid resolution with fixed particle number and varying both while holding $N_p(\Delta x/M)^3$ constant. For a fixed initial configuration and sampling prescription, this last sequence approximately preserves the initial number of particles per cell. These evolution tests use Monte Carlo sampling; we separately compare the initial density errors produced by Monte Carlo and mass-weighted lattice initialization. The simulation parameters and convergence results are presented in Appendix \ref{sec:os-convergence}.

Figure \ref{fig:os_particles_3d_amr} shows three-dimensional snapshots of the collapse, with particles colored by their initial distance from the center. The snapshots illustrate the approximately spherical contraction of the dust, the accompanying increase in mesh refinement, and the formation of the black-hole remnant.

In Fig.~\ref{fig:os_central_lapse}, we show that the central lapse $\alpha_c$ follows Eq.~\eqref{eq:os-staley-lapse} until $\tau_c\simeq3.5M$, when the surface gauge characteristic reaches the center, and subsequently remains above the analytic lower limit, reproducing the behavior reported by Staley et al.~\cite{Staley:2011ss}. Before particle excision, the deposited central energy density also follows the cycloidal evolution given in Eqs.~\eqref{eq:os-cycloid-radius}--\eqref{eq:os-cycloid-time-density}, as examined further in Appendix \ref{sec:os-convergence}. Figure \ref{fig:os_lapse_energy_profiles} shows the radial lapse and energy-density profiles as functions of areal radius $R$. At late times, the lapse profile approaches the stationary $1+\log$ Schwarzschild trumpet solution \cite{Hannam:2008sg}, while the outer edge of the matter distribution remains identifiable during the resolved collapse preceding excision.

We measure the final black-hole mass after particle excision using both the built-in \texttt{Fastflow} module in \texttt{AthenaK} and the external \texttt{BHaHAHA} apparent-horizon finder \cite{Etienne:2025auk}. Comparing the final horizon mass with the initial ADM mass tests the recovery of the total gravitational mass through collapse and excision. The fiducial run gives a relative error of $\Delta M/M_{ADM}\lesssim 10^{-3}$. The convergence study shows that the final-mass error decreases with increasing particle number at fixed mesh and AMR prescription over the range considered. Increasing spatial resolution at fixed particle number, however, does not produce a monotonic reduction in this error. These results emphasize the need to control both particle-sampling errors and spacetime-discretization errors, as discussed in Appendix \ref{sec:os-convergence}.

\begin{figure*}[t]
    \centering
 \includegraphics[width=\linewidth]{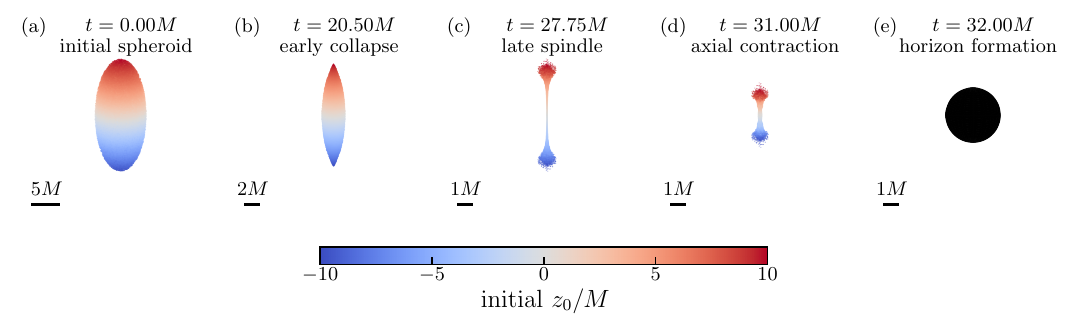}
 \caption{Collapse of an initially cold prolate spheroid of collisionless matter. The snapshots show (a) the initial configuration at $t=0$, (b) early collapse and density-caustic formation at $t=20.50M$, (c) a thin spindle at $t=27.75M$, (d) contraction along the symmetry axis at $t=31.00M$, and (e) the apparent horizon at $t=32.00M$. Particles are colored by their initial axial coordinate $z_0/M$, using the same color scale in all panels. The spatial scale of each panel is indicated by its scale bar.}
 \label{fig:singularity_particles_3d_2x2}
\end{figure*}

\subsection{Weak cosmic censorship test}
\label{sec:censorship}

{We next consider the collapse of a highly prolate, initially cold distribution of collisionless matter, which is a longstanding numerical test of weak cosmic censorship. The conjecture states that, from an asymptotically flat setting, singularities arising from generic, regular initial data with suitable matter should remain hidden from distant observers behind event horizons \cite{Penrose:1969pc}. Prolate collapse is particularly relevant here because strong transverse contraction can produce large curvature while the matter remains extended along the symmetry axis. The hoop conjecture provides a heuristic criterion for horizon formation that the mass distribution must be compact enough in every direction such that the maximum circumference satisfies $C_{\rm max}\lesssim 4\pi M$ \cite{1972mwm..book..231T}, which is in tension with horizon formation for this prolate configuration.}

{This configuration was first simulated in Ref. \cite{Shapiro:1991zza}, which reported rapidly growing curvature in cold prolate collapse without detecting an apparent horizon before the numerical evolution became unreliable, suggesting a possible naked singularity \cite{Shapiro:1991zza}. This interpretation requires caution because the absence of an apparent horizon on a chosen slicing does not establish the absence of an event horizon \cite{Wald:1991zz}. Ref. \cite{Yoo:2016kzu} subsequently revisited the problem in three dimensions, finding resolution-dependent curvature peaks inside the matter and no apparent horizon during the simulated evolution of the fiducial configuration \cite{Yoo:2016kzu}. Ref. \cite{East:2019bwu} later used generalized harmonic evolution to follow the collapse through the formation of a matter caustic and observed subsequent black-hole formation in all configurations studied. Here we revisit this classic problem using the Z4c formulation of the Einstein equations and moving-puncture gauge. This configuration provides a demanding test of the coupled particle–spacetime evolution, probing the code’s ability to follow strongly anisotropic contraction, continue through density-caustic formation, and capture the subsequent formation of an apparent horizon.}

We adopt the same continuum initial-data family used in the earlier cold-collapse studies \cite{Shapiro:1991zza,Yoo:2016kzu, East:2019bwu}, and choose the fiducial eccentricity $e=0.9$ and polar semiaxis $b=10M$, where $M$ denotes the initial ADM mass. In the initial conformally flat Cartesian coordinates, the equatorial semiaxis is $a=b\sqrt{1-e^2}\simeq 4.3589M$, and the particles initially occupy the spheroidal region
\begin{equation}
    \frac{x^2+y^2}{a^2}+\frac{z^2}{b^2}\leq 1.
\end{equation}
All particles are initially at rest relative to the observers normal to the initial slice, with $u_{i,p}=0$ and $W_p=1$, so there is no velocity dispersion impeding collapse. The initial spatial geometry is conformally flat and time symmetric.

We construct the initial data following the method in Ref.~\cite{Yoo:2016kzu}. The particle distribution is generated from a uniform auxiliary density $\rho_N=3M/(8\pi a^2b)$ inside the spheroid and vanishes outside. Solving the flat-space Poisson equation $\nabla^2\Phi_N=4\pi\rho_N$ defines the auxiliary conformal factor $\bar\psi=1-\Phi_N$, from which the corresponding ADM energy density is $\bar\rho_{\rm ADM}=2\rho_N\bar\psi^{-5}$. Equal-rest-mass particles are then sampled from the associated proper rest-mass distribution, whose coordinate-space probability density is proportional to $\bar\psi$ within the spheroid. In the weak-field limit, $\bar\psi\rightarrow1$, and the physical matter distribution therefore approaches a uniform-density spheroid. The continuum rest mass of the fiducial configuration is $M_{\rm rest}\simeq1.09815M$, which determines the initial mass assigned to each particle. For this test, we sample a total number of $N_p=2\times10^6$ particles. To reduce low-order sampling asymmetries, we impose reflection symmetry across the equatorial plane together with invariance under a rotation by $\pi$ about the polar axis. This suppresses the leading dipolar sampling fluctuation while retaining a fully three-dimensional particle representation.

The simulation domain is $[-128M,128M]^3$ and has a $128^3$ root grid and $32^3$-cell MeshBlocks. The initial refinement hierarchy resolves the entire matter distribution with $\Delta x=M/8$, while adaptive mesh refinement permits a finest spacing of $\Delta x=M/64$ as the spindle contracts. Apparent-horizon searches and the particle excision scheme are also activated to test mass conservation. Particles are removed when their interpolated lapse falls below $\alpha_{\rm exc}=0.05$, independently of the horizon searches.

Fig.~\ref{fig:singularity_particles_3d_2x2} follows the particle morphology from the initial spheroid through early contraction and the onset of strong axial focusing at $t=20.5M$, the formation of a narrow axial spindle at $t=27.75M$, continued contraction along the symmetry axis at $t=31M$, and the configuration after apparent-horizon detection at $t=32M$. The progressively smaller plotting windows in the later panels emphasize the large hierarchy of spatial scales generated during the collapse. As in the previous cold-collapse studies \cite{Shapiro:1991zza,Yoo:2016kzu,East:2019bwu}, the initially smooth distribution develops a collisionless density caustic and subsequently reorganizes primarily along the symmetry axis.

\begin{figure}[t]
    \centering
    \includegraphics[width=\linewidth]{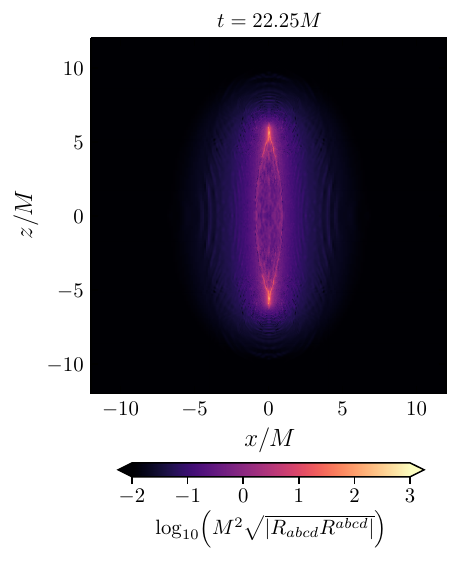}
    \caption{Curvature in the meridional $x$–$z$ plane during the caustic stage of prolate collapse. Color shows the logarithm of the dimensionless curvature magnitude. Two regions of high curvature lie within the matter near opposite ends of the forming spindle along the symmetry axis.}
    \label{fig:singularity_curvature_xz}
\end{figure}

To characterize the focusing, we monitor the full Kretschmann invariant $\mathcal I=R_{abcd}R^{abcd}$ together with the deposited matter density and the locations of their maxima. Fig.~\ref{fig:singularity_curvature_xz} shows the meridional slice of the curvature scalar during axial focusing. Two localized high-curvature regions develop near opposite ends of the spindle and remain embedded within the matter distribution.

{We also perform convergence tests using controlled runs with finest allowed spacings $M/32$, $M/64$, and $M/128$, and particle counts from $5\times10^5$ to $8\times10^6$. At every finite resolution considered here, the curvature remains finite throughout the numerical evolution, while the maximum attained during the caustic stage increases with resolution, resembling the behavior reported in Ref.~\cite{Yoo:2016kzu}. We therefore do not assign a continuum value to the peak curvature, and we do not infer from its finite-resolution behavior alone whether the idealized cold distribution develops a singular caustic in the continuum limit. Instead, the robust dynamical result is that the evolution can be continued past the strongly focused stage and subsequently forms a common apparent horizon. This is qualitatively consistent with the result in Ref.~\cite{East:2019bwu}. Moreover, we find that the first-horizon detection times vary by less than $0.1M$, indicating that the transition to black-hole formation is substantially more robust than the peak curvature attained at the unresolved caustic. Therefore, we demonstrate that NRPIC can evolve through the numerically challenging focusing stage and recover black hole formation.}

\subsection{Relativistic cluster}
\label{sec:st-cluster}

{Relativistic collisionless clusters provide a controlled setting for studying self-gravitating many-body systems in general relativity in which support against gravity is supplied by orbital motion rather than fluid pressure \cite{Andreasson:2011ng}. In the collisionless regime, the constituent particles interact predominantly through the collective gravitational field and are described in the continuum limit by the Einstein–Vlasov system. Stationary configurations can therefore be constructed in which the velocity dispersion of the distribution function balances self-gravity, providing relativistic analogues of equilibrium stellar systems \cite{Andreasson:2006ja}. As the compactness is increased, however, general-relativistic effects can destabilize this balance \cite{1968ApJ...154..251I}. Sufficiently relativistic equilibrium sequences develop unstable branches whose nonlinear evolution can lead either to substantial redistribution of the collisionless matter or to gravitational collapse and black-hole formation \cite{1985ApJ...298...34S,1985ApJ...298...58S}. Numerical studies have demonstrated such behavior for both spherical and rotating relativistic clusters \cite{1985ApJ...298...58S,Abrahams:1994ge}.}

These properties also make relativistic clusters stringent tests of the coupled evolution of particles and spacetime. An equilibrium requires the phase-space distribution and spacetime geometry to satisfy the Vlasov and Einstein equations self-consistently. Preserving this balance over many orbital periods therefore tests the accuracy of spacetime evolution, particle transport, and stress-energy deposition, as well as the secular accumulation of errors introduced by finite particle sampling, which should be sufficiently damped (see the discussion in Sections~\ref{sec:z4c} and \ref{sec:numerical_implementation}). Evolution away from equilibrium provides a complementary test of whether the numerical scheme can accurately follow collective redistribution in a dynamically changing gravitational field.

Following the series of studies by Shapiro and Teukolsky on the dynamics of relativistic collisionless clusters (hereafter ST85) \cite{1985ApJ...298...34S,1985ApJ...298...58S}, we first validate our code by conducting a compactness parameter scan of the monoenergetic distribution function in Section~\ref{sec:monoenergetic}. This model provides a simple isotropic distribution function and a well-studied family of relativistic equilibria with different compactness and stability properties, allowing us to compare our fully three-dimensional evolutions qualitatively with the original spherically symmetric calculations. In Section~\ref{sec:cluster-migration}, we then initialize an unstable member of the same monoenergetic family and follow its nonlinear migration toward a less compact, stable configuration, analogous to the migration tests widely used for unstable relativistic stars in numerical hydrodynamics \cite{Font:2001ew}. Finally, we apply a global density perturbation and observe a more general reconfiguration into a persistent core–halo distribution in Section~\ref{sec:cluster-reconfig}. We characterize the resulting radial profile using a generalized form inspired by the classical Newtonian Plummer model \cite{Plummer:1911zza} and compare it with relativistic extensions of the Plummer and hypervirial families \cite{1964ApJ...140.1512B,Nguyen:2013nka,Gauy:2017cyu}. Together, these tests probe equilibrium preservation, relativistic instability, nonlinear migration, and large-scale phase-space reconfiguration in fully three-dimensional Cartesian evolutions.

\subsubsection{Monoenergetic distribution}
\label{sec:monoenergetic}

\begin{figure}[t]
 \centering
 \includegraphics[width=\linewidth]{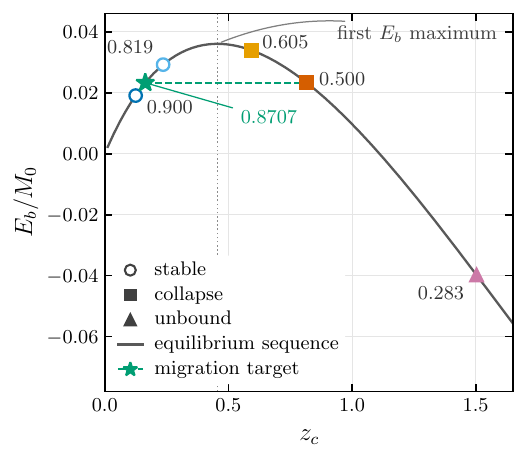}
 \caption{Fractional binding energy $E_b/M_0=(M_0-M)/M_0$ as a function of the initial central gravitational redshift $z_c$, where $M_0$ and $M$ denote the total rest mass and ADM mass, respectively. The curve represents the equilibrium sequence, with labels indicating the corresponding central parameter $y_c$. Open circles identify models that remain stable, filled squares indicate models that undergo gravitational collapse and form black holes, and the triangle denotes the unbound model that expands without forming a black hole. The vertical dotted line marks the first maximum of the fractional binding energy. The star marker connected by the horizontal dashed line indicates the predicted migration target discussed in Section~\ref{sec:cluster-migration}.}
 \label{fig:st-fate}
\end{figure}

The dynamical stability of relativistic collisionless systems is closely connected to the behavior of their binding energy along equilibrium sequences. Building on the variational stability theory of Ipser and Thorne \cite{1968ApJ...154..251I}, a sufficient criterion for the stability of spherical, isotropic collisionless systems was subsequently derived: along appropriate equilibrium sequences, configurations remain stable against radial perturbations at least up to the first maximum of the fractional binding energy $\epsilon_{\rm b}=(M_{\rm rest}-M_{\rm ADM})/M_{\rm rest}$ \cite{1980ApJ...238.1101I}. Subsequent numerical evolutions in ST85 provided strong evidence that this maximum also marks the onset of dynamical instability for several families of equilibria, demonstrating that unstable configurations can undergo gravitational collapse and form black holes \cite{1985ApJ...298...58S}. More recent studies have nevertheless identified equilibrium families that remain stable beyond the first binding-energy maximum, indicating that this criterion is not a universal condition for the onset of instability \cite{Gunther:2021pbf}. These results connect the properties of stationary solutions to their nonlinear evolution and provide a physically motivated benchmark for relativistic particle methods.

Neutron star models provide a fluid (collisional) analogue. Along a nonrotating equilibrium sequence with a fixed barotropic equation of state, the fundamental radial mode reaches zero frequency at the first maximum of the gravitational mass, $dM/d\rho_c=0$, where $\rho_c$ is the central density \cite{Chandrasekhar:1964zz,Friedman:1988er}. Depending on the perturbation, a star on the unstable branch can collapse to a black hole or expand and migrate toward a lower-density stable configuration \cite{Font:2001ew}.

For validation of the code, we aim to recover this established stability behavior, including the long-term preservation of stable equilibria and the development of relativistic instability in sufficiently compact configurations. Specifically, we construct stationary, spherically symmetric, isotropic solutions of the Einstein–Vlasov system following the monoenergetic equilibrium sequence investigated by ST85 \cite{1985ApJ...298...58S},
\begin{equation}
    f(E)=K\delta(E-E_0), \label{eq:monoenergetic_distrib}
\end{equation}
where $E\equiv-p_t$ is the conserved Killing energy measured at spatial infinity, {with $p_t=u_t$,} and $E_0$ is its prescribed value. The normalization $K$ fixes the overall density scale; that is, after solving the dimensionless equilibrium, we set this scale by choosing the total ADM mass $M=1$. In ST85, the matter is sampled as spherical particle shells, whereas we sample the same distribution in full three dimensions. This calculation therefore simultaneously tests the equilibrium initial data, Z4c evolution, geodesic pusher, stress-energy deposition, adaptive mesh refinement, particle excision, apparent-horizon finder, and long-term error accumulation in one calculation.

\begin{table*}[t]
 \caption{Summary of the monoenergetic equilibrium models and their subsequent dynamical evolution. All simulations use a total particle number $N=2\times10^6$, a $128^3$ root mesh, and $32^3$ cells per MeshBlock. For each setup, $y_c$ and $z_c$ denote the initial compactness and central redshift. $R$ is the initial areal surface radius. $M_0$ is the initial total rest mass. The domain is $[-L,L]^3$. $\Delta x_{\rm init}$ is the initial finest spacing and $\Delta x_{\rm min}$ is the finest spacing permitted by AMR. For collapsing models, $t_{BH}$ is the time of apparent horizon formation.}
 \label{tab:st-suite}
 \begin{ruledtabular}
 \begin{tabular}{cccccccc}
 $y_c$ & $z_c$ & $R/M$ & $M_0/M$ & $L/M$ &
 $\Delta x_{\rm init}/M$ & $\Delta x_{\rm min}/M$ & $t_{\rm BH}/M$ \\
 \colrule
 0.900 & 0.125 & 16.4282 & 1.019453 & 1024 & $1/2$ & $1/4$  & N/A \\
 0.819 & 0.236 &  9.9642 & 1.030106 & 1024 & $1/2$ & $1/4$  & N/A \\
 0.605 & 0.592 &  5.7502 & 1.035171 &  512 & $1/4$ & $1/16$ & 109 \\
 0.500 & 0.817 &  5.0758 & 1.023830 &  512 & $1/4$ & $1/16$ & 73 \\
 0.283 & 1.504 &  4.5846 & 0.961732 &  512 & $1/4$ & $1/16$ & N/A \\
 \end{tabular}
 \end{ruledtabular}
\end{table*}

We refer to Appendix A of Ref.~\cite{Bamber:2025gxj} for a detailed derivation of the initial metric and matter profile of the relativistic monoenergetic equilibria. We briefly outline the equilibrium equations and their transformation to the numerical initial data. In Schwarzschild areal coordinates, the static, spherically symmetric line element is
\begin{equation}
    ds^2 =-e^{2\Phi}dt^2
 +\left(1-\frac{2m}{r_s}\right)^{-1}dr_s^2+r_s^2d\Omega^2,
 \label{eq:st-metric}
\end{equation}
where $\Phi(r_s)$ is the gravitational potential and $m(r_s)$ is the enclosed gravitational mass at radius $r_s$. The rest-mass density, proper density, and pressure can be obtained from the distribution function Eq.~\eqref{eq:monoenergetic_distrib} using the described method of moments in Eqs.~\eqref{eq:kinetic-moments}--\eqref{eq:vlasov-stress-energy}
\begin{equation}
 \begin{aligned}
 \rho_0&=\int f\,d^3q,\qquad \rho=\int Wf\,d^3q,\\
 P&=\frac13\int\frac{q^2}{W}f\,d^3q.
 \end{aligned}
 \label{eq:st-moments}
\end{equation}
{Here $\boldsymbol q$ is the orthonormal spatial four-velocity and $q=|\boldsymbol q|$.} Einstein's equilibrium radial structure equations and the proper rest-mass integral give
\begin{align}
 \frac{dm}{dr_s}&=4\pi r_s^2\rho,\qquad
 \frac{dm_0}{dr_s}=4\pi r_s^2B\rho_0,\nonumber\\
 \frac{d\Phi}{dr_s}&=\frac{m+4\pi r_s^3P}{r_s(r_s-2m)}.
 \label{eq:st-physical-structure}
\end{align}
where $m_0(r_s)$ is the enclosed rest mass and $B=(1-2m/r_s)^{-1/2}$ accounts for the proper radial volume element. For numerical integration, the cluster is constructed using dimensionless quantities defined as
\begin{align}
     y=e^{2(\Phi-\Phi_s)},\qquad& x=\frac{r_s}{L},\notag\\
     \zeta=\frac{P_c}{\rho_c}=\frac{1-y_c}{3},\quad & L=\left(\frac{P_c}{4\pi\rho_c^2}\right)^{1/2},\notag\\
     v=m\rho_c/(LP_c),\quad & v_0=m_0\rho_c/(LP_c).
     \label{eq:st-dimensionless}
\end{align}
Here $\Phi_s$ denotes the potential at the cluster surface, while $\rho_c$ and $P_c$ are the central energy density and pressure. The parameter $y_c=y(0)$ labels the equilibrium sequence and controls its relativistic compactness, with $y=1$ defining the surface. For the monoenergetic distribution, the energy density, rest-mass density, and pressure can be expressed analytically in terms of $y$. Defining $\bar\rho=\rho/\rho_c$, $\bar\rho_0=\rho_0/\rho_c$, and $\bar P=P/P_c$, we obtain
\begin{align}
 \bar\rho &={\left(\frac{y}{y_c}\right)}^{-3/2}
 \left(\frac{y^{-1}-1}{y_c^{-1}-1}\right)^{1/2},
 \nonumber\\
 \bar\rho_0 &=y_c^{1/2}{\left(\frac{y}{y_c}\right)}^{-1}
 \left(\frac{y^{-1}-1}{y_c^{-1}-1}\right)^{1/2},
 \nonumber\\
 \bar P &={\left(\frac{y}{y_c}\right)}^{-1/2}
 \left(\frac{y^{-1}-1}{y_c^{-1}-1}\right)^{3/2}.
 \label{eq:st-closures}
\end{align}
The equilibrium structure equations then reduce to
\begin{align}
 \frac{dy}{dx}
 &=2\zeta\frac{y}{x^2}
 \frac{v+\zeta x^3\bar P}{1-2\zeta v/x},
 &\frac{dv}{dx}&=x^2\bar\rho,
 \nonumber\\
 \frac{dv_0}{dx}
 &=x^2\bar\rho_0\left(1-\frac{2\zeta v}{x}\right)^{-1/2}.
 \label{eq:st-structure}
\end{align}
We integrate Eq.~\eqref{eq:st-structure} outward from a regular central series until $y=1$.  If the surface values are denoted by $(x_s,v_s,v_{0,s})$, the scale-independent observables are
\begin{align}
    \frac{R}{M}&=\frac{x_s}{\zeta v_s},\qquad
 \frac{E_b}{M_0}=1-\frac{v_s}{v_{0,s}},
 \label{eq:st-global}\\
 z_c&=\left[y_c\left(1-\frac{2M}{R}\right)\right]^{-1/2}-1.
 \label{eq:st-redshift}
\end{align}
Thus $y_c$ labels a one-parameter equilibrium sequence.  Along this sequence, decreasing $y_c$ increases the central particle speed $\hat v_c=\sqrt{1-y_c}$, the central redshift $z_c$, and the compactness $M/R$. It therefore moves the models from the weakly relativistic stable branch toward and beyond the first binding-energy maximum. For this particular equilibrium family, the maximum closely marks the onset of dynamical instability found in ST85 \cite{1985ApJ...298...58S}, as illustrated in Fig.~\ref{fig:st-fate}.

\begin{figure*}[t]
 \centering
 \includegraphics[width=\textwidth]{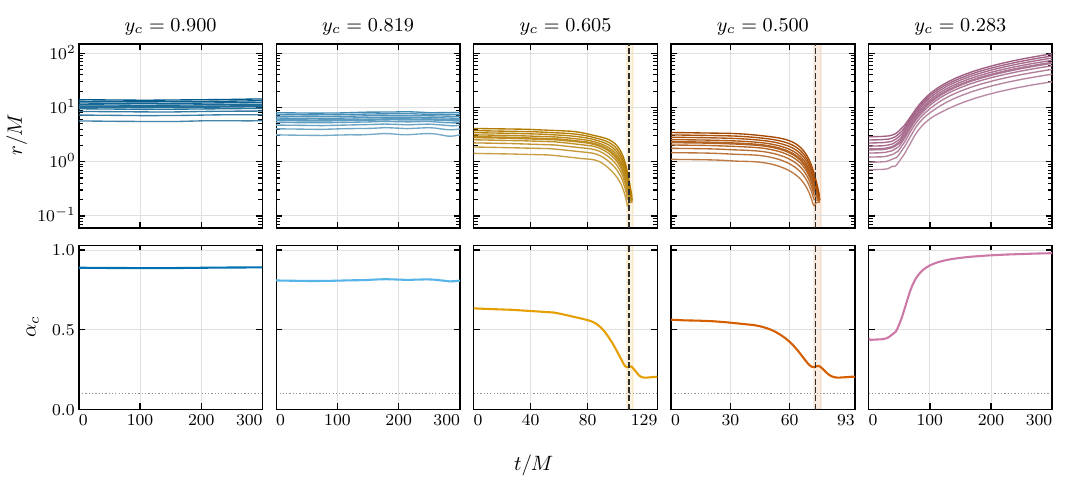}
 \caption{Dynamical evolution of five monoenergetic Shapiro–Teukolsky cluster models, each initialized with $2\times10^6$ equal-rest-mass particles. \textbf{Top}: isotropic-coordinate radii measured about the instantaneous particle center of mass that enclose 10–90\% of the surviving rest mass, in increments of 10\%. The median radius is emphasized by a thicker line. \textbf{Bottom}: central lapse $\alpha_c$. Shaded regions indicate intervals of particle excision, vertical dashed lines mark first accepted apparent-horizon detection, and horizontal dotted lines indicate the local particle-excision threshold $\alpha_{\rm exc}=0.1$ for reference. The $y_c=0.605$ and $0.500$ models are plotted through their respective horizon formation times, and the remaining models are shown through $300M$.}
 \label{fig:st-dynamics}
\end{figure*}

The numerical slice is initialized in isotropic radius $\bar r$.  We integrate
\begin{equation}
 \frac{d\ln\bar r}{d\ln r_s}
 =\left(1-\frac{2m}{r_s}\right)^{-1/2},
 \qquad A(\bar r)=\frac{r_s}{\bar r}.
 \label{eq:st-isotropic-map}
\end{equation}
The corresponding initial ADM variables are
\begin{align}
 \gamma_{ij}&=A^2\delta_{ij},
 &\alpha&=\sqrt{(1-2M/R)y},
 \nonumber\\
 \beta^i&=0,
 &K_{ij}&=0.
 \label{eq:st-adm-data}
\end{align}
The interior fields are joined to isotropic exterior Schwarzschild data at the surface. We construct five equilibrium configurations by varying \(y_c\), with their initial properties and subsequent dynamical outcomes summarized in Table \ref{tab:st-suite}. Fig.~\ref{fig:st-fate} shows the fractional binding energy along the equilibrium sequence as a function of the central redshift, with the simulated models identified by their corresponding $y_c$. Hollow circles mark stable configurations, while solid squares and the triangle mark unstable configurations that either collapse to a BH or undergo unbound expansion. All simulations use a total particle number $N=2\times 10^6$ and activate adaptive mesh refinement to track black hole formation. In collapsing models, particles are removed once their interpolated local lapse falls below $\alpha_{\rm exc}=0.1$.

The particles are sampled using a stratified particle sampler for both the radial and angular distributions and for the isotropic velocity distribution, to minimize center-of-mass drift as well as initial momentum-constraint violations. This method assumes $N$ equal-mass particles and samples the coordinates $\overline{r},\theta,\phi$ and the two momentum directions $\vartheta, \varphi$ using a randomized stratification. Mathematically, for particle $a$ and variable $d$, we define
\begin{equation}
 U_{ad}=\frac{\pi_d(a)+\xi_{ad}}{N}, \qquad \xi_{ad}\sim{\cal U}[0,1),
\end{equation}
where each $\pi_d$ is an independent random permutation of $\{0,\ldots,N-1\}$ and the offsets $\xi_{ad}$ are independent. This construction places exactly one sample within each of the $N$ equal-probability intervals of every marginal distribution. The particle positions and momentum directions are then obtained from
\begin{align}
 &\bar r_a = F^{-1}(U_{a1}),
 &\cos\theta_a &= 2U_{a2}-1, & \phi_a &= 2\pi U_{a3}, \nonumber\\
 & &\cos\vartheta_a &= 2U_{a4}-1, & \varphi_a&= 2\pi U_{a5}.
\end{align}
where $F(\bar r)=m_0[r_s(\bar r)]/M_0$ is the cumulative rest-mass distribution. The momentum directions are isotropic in the local orthonormal frame, while their magnitudes are determined by the monoenergetic condition. Compared with unstratified Monte Carlo sampling, this method is intended to reduce fluctuations in the initial matter distribution, center-of-mass motion, and momentum-constraint violations. We separately quantify particle-number convergence using standard Monte Carlo sampling in Appendix \ref{sec:st-mc-convergence}. These tests show that the initial deposited matter sources seed Hamiltonian and momentum-constraint fluctuations scaling approximately as $N^{-1/2}$, which are subsequently propagated and damped by the Z4c evolution.

\begin{figure*}[t]
 \centering
 \includegraphics[width=\linewidth]{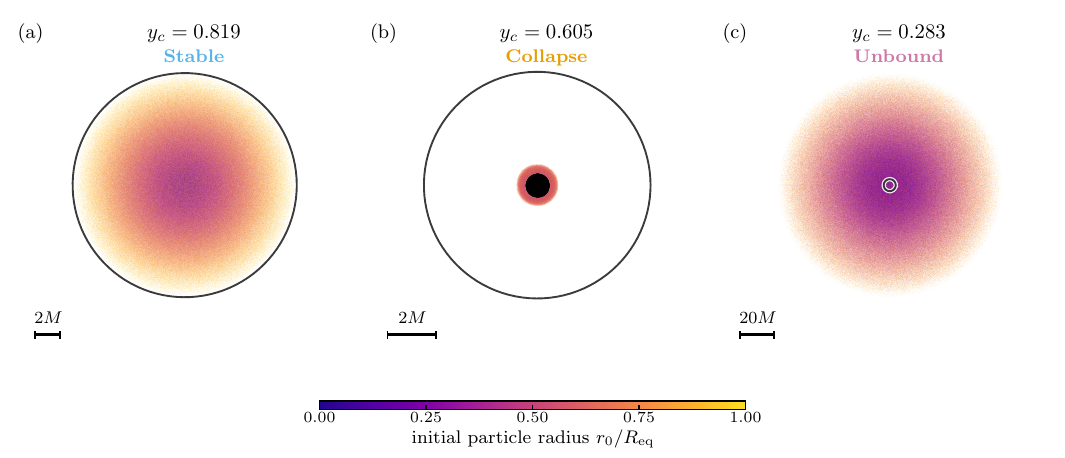}
 \caption{Three-dimensional particle configurations of the stable, collapsing, and unbound models. The panels are (a) $y_c=0.819$ at $t=300M$, (b) $y_c=0.605$ at $t=108M$, and (c) $y_c=0.283$ at $t=200M$, from left to right, respectively. Particle colors encode their initial isotropic radii, normalized by the initial equilibrium surface radius of each model. Reference circles indicate the corresponding initial cluster extents, while the black sphere in panel (b) represents the detected apparent horizon. All panels share a common viewing direction, with individual scale bars indicating their different coordinate extents.}
 \label{fig:st-3D}
\end{figure*}

Figure~\ref{fig:st-dynamics} compares the dynamics of the five clusters with different $y_c$, showing the time evolution of the radii enclosing 10--90\% of the surviving rest mass in increments of 10\%, together with the central lapse. To probe their dynamical response, we impose a coherent inward radial velocity perturbation for all five models with perturbation amplitude $\lesssim 1\%$ of the unperturbed equilibrium local velocity. The two less compact configurations, $y_c=0.900$ and $0.819$, preserve their overall radial structure through $t=300M$, exhibiting only small-amplitude oscillations around equilibrium. In contrast, the more compact $y_c=0.605$ and $0.500$ models undergo rapid contraction, central lapse collapse, and eventual apparent-horizon formation at approximately $t=109M$ and $73M$, respectively. Their active particle-excision windows are shown as shaded vertical regions, consistent with their AH formation times. We find that the more compact configuration collapses earlier than the other, consistent with its stronger relativistic self-gravity. The detected AH mass is consistent with the initial ADM mass of the cluster, showing the mass conservation of our code. The collapse time should be treated as a proof of principle, as it could depend on the number of particles sampled and therefore the initial numerical noise.

{A notable difference from Ref. \cite{1985ApJ...298...58S} arises for the most compact model, $y_c=0.283$, which has negative fractional binding energy, $E_b/M_0\simeq-0.0398$. Whereas Ref. \cite{1985ApJ...298...58S} reported black-hole formation for this configuration, our simulation instead exhibits sustained expansion, with the central lapse approaching unity and the enclosing-mass radii increasing substantially. The negative binding energy indicates that the total ADM mass exceeds the rest mass and that the configuration is not globally bound in the usual energetic sense, but the discrepancy between simulation results highlights the need for further investigation of the highly relativistic, nonlinear evolution of collisionless clusters.}

Finally, Fig.~\ref{fig:st-3D} illustrates the spatial structure associated with these three outcomes. At $t=300M$, the stable $y_c=0.819$ cluster remains nearly spherical and close to its initial equilibrium extent. The collapsing $y_c=0.605$ model has contracted well inside its initial outer radius. The snapshot captures a newly detected apparent horizon while an exterior particle envelope remains. The unbound $y_c=0.283$ model instead develops a diffuse, expanding cloud whose extent at $t=200M$ greatly exceeds its initial size. The reference circles and scale bars highlight these differences in coordinate extent, while the particle colors track the initial radial positions of the corresponding matter populations.

\subsubsection{Cluster Migration}
\label{sec:cluster-migration}
{The nonlinear migration of an unstable relativistic star toward a stable configuration provides a well-established benchmark for numerical relativity. In the standard neutron-star migration problem, a model on the unstable equilibrium branch expands and undergoes large radial oscillations about a configuration on the stable branch with approximately the same rest mass \cite{Font:2001ew}. It probes the coupling between matter and spacetime through repeated expansion and contraction, beyond the small perturbations examined in equilibrium-preservation tests. Migration has consequently been adopted as a standard benchmark test in many GRMHD codes (e.g., \cite{Bucciantini:2010ax,Lam:2025pmz}). In particular, simulations of unstable spherical $\Gamma=2$ polytropes have demonstrated that numerical truncation errors alone can trigger migration toward the stable branch. With a cold polytropic equation of state, the resulting radial oscillations remain nearly undamped, whereas an ideal-fluid evolution permits shock heating to convert bulk kinetic energy into thermal energy, progressively damping the pulsations \cite{Bucciantini:2010ax}. More recently, migration tests have been extended to uniformly rotating neutron stars, including a proposed standardized benchmark for fully three-dimensional numerical-relativity codes \cite{Petit:2026stp}.}

{A collisionless relativistic cluster provides an analogous migration problem, but with a fundamentally different mechanism of matter support and relaxation. Unlike a fluid star, whose pressure arises from local thermodynamic properties, a collisionless cluster is supported by the orbital velocity dispersion of its constituent particles and does not undergo hydrodynamic shock heating. Nevertheless, a time-dependent gravitational field can redistribute individual particle energies, while differences in orbital frequencies can suppress coherent collective oscillations through phase mixing. These processes provide mechanisms for collisionless relaxation without direct particle collisions \cite{Lynden-Bell:1966zjv,1986MNRAS.219..285T}. Consequently, an initially monoenergetic cluster need not recover its original distribution function or settle into another exact member of the same equilibrium family.} Following the matter distribution and spacetime geometry over multiple oscillation periods therefore provides a demanding test of long-term particle–spacetime coupling, conservation properties, and nonlinear redistribution in a time-dependent gravitational field.

In this section, we investigate the migration of an unstable monoenergetic cluster toward a more extended, stable configuration.  We first identify a candidate equilibrium within the same monoenergetic family by requiring conservation of the total rest mass and gravitational mass. In the following subsection, we examine a more general reconfiguration in which the system develops a different radial density profile. For the current test, we select the monoenergetic ST model with $y_c=0.500$, on the high-redshift side of the first binding-energy maximum in Fig.~\ref{fig:st-fate}. Its total rest mass and central equilibrium lapse are $M_0=1.02383M$ and $\alpha_c^{\rm eq}=0.550442$, corresponding to $z_c^{\rm eq}=0.816723$. The initial fractional binding energy is
\begin{equation}
 \epsilon_{\rm b}\equiv\frac{E_b}{M_0}
 =\frac{M_0-M}{M_0}=0.0232758.
 \label{eq:mig-binding}
\end{equation}
If both total gravitational mass and rest mass are conserved, another monoenergetic equilibrium with the same masses must have this same binding fraction. The dashed horizontal construction in Fig.~\ref{fig:st-fate} selects the lower-redshift intersection,
\begin{equation}
 \begin{aligned}
 y_c^{(E_b)}&=0.870743,\\
 \alpha_c^{(E_b)}&=0.859266,\qquad z_c^{(E_b)}=0.163784.
 \end{aligned}
 \label{eq:mig-binding-target}
\end{equation}
The star in Fig.~\ref{fig:st-fate} marks this candidate on the stable branch.

We represent the distribution by $N=2\times10^6$ equal-rest-mass particles with stratified sampling of both spatial positions and isotropic velocities, as described in Section~\ref{sec:monoenergetic}. No coherent density or radial velocity perturbation is imposed. Instead, finite particle sampling and numerical discretization errors provide the initial perturbations that trigger the unstable mode. The computational domain is $[-512M,512M]^3$, with a $128^3$ root grid and $32^3$ cells per MeshBlock. Five nested refinement regions give an initial finest spacing $\Delta x=M/4$ inside $[-16M,16M]^3$, while adaptive refinement permits spacings down to $M/16$. Particles are excised when their interpolated local lapse falls below $\alpha_{\rm exc}=0.1$. We follow the evolution through $t=2500M$, by which time the cluster approaches a long-lived, quasi-stationary configuration with residual oscillations.

\begin{figure}[t]
    \centering
    \includegraphics[width=\linewidth]{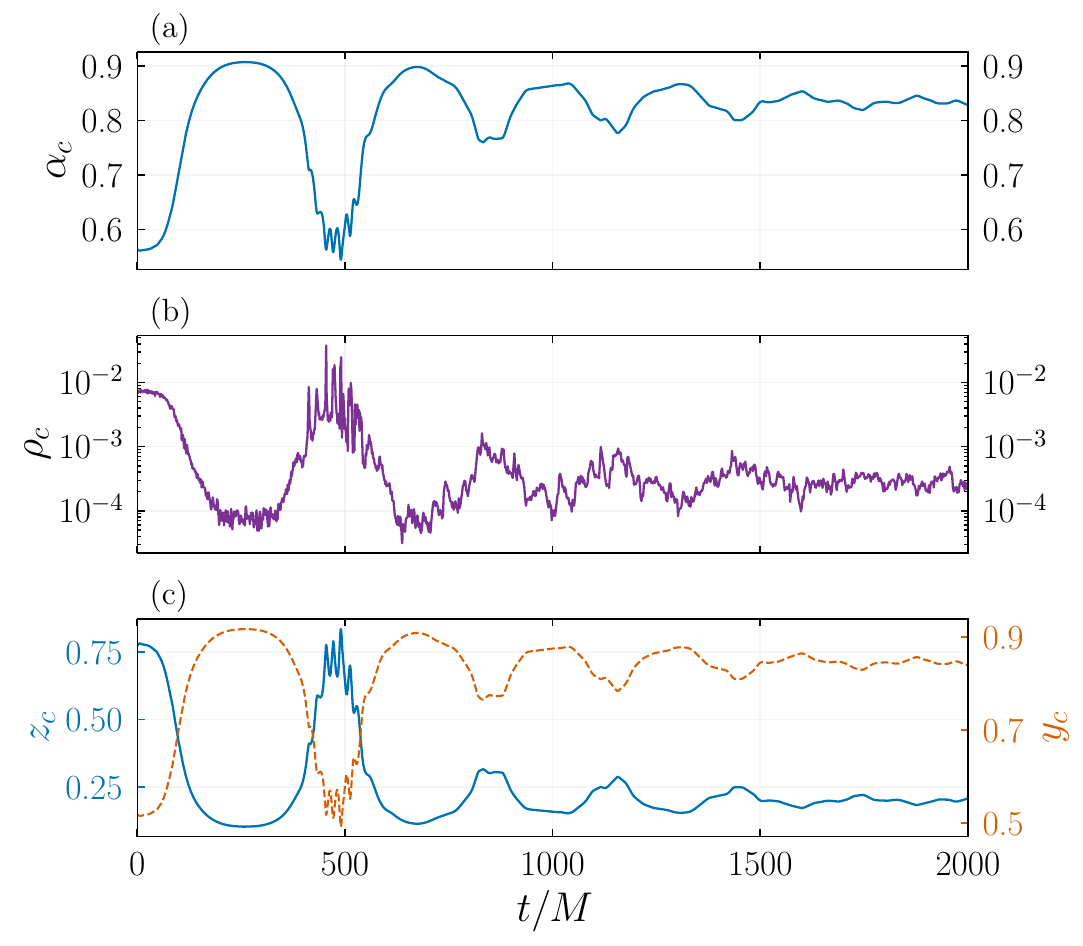}
    \caption{Evolution of the initially $y_c=0.500$ cluster. \textbf{top}: Central lapse $\alpha_c$. \textbf{middle}: Near-center Eulerian energy density $\rho_c=T_{\mu\nu}n^\mu n^\nu$, evaluated from the deposited stress-energy tensor. \textbf{bottom:} Lapse-derived central redshift proxy $z_c=\alpha_c^{-1}-1$ (blue, left axis) and the corresponding equilibrium-equivalent parameter $y_c$ (orange dashed, right axis), obtained by mapping onto the monoenergetic equilibrium sequence.}
    \label{fig:migration_central_evolution}
\end{figure}

\begin{figure}[t]
    \centering
    \includegraphics[width=\linewidth]{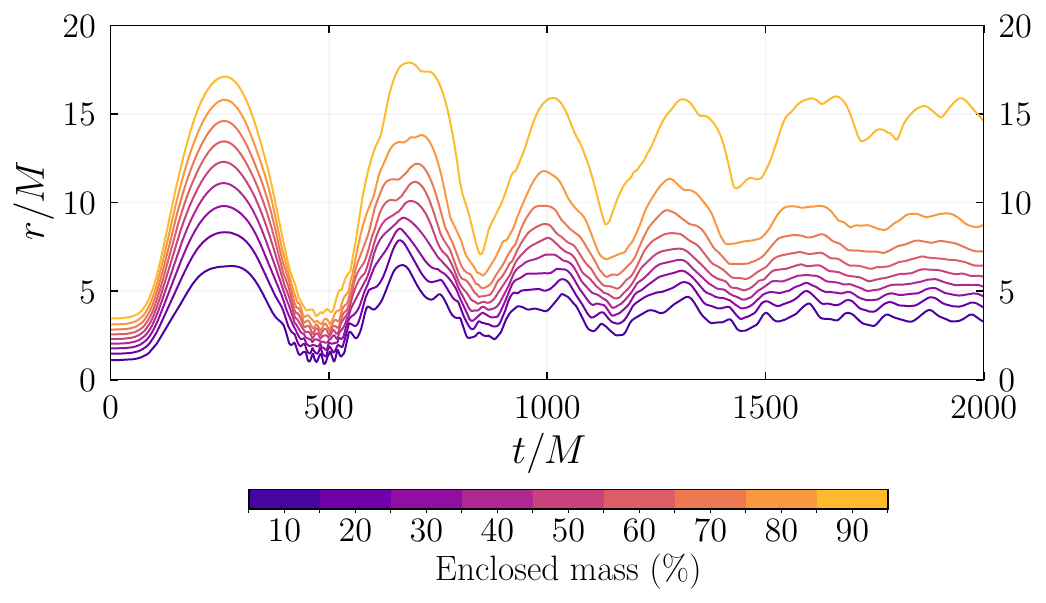}
    \caption{Evolution of the isotropic-coordinate radii enclosing $10\%\hbox{--}90\%$ of the surviving rest mass for the initially $y_c=0.500$ cluster, measured about its instantaneous center of mass. Each curve represents an instantaneous enclosed-mass fraction. The cluster undergoes large initial expansion and re-contraction, followed by oscillations of progressively smaller amplitude as the matter redistributes into a more extended configuration.}
    \label{fig:migration_mass_shells}
\end{figure}

\begin{figure*}[t]
 \centering
 \includegraphics[width=\linewidth]{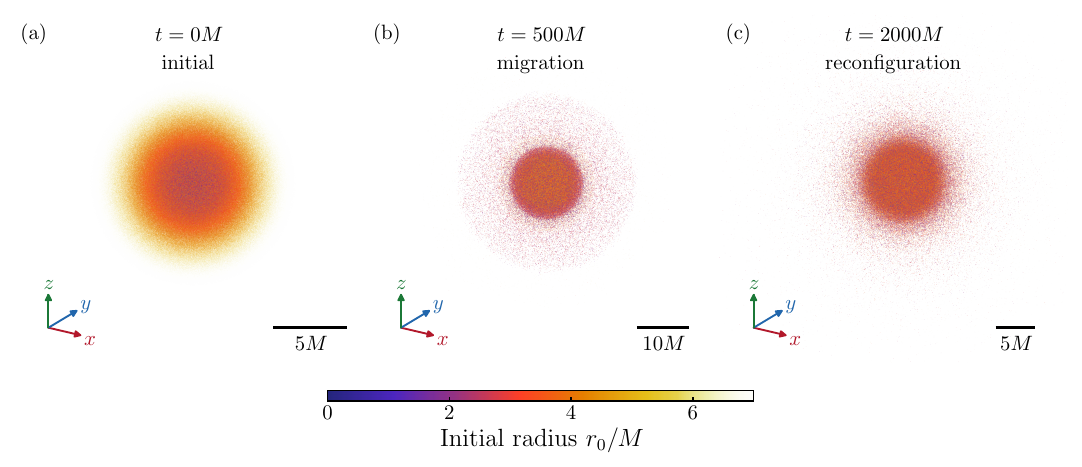}
 \caption{Three-dimensional particle distributions illustrating the nonlinear reconfiguration of an initially monoenergetic cluster into a core–halo structure at $t/M=0$, $500$, and $2000$. Particles are colored by their initial radius $r_0$, from dark purple for the innermost population to light yellow for the outermost population, with colors remaining attached to particles throughout the evolution. The overlap of initially stratified color populations illustrates substantial radial mixing during the development of the concentrated core and extended halo. The viewing angle is fixed, with the length scale labeled for each panel individually.}
 \label{fig:migration_particle}
\end{figure*}

We present the migration process through the evolution of the central lapse $\alpha_c$, density $\rho_c=T_{\mu\nu}n^\mu n^\nu$, and redshift in Fig.~\ref{fig:migration_central_evolution}. During the first expansion, the lapse rises from approximately 0.56 to 0.91 and the central energy density falls by roughly two orders of magnitude. The cluster subsequently recontracts, reaching a lapse minimum and a pronounced density maximum near $t\simeq450\hbox{--}500M$. Further expansion and contraction cycles follow, with progressively smaller amplitudes. At late times, the mean central lapse approaches approximately 0.83, while residual density fluctuations persist. This behavior qualitatively resembles the nonlinear migration of unstable neutron stars, although the reduction of coherent oscillations in the collisionless system does not require shock heating.

For comparison with the equilibrium sequence, we follow the definition in Section \ref{sec:monoenergetic} and define the central redshift as
\begin{equation}
 z_c(t)=\alpha_c(t)^{-1}-1
 \label{eq:mig-redshift-proxy}
\end{equation}
and an equilibrium-equivalent parameter $y_c(t)$ by inverting Eq.~\eqref{eq:st-redshift}. The time-averaged mean core lapse over the late stable stage gives
\begin{equation}
 \begin{aligned}
 \langle\alpha_c\rangle&=0.833530,\\
 z_c&=\langle\alpha_c\rangle^{-1}-1=0.199717,\\
 y_c&=0.844603.
 \end{aligned}
 \label{eq:mig-core-reference}
\end{equation}
These values are broadly consistent with migration toward the stable, lower-redshift branch and the prediction in Eq.~\eqref{eq:mig-binding-target}. The remaining difference may reflect residual oscillations, departures from the original monoenergetic distribution, or numerical effects. Nevertheless, the substantial increase in the central lapse and reduction of the central density demonstrate that the initially compact configuration has undergone nonlinear migration toward a less relativistic state.

We further characterize the structural evolution in Fig. \ref{fig:migration_mass_shells} using the isotropic-coordinate radii enclosing fixed fractions of the surviving particle rest mass. Specifically, $r_q(t)$ denotes the radius enclosing a fraction $q$ of the instantaneous rest mass, measured about the particle center of mass. The evolution is qualitatively similar to the central diagnostics and follows the same oscillation phase. The first expansion reaches its largest radii at approximately $t/M=250\hbox{--}300$, followed by a deep contraction near $450\hbox{--}500$ and another expansion near $650\hbox{--}700$. Subsequent pulsations have smaller amplitudes, and the inner mass scales become comparatively steady after about $t=1500M$, while the outer shells retain more visible variability. This is likely to be attributed to insufficient sampling and numerical diffusion.

{This test demonstrates that NRPIC can follow the nonlinear migration of a relativistic collisionless cluster through large-amplitude expansion, repeated recontraction, and substantial redistribution of its matter. As in the corresponding neutron-star benchmark, the evolution proceeds from an unstable relativistic equilibrium toward a less compact configuration exhibiting progressively weaker collective oscillations. Unlike a fluid star, however, the collisionless system cannot dissipate coherent motion through hydrodynamic shocks. The observed reduction of pulsation amplitudes is instead consistent with collisionless phase mixing and redistribution of orbital energies, potentially accompanied by numerical dissipation.}

\subsubsection{Cluster Reconfiguration}
\label{sec:cluster-reconfig}
{In this section, we investigate a different outcome of nonlinear evolution, in which a monoenergetic cluster near the critical binding-energy turning point undergoes substantial matter redistribution and develops a core–halo structure, which is not a member of the same monoenergetic family. Such redistribution is qualitatively consistent with collisionless relaxation, in which time-dependent collective gravitational fields and orbital phase mixing can alter the macroscopic matter distribution without direct particle collisions \cite{Lynden-Bell:1966zjv,1986MNRAS.219..285T}.}

We again use the monoenergetic configuration with near-critical $y_c=0.756$ and initial ADM mass $M=1$, represented by $N=10^6$ equal-rest-mass particles. To trigger reconfiguration, we add an initial density perturbation of $10\%$. The computational domain is $[-1024M,1024M]^3$, with a $128^3$ root grid and $32^3$ cells per MeshBlock. The initial nested hierarchy has a finest spacing $\Delta x=M/8$ and is subsequently evolved with adaptive mesh refinement.

Figure~\ref{fig:migration_particle} shows the particle distribution at $t/M=0$, $500$, and $2000$. The initially compact cluster evolves into a configuration comprising a concentrated central core surrounded by an extended, dilute halo. Each particle retains a color corresponding to its initial radius, so the subsequent overlap of colors reveals radial mixing of populations. Particles initially in the inner region can populate both the core and the envelope. By $2000M$, the cluster has reorganized into a persistent core--halo structure that differs substantially from its initial monoenergetic configuration.

\begin{figure}[t]
 \centering
 \includegraphics[width=\linewidth]{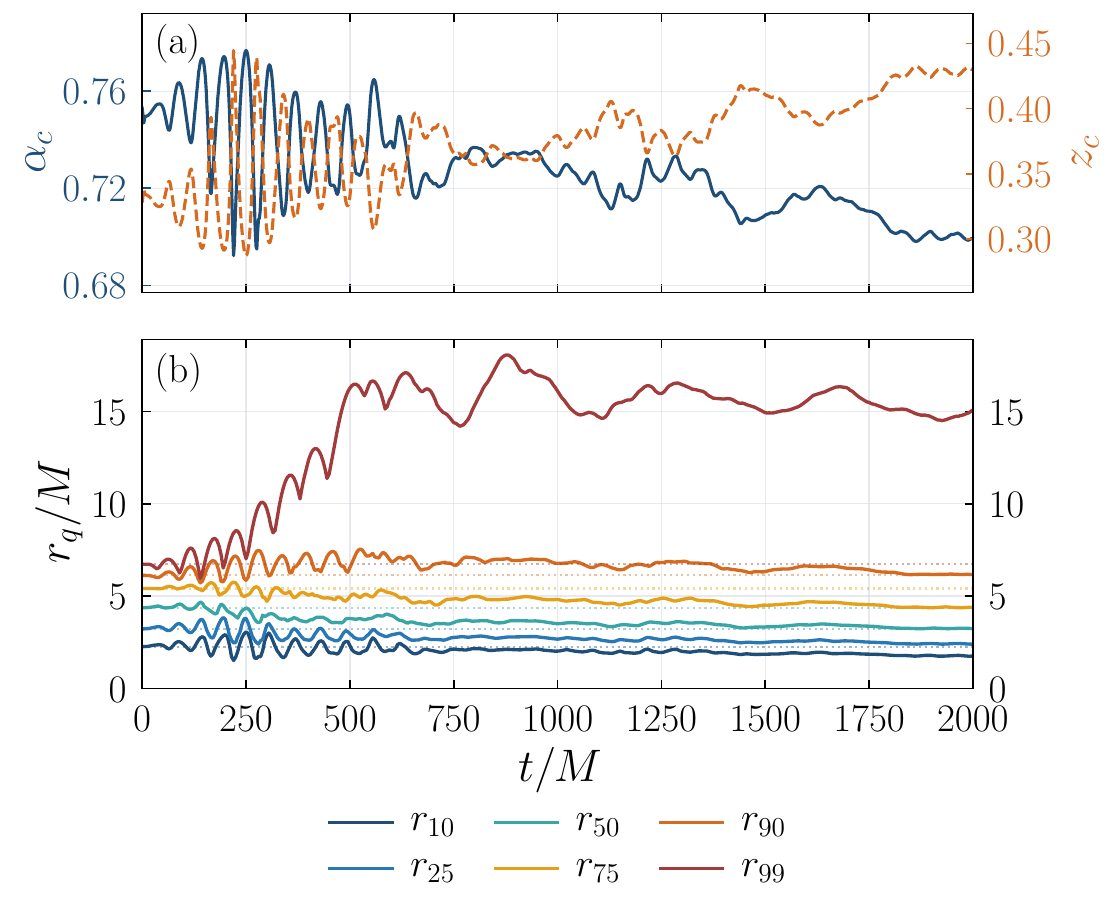}
 \caption{Spacetime and matter evolution of the cluster reconfiguration simulation through $t=2000M$. \textbf{Top}: central lapse $\alpha_c$ (solid blue, left axis), and the derived redshift diagnostic $z_c$ (dashed orange, right axis). \textbf{Bottom}: coordinate radii enclosing $10\%$, $25\%$, $50\%$, $75\%$, $90\%$, and $99\%$ of the total rest mass. Dotted horizontal lines show their initial values.}
 \label{fig:migration_mass_shell}
\end{figure}

The spacetime and matter diagnostics in Fig.~\ref{fig:migration_mass_shell} complement this spatial picture. We show the central lapse $\alpha_c$ and effective central redshift $z_c$ calculated as in Eq.~\eqref{eq:mig-redshift-proxy}. The early response contains pronounced lapse oscillations and corresponding opposite variations in $z_c$. Their amplitudes become smaller as the core--halo structure develops. At $t=2000M$, the central lapse reaches approximately $\alpha_c\simeq0.70$, corresponding to $z_c\simeq0.43$, compared with the initial values $\alpha_c\simeq0.75$ and $z_c\simeq0.33$. The decreasing lapse and increasing redshift proxy are consistent with the development of a more centrally concentrated matter distribution.

To quantify the redistribution, the lower panel of Fig. \ref{fig:migration_mass_shell} shows the coordinate radii $r_q(t)$ enclosing specified fractions $q$ of the total rest mass $M_0$, defined as
\begin{equation}
 M_0(<r_q,t)=qM_0.
 \label{eq:migration-mass-shell}
\end{equation}
All particles remain present throughout the displayed interval, and their total rest mass is conserved. The radii are measured about the coordinate origin and represent instantaneous enclosed-mass quantiles rather than trajectories of fixed particle populations. During the first several hundred $M$, the mass radii undergo large-amplitude oscillations, accompanied by an increasing separation between the inner and outer radial scales, also as demonstrated in the middle panel of Fig.~\ref{fig:migration_particle}. The inner quantiles contract below their initial values, while the radius enclosing 99\% of the rest mass, $r_{99}$, grows from approximately $7M$ to excursions near $18M$ before settling around $15M$ at $t=2000M$. The subsequent reduction in oscillation amplitudes, together with the separation of the inner and outer mass scales, characterizes the emergence of a persistent, centrally concentrated core and extended halo.

The formation of this core–halo structure motivates a comparison with the Plummer family, a classical analytic model of self-gravitating stellar systems \cite{Plummer:1911zza}. The Newtonian Plummer model consists of a regular central core and an extended power-law envelope, with finite total mass despite its infinite spatial extent. Its gravitational potential and density are
\begin{align}
 \Phi_{\rm P}(r)&=-\frac{M}{\sqrt{r^2+a^2}},
 \label{eq:plummer-newton-potential}\\
 \rho_{\rm P}(r)&=\frac{1}{4\pi}\nabla^2\Phi_{\rm P}=\frac{3Ma^2}{4\pi(r^2+a^2)^{5/2}},
 \label{eq:plummer-newton-density}
\end{align}
where $a$ is the characteristic scale radius. The density is finite at the origin and falls as $r^{-5}$ at large radius. The corresponding enclosed mass, $M_{\rm P}(<r)=Mr^3/(r^2+a^2)^{3/2}$, approaches $M$ at infinity. The regular central density and extended envelope make the Plummer family a useful reference for characterizing the radial structure of our reconfigured collisionless cluster.

A relativistic analogue of the Plummer model was first constructed as an exact, static, spherically symmetric perfect-fluid solution \cite{1964ApJ...140.1512B}. Its collisionless interpretation was subsequently investigated in Ref.~\cite{1971ApJ...165..489F}. We use the isotropic Plummer member of the relativistic hypervirial family
\cite{Nguyen:2013nka,Gauy:2017cyu},
\begin{align}
 ds^2 &=-\alpha^2dt^2+\psi^4(dr^2+r^2d\Omega^2),
 \label{eq:plummer-metric}\\
 u(r)&=\frac{M}{2\sqrt{r^2+a^2}}, \qquad \psi(r)=1+u(r), \nonumber\\
 \alpha(r)&=\frac{1-u(r)}{1+u(r)}.
 \label{eq:plummer-metric-functions}
\end{align}
Here $r$ is an isotropic coordinate radius, related to the areal radius by $R=r\psi^2$. For this family, the central lapse is fixed by the mass-to-scale-radius ratio,
\begin{equation}
\alpha_c=\frac{1-M/(2a)}{1+M/(2a)},
\end{equation}
with $a>M/2$ ensuring a positive, regular lapse. In the weak-field limit $M/a\ll1$, we recover $\alpha(r)=1+\Phi_{\rm P}(r)+O[(M/a)^2]$ as expected from the Newtonian Plummer potential Eq.~\eqref{eq:plummer-newton-potential}.

With zero shift and $K_{ij}=0$, the Hamiltonian constraint becomes
\begin{equation}
 \nabla^2\psi=-2\pi\psi^5\rho,
 \label{eq:plummer-hamiltonian}
\end{equation}
where $\nabla^2$ is the flat-space Laplacian and $\rho=n_\mu n_\nu T^{\mu\nu}$ is the energy density measured by the static Eulerian observer. Using
\begin{equation}
 \nabla^2u=-\frac{3Ma^2}{2(r^2+a^2)^{5/2}},
\end{equation}
we obtain the required density
\begin{equation}
 \rho(r)
 =\frac{3Ma^2}
 {4\pi(r^2+a^2)^{5/2}[1+u(r)]^5}.
 \label{eq:plummer-rel-density}
\end{equation}
The radial and angular components of Einstein's equations give equal orthonormal stresses,
\begin{equation}
 P_r=P_\theta=P_\phi=P,
 \qquad
 P(r)=\frac{\rho(r)u(r)}{3[1-u(r)]}.
 \label{eq:plummer-pressure}
\end{equation}
Thus the relativistic model retains a regular core and an asymptotic $r^{-5}$ envelope, while its interior energy density differs from the Newtonian Plummer profile by the factor $(1+u)^{-5}$. The isotropy of the stresses also makes this solution a natural reference for velocity-supported collisionless configurations.

{For a collisionless interpretation, the stress-energy tensor must be reproduced by a nonnegative particle distribution satisfying the stationary Vlasov equation. In a static, spherically symmetric configuration with isotropic momentum distribution, the distribution function can depend on the conserved specific Killing energy, i.e., $f(e)$, where $e=-u_t=\alpha W$.} The corresponding moments can again be constructed using Eqs.~\eqref{eq:kinetic-moments}--\eqref{eq:vlasov-stress-energy}
\begin{align}
 \rho_0
 &=4\pi\int_0^{q_{\max}}f(\alpha W)q^2\,dq,
 \nonumber\\
 \rho
 &=4\pi\int_0^{q_{\max}}f(\alpha W)Wq^2\,dq,
 \nonumber\\
 P
 &=\frac{4\pi}{3}\int_0^{q_{\max}}
          f(\alpha W)\frac{q^4}{W}\,dq.
 \label{eq:plummer-moments}
\end{align}
where $q_{\max}=\sqrt{\alpha^{-2}-1}$ for bound particles with $e\leq 1$. {The existence and physical admissibility of such relativistic distribution functions have been investigated \cite{1971ApJ...165..489F,Gauy:2017cyu}, and are used here as a motivation to characterize the rest-mass distribution of the reconfigured cluster using a generalized Plummer-like profile ansatz}
\begin{equation}
    D_{\rm fit}(R)\equiv\frac{1}{4\pi R^2}\frac{dM_0}{dR} =D_*\left[1+(R/a_{\rm fit})^2\right]^{-p},
 \label{eq:migration-profile-fit}
\end{equation}
where $p$ is the power-law index, $a_{\rm fit}$ is the fitted scale radius, and $D_\ast$ normalizes the total rest mass. A least-squares fit gives
\begin{equation}
 a_{\rm fit}\simeq6.52M,\qquad p\simeq4.17.
 \label{eq:migration-fit-parameters}
\end{equation}
These parameters provide an approximate description of the mass distribution.

{The classical Plummer model corresponds to $p=5/2$, whereas the fitted profile has a substantially steeper formal outer falloff, $D_{\rm fit}\propto R^{-2p}\simeq R^{-8.34}$ for $R\gg a_{\rm fit}$. The comparison places the reconfigured cluster in a different radial-profile family from the initial monoenergetic model. Its core--halo structure has the same broad physical organization as idealized stellar-cluster models supported by velocity dispersion. Collisionless redistribution, phase mixing, and potential numerical dissipation provide a natural interpretation of the observed mixing and reduced coherent pulsations.}

\section{Jeans collapse and binary evolution in a relativistic cluster}
\label{sec:Jeans}

\begin{figure*}[t]
    \centering
    \includegraphics[width=\linewidth]{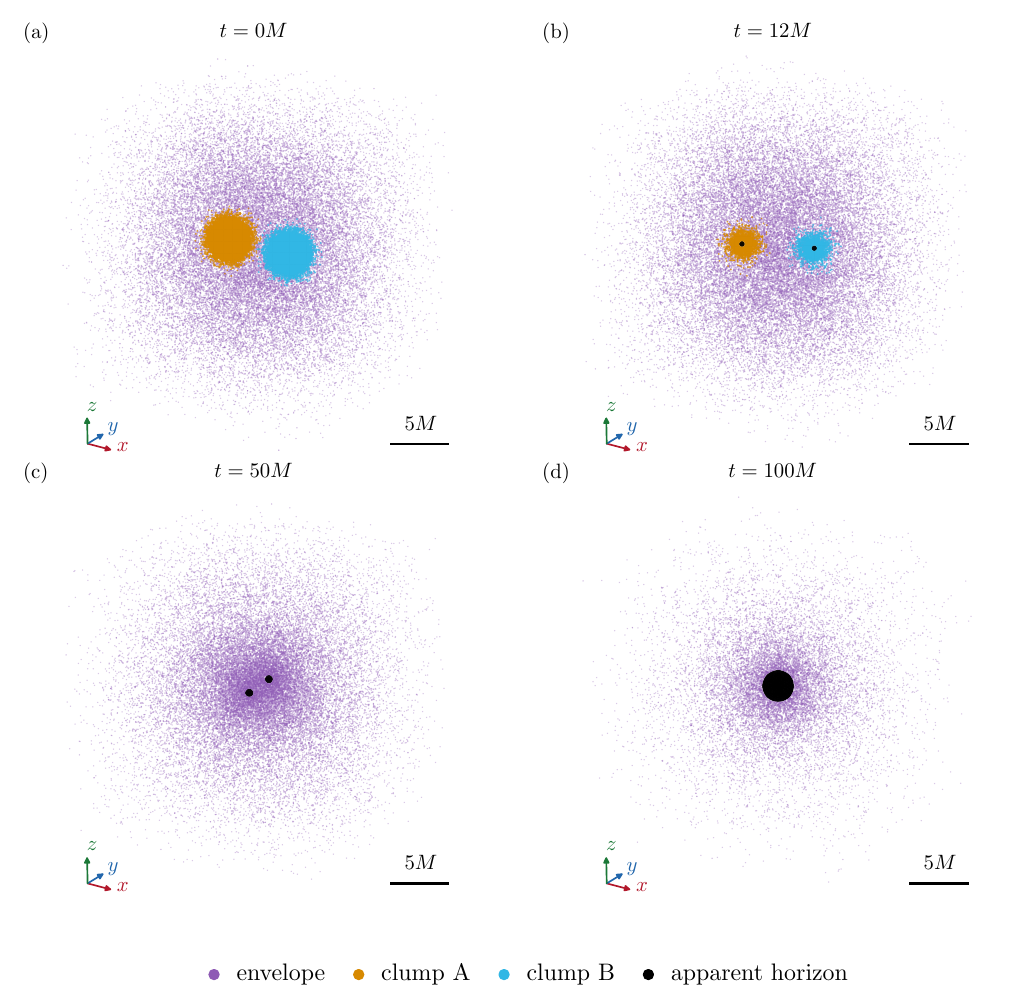}
    \caption{Three-dimensional Jeans instability and black hole formation inside a self-gravitating cluster. Shown are particle snapshots at \(t/M=0,12,50,100\). Purple, orange, and blue identify the envelope and original clumps A and B, respectively. Opaque black spheres mark accepted apparent horizons using area-equivalent radii \(\sqrt{A_{\rm AH}/(4\pi)}\). Shown are a) initial Gaussian overdensities, b) collapse and black hole formation, c) binary inspiral inside the self-gravitating cluster, d) final merger product.}
    \label{fig:Jeans_particles}
\end{figure*}

One important aspect of collisional dynamics is the dynamical formation of black holes in nontrivial environments.
As such, we now consider the case of gravitational instability and black hole formation from overdensities in a self-gravitating cloud. This is a relativistic version of the standard Jeans collapse paradigm \cite{2008gady.book.....B}.

Jeans instability arises when the self-gravity of a density perturbation overcomes the dispersive effect of random particle motions, providing a basic mechanism for gravitational structure formation in collisionless matter. In a homogeneous Newtonian medium with an isotropic Maxwellian velocity distribution, perturbations are unstable on wavelengths \cite{2008gady.book.....B}
\begin{equation}
    \lambda>\lambda_{\rm J},
\qquad
\lambda_{\rm J}
=\left(\frac{\pi\sigma^{2}}{G\rho_{0}}\right)^{1/2},
\end{equation}
where \(\rho_{0}\) is the unperturbed mass density, \(\sigma\) is the one-dimensional velocity dispersion, and \(G\) is Newton’s gravitational constant. At fixed velocity dispersion, increasing the density therefore lowers the characteristic scale above which self-gravity can drive contraction. This competition motivates the construction of overdense clumps within an otherwise stable cluster. For a finite, inhomogeneous relativistic system, the Newtonian criterion provides qualitative guidance, while the subsequent collapse must be followed through the coupled evolution of the particle distribution and spacetime. Sufficiently compact configurations can form black holes while leaving an extended collisionless distribution outside the horizons. Introducing multiple clumps thus provides an idealized setting for studying localized gravitational collapse and the subsequent interaction of black holes within a common matter environment. Numerically, this tests the ability of our method to follow several collapsing regions and their surrounding collisionless matter self-consistently through black-hole formation and evolution.

Specifically, we reuse a stable configuration of the monoenergetic distribution in Section \ref{sec:st-cluster}. In the following part, we use the system's normalized ADM total mass $M=1$. The equilibrium cluster has gravitational mass $M_c=0.76M$ and an areal surface radius of $30M$. We also place two Gaussian clumps as density perturbations inside the cluster, each with mass $0.12M$ and Gaussian width $0.70M$. The centers of the two clumps are placed at $\boldsymbol{x}_A=(-3,0,0)M$ and
$\boldsymbol{x}_B=(3,0,0)M$. We set the bulk velocities of the cluster to $v_A^{\hat y}=-0.133215$ and $v_B^{\hat y}=+0.133215$; therefore, the clumps acquire positive angular momentum and initial circular {orbits} to leading order, calculated from both the central potential of the cluster and the perturbation from the other clump.

This simulation uses $5\times10^6$ particles: $3\times10^6$ in the envelope and $10^6$ in each clump. Therefore, the particle resolution in each clump is approximately twice that in the envelope, providing a more precise evolution of the collapse. The computational domain is $[-512M,512M]^3$, with a $256^3$ root grid, $32^3$-cell mesh blocks, and eleven permitted refinement levels. The initial finest spacing is $M/64$, with refinement permitted down to $M/256$. Particles are removed when the lapse falls below $0.05$. We also activate the AH finder for each initial clump and the common horizon after merger, and record the gravitational wave data.

In Fig.~\ref{fig:Jeans_particles}, we visualize the particle distribution in 3D and color the particles using purple, orange, and blue to identify membership in the envelope and the two clumps. We show four snapshots including the initial configuration, the gravitational instability and formation of black holes, binary inspiral in a collisionless environment, and post-merger accretion and growth of the merger product.  We observe black hole formation in clumps A and B at $t=8.475M$ and $8.525M$, as evidenced by apparent horizon formation. A common enclosing apparent horizon at merger is found at $t_{\rm m}=64.15625M$.

\begin{figure}[t]
    \centering
    \includegraphics[width=\linewidth]{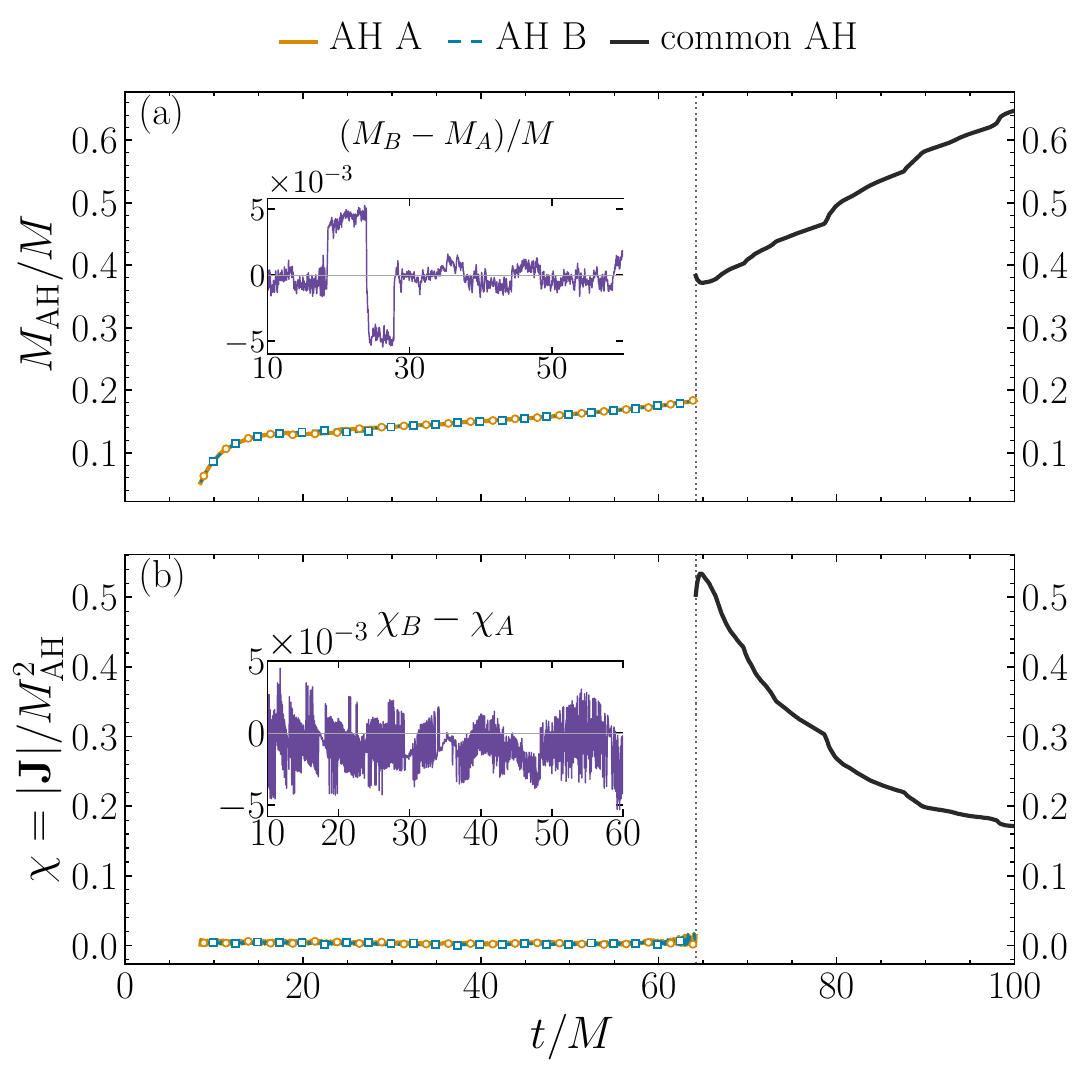}
    \caption{Mass and spin diagnostics for the Jeans collapse simulations. Shown are a) masses, $M_{\rm AH}$, and b) dimensionless spins \(\chi=|\boldsymbol{J}|/M_{\rm AH}^{2}\). Orange solid curves with open circles and blue dashed curves with open squares identify the two individual horizons; black curves identify the common horizon. Insets show the differences \(B-A\). The vertical dotted line marks the first accepted common horizon at \(t/M=64\), where the individual-horizon curves end.}
    \label{fig:Jeans_AH}
\end{figure}

\begin{figure}[t]
    \centering
    \includegraphics[width=\linewidth]{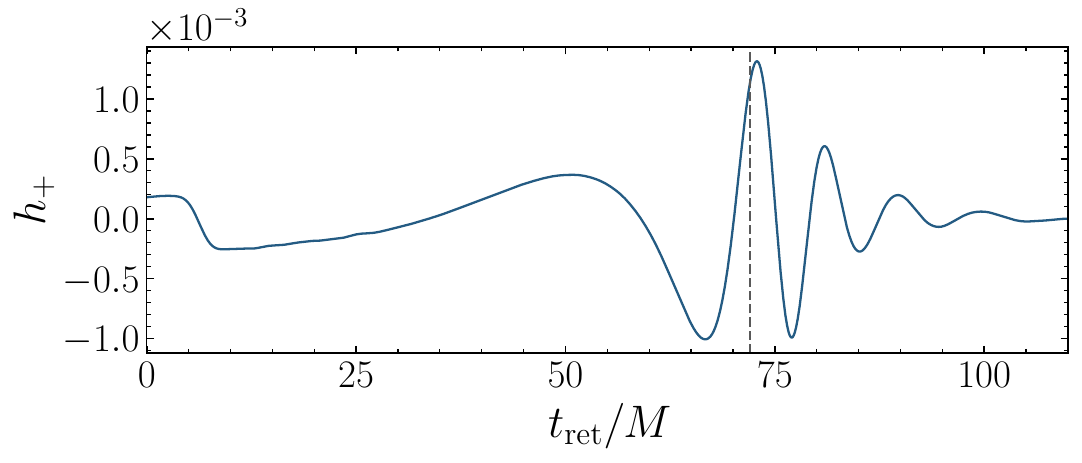}
    \caption{Gravitational-wave strain \(h_+\) extracted at coordinate radius \(R=40M\), plotted against approximate retarded time \(t_{\rm ret}=t-R\). The vertical dashed line marks the maximum total strain amplitude, \(|h|=\sqrt{h_+^2+h_\times^2}\), used as an approximate waveform-based merger-time marker.}
    \label{fig:Jeans_GW}
\end{figure}

We show the detailed AH mass and dimensionless spin ($\chi=|\boldsymbol{J}|/M_{\rm AH}^2$) evolution in Fig.~\ref{fig:Jeans_AH}. We also show the relative difference in mass and dimensionless spin between the two BHs during their formation and inspiral, which is negligible and shows rather symmetric evolution paths. After the first AH formation, the BHs continue to accrete from the clump, characterized by a relatively steep growth curve before $t=15M$, when most of the clump material is accreted into the BH. Later, the mass of each BH continues to grow, but at a relatively slow rate, similar to Bondi-Hoyle-Lyttleton accretion from the envelope material while simultaneously inspiraling inwards \cite{Bondi:1944rnk}. At \(t=10M\), the individual masses are $0.08802M$ and $0.08772M$; by $60M$, they have increased to $0.17421M$ and $0.17605M$, exceeding the initial clump mass of $0.12M$. The BHs during this stage have negligible spins (of order $10^{-5}$). On the one hand, we only impose bulk velocities on the clumps treated as point masses but do not add any relative shear motion around the clump center. On the other hand, the background envelope uses the isotropic monoenergetic distribution and results in zero net angular momentum. The accretion of zero net angular momentum envelope material during the inspiral phase also results in more rapid orbital shrinkage and subsequent merger on a much shorter timescale.

We want to caution here that the drag force from envelope matter could also transfer angular momentum between the black-hole binary and the environment, which drives faster coalescence \cite{Chandrasekhar:1943ys}. To assess and quantify the redistribution during the inspiral, we compare the orbital angular momentum inferred from the individual horizon momenta with the angular momentum of the surviving envelope. Between $t=20M$ and $60M$, the summed BH orbital angular momentum decreases by $27.3\%$, while the envelope gains a comparable amount, when regions enclosing both horizons are excluded. Thus, although the envelope initially has zero net angular momentum, it acquires positive angular momentum as the binary inspirals.

Merger occurs roughly at $t=64.175M$ when a common horizon is found. Initially, it has $M_{\rm AH}=0.38326M$ and $\chi=0.50355$, which is mostly inherited from the orbital angular momentum. We observe that the merger remnant continues to accrete from the envelope at a nearly constant rate. By $t=100M$, the mass has grown to $0.64740M$. Since the accreted envelope material has zero net angular momentum, accretion raises only the mass of the merged BH, without changing its absolute spin magnitude. Consequently, the dimensionless spin of the final remnant decreases to $\chi=0.17132$ at $t=100M$. We fit the decrease in dimensionless spin as proportional to $M^{-2}_{\rm AH}$ and check the conservation of total angular momentum. Indeed, we find that $J=\chi M^2_{AH}$ is conserved to within 2\% during the evolution. Therefore, a decrease in $\chi$ alone does not measure a corresponding fractional loss of spin angular momentum.

Finally, we show the gravitational wave strain extracted at finite coordinate radius $R=40M$ in Fig.~\ref{fig:Jeans_GW}. We reconstruct the observer signal from all available $2\leq\ell\leq8$ modes and use fixed-frequency integration \cite{Reisswig:2010di} to get the strain from the $\Psi_4$ Penrose scalar. The waveform contains a pronounced late oscillatory pulse and a clear ring-down signature \cite{Berti:2009kk}, but no inspiral signal, consistent with the fact that the cloud interaction leads to rapid orbital decay.

\section{Conclusions}

Recent interest in extremal black hole formation from collisionless matter \cite{Kehle:2024vyt,East:2025nfb} and questions about compact-object formation via gravitational instability \cite{Toomre:1964zx} in a self-gravitating relativistic environment \cite{Chen:2025uwd} have raised the need for first-principles particle-based models of numerical relativity, including accurate solutions of the Einstein-Vlasov system. While only a small number of studies have made progress on this topic over the past decades \cite{1985ApJ...298...34S,1985ApJ...298...58S,Shibata:1999va,Shibata:1999wi,Yoo:2016kzu,East:2019bwu} (see also Refs. \cite{Diener:2022hui,Rosswog:2023nnl} for smoothed particle hydrodynamics in dynamical GR, and Ref. \cite{Andreasson:2011ng} for a review of mathematical progress), the corresponding Maxwell-Vlasov system in general-relativistic plasma astrophysics is now routinely solved at high computational fidelity \cite{Parfrey:2018dnc,Chen:2019osy,Hirotani:2020ubl,Torres:2024usc,Chen:2025uwd,Galishnikova:2025pfi}.
Here, by using some of the insights and modern techniques gained from the plasma physics community, we present a Numerical-Relativity Particle-In-Cell (NRPIC) algorithm for the Einstein-Vlasov system implemented in a performance-portable way on top of the adaptive mesh-refinement \texttt{AthenaK} framework \cite{2026ApJS..283...27S}. By exploiting the electrodynamic structure of the Einstein equations \cite{Olivares:2021cfa,Boyeneni:2025tsx}, we have also constructed a Boris-type particle pusher \cite{boris1970relativistic}, akin to those routinely used in plasma physics problems \cite{Bacchini:2020cze,Zou:2024prf}.

We present a series of test problems in spherical, axisymmetric, and nonsymmetric settings to assess the validity and accuracy of the code.
For Oppenheimer--Snyder dust collapse,
the deposited density and central lapse reproduce the analytic and
moving-puncture benchmarks, and the fiducial final horizon mass agrees with
the initial ADM mass to within $10^{-3}$. We confirm recent simulations \cite{East:2019bwu} that the collapse of prolate spheroids does not lead to naked singularity formation \cite{Shapiro:1991zza}, and provide a detailed analysis of its time evolution.

Relativistic clusters provide a complementary test of sustained kinetic
support and nonlinear matter redistribution. Weakly relativistic
monoenergetic models \cite{1985ApJ...298...34S} remain close to equilibrium, while inward-perturbed
compact models collapse to black holes. The negative-binding-energy model
instead undergoes expansion. 
Direct NRPIC simulations demonstrate that the method
can follow substantial changes in the phase-space distribution beyond
equilibrium preservation in all these cases.

As a first step towards more detailed studies of the gravitational instability, we have followed the collapse of two localized
overdensities inside a common collisionless self-gravitating envelope. Both overdensities are Jeans unstable and form black holes, which accrete, inspiral, and merge while remaining coupled to the
surrounding matter. The envelope acquires angular momentum during inspiral,
consistent with redistribution between the binary and its environment.
In addition, the merged black hole undergoes a mass accretion and growth period.
The calculation also produces a gravitational-wave signal of the resulting dynamics, illustrating its potential in hierarchical merger scenarios \cite{Wu:2026hth}. This idealized example demonstrates that
black-hole formation, binary evolution, and accretion from collisionless
matter can be followed within a single dynamical-spacetime calculation.

Future work should be devoted to the inclusion of higher-order methods \cite{Rosswog:2023nnl,Boss:2026fyj}, extremal black hole formation \cite{Gelles:2025gxi,East:2025nfb} (see also Ref. \cite{Kehle:2024vyt}), and cosmic censorship \cite{Shapiro:1991zza,East:2019bwu}, as well as gravitational instability and collapse more generally.

\begin{acknowledgments}
The authors gratefully acknowledge insightful discussions with Anatoly Spitkovsky and Saul Teukolsky.
ERM and JW acknowledge support from the U.S. National Science Foundation {(NSF)} under Grant No. PHY-2541792 and through a Research Fellowship from the Sloan Foundation. ERM is also supported by a William H. Hurt Scholarship at Caltech.
The simulations were performed on the NSF Vista supercomputer under grant AST21006. The authors also acknowledge the use of Delta at the National Center for Supercomputing Applications (NCSA) through allocation PHY210074 from the Advanced Cyberinfrastructure Coordination Ecosystem: Services \& Support (ACCESS) program, which is supported by National Science Foundation grants \#2138259, \#2138286, \#2138307, \#2137603, and \#2138296. Support also comes from the Resnick High Performance Computing Center, a facility supported by the Resnick Sustainability Institute at the California Institute of Technology. We acknowledge support from OpenAI’s Researcher Access Program for providing access to ChatGPT models to support this work. Work in this paper primarily used OpenAI's ChatGPT 5.6-sol, 5.5, and 6-astra, with additional tasks carried out by Anthropic's Opus 4.8 and 5, as well as Fable 5 and 5.1.
\end{acknowledgments}

\appendix

\begin{figure*}[t]
\centering
\includegraphics[width=\textwidth]{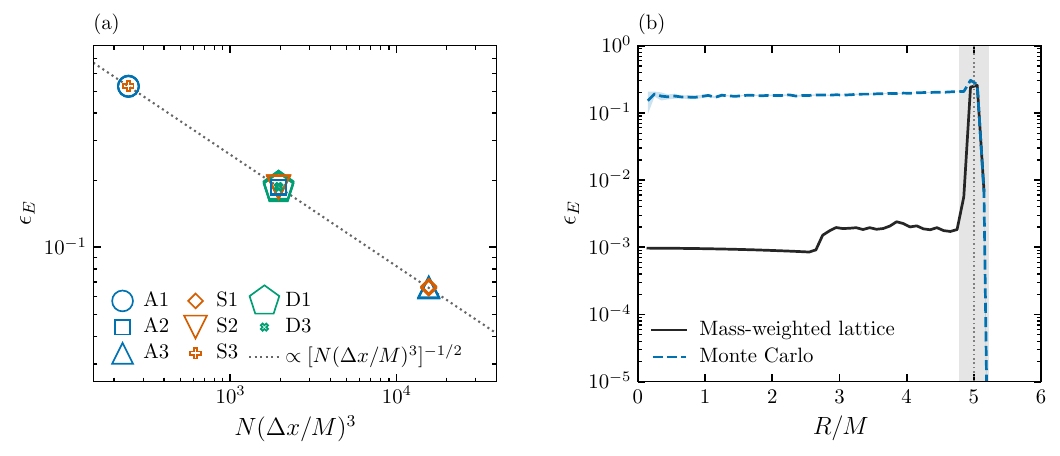}
\caption{Initial Oppenheimer-Snyder sampling error. (a) Interior proper-volume-weighted energy-density error for all eight cases. Markers are labeled in the bottom left corner, and the same color is used for each sequence. The dotted line has log-log slope $-1/2$. (b) Shell-average error at matched particle count, using analytic initial areal-radius bins of width $0.1M$. The Monte Carlo curve is the five-seed mean, with a band of one sample standard deviation. The vertical line marks $R_0=5M$ and the gray band indicates the region where the deposition kernel overlaps with the stellar surface, where truncation error and sampling discontinuity dominate the error.}
\label{fig:os-initial-new}
\end{figure*}

\section{Oppenheimer--Snyder convergence}
\label{sec:os-convergence}

The Oppenheimer--Snyder (OS) problem is a well-understood problem with an exact reference solution. As such, it provides an ideal testbed for assessing the accuracy and convergence properties of our algorithm and implementation. This allows us to separate errors in particle sampling, spacetime resolution, and the final black-hole mass, which should be perfectly conserved in the absence of gravitational wave emission. We use the same initial geometric configuration as in the main text, with ADM mass $M=1$, initial areal radius $R_0=5M$, initially unit lapse, and vanishing particle momentum. All evolved convergence cases use Monte Carlo sampling (see Section~\ref{sec:os-collapse}, see also the discussion below for implementation details) and are followed to $t=60M$. Table~\ref{tab:os-convergence-new} summarizes eight configurations, each with five different seeds to avoid errors from a specific random sample.

In Table~\ref{tab:os-convergence-new}, we present three sequences probing different numerical errors. A1\hbox{--}A3 increase $N$ at fixed mesh and AMR prescription, with initial finest spacing $M/8$ and minimum permitted spacing $M/32$. Their actual refinement histories can depend on the particle realization, but all simulations reach the maximum refinement level. S1\hbox{--}S3 fix particle number at $N=10^6$ and perform static mesh refinement at different maximum resolutions. The full cluster is covered by the uniform maximum level for each of these runs. The diagonal D1\hbox{--}S2\hbox{--}D3 increases $N$ by a factor of eight whenever the finest spacing is halved, keeping $N(\Delta x/M)^3$ constant. This combination measures the error scaling of resolution and particle number while keeping the particle number per cell $N_{\rm ppc}$ constant.

\begin{table*}[t]
\centering
\caption{OS convergence configurations. The columns are as follows: {\it Mesh} refers to the use of adaptive mesh refinement (AMR) or static mesh refinement (SMR), $N$ denotes particle number, $\Delta x/M$ denotes the finest level resolution, $\epsilon_E$ is the initial interior RMS error defined in Eq.~\eqref{eq:os-initial-error-new}, quantifying the initial error from sampling noise, and $M_{\rm irr}$ is the apparent-horizon mass measured at the simulation termination time $t=60M$.}
\label{tab:os-convergence-new}
\begin{tabular}{llrrcc}
\toprule
Case & Mesh & $N$ & $\Delta x/M$ & $\epsilon_E$ &
$\langle M_{\rm irr}\rangle/M$ \\
\midrule
A1 & AMR & $125\,000$ & $1/8\rightarrow1/32$ & 0.52643 & $0.997943\pm0.000140$\\
A2 & AMR & $1\,000\,000$ & $1/8\rightarrow1/32$ & 0.18617 & $0.999663\pm0.000976$\\
A3 & AMR & $8\,000\,000$ & $1/8\rightarrow1/32$ & 0.06607 & $0.999535\pm0.000636$\\
\midrule
S1 & SMR & $1\,000\,000$ & $1/4$ & 0.06629 & $1.000019\pm0.000105$\\
S2 & SMR & $1\,000\,000$ & $1/8$ & 0.18617 & $1.000655\pm0.000090$\\
S3 & SMR & $1\,000\,000$ & $1/16$ & 0.52741 & $1.000549\pm0.000056$\\
\midrule
D1 & SMR & $125\,000$ & $1/4$ & 0.18577 & $0.999650\pm0.000308$\\
D3 & SMR & $8\,000\,000$ & $1/16$ & 0.18657 & $1.000889\pm0.000012$\\
\bottomrule
\end{tabular}
\end{table*}

For Monte Carlo sampling of equal-mass particles, the probability density in isotropic coordinate volume is proportional to $\rho_0\psi^6$, where $\rho_0=3M/(4\pi R_0^3)$ and $\psi$ is the conformal factor. Each particle carries rest mass $M_0/N$, with $M_0=\int\rho_0\psi^6\,d^3x$. For the lattice initialization used in the fiducial evolution, we retain cell centers of a uniform Cartesian lattice inside the initial isotropic surface $\bar r_0$ and assign
\begin{equation}
m_p=\rho_0\psi^6(\boldsymbol{x}_p)h^3,
\label{eq:os-lattice-mass-new}
\end{equation}
where $h$ is the particle-lattice spacing.

\begin{figure*}[t]
\centering
\includegraphics[width=\textwidth]{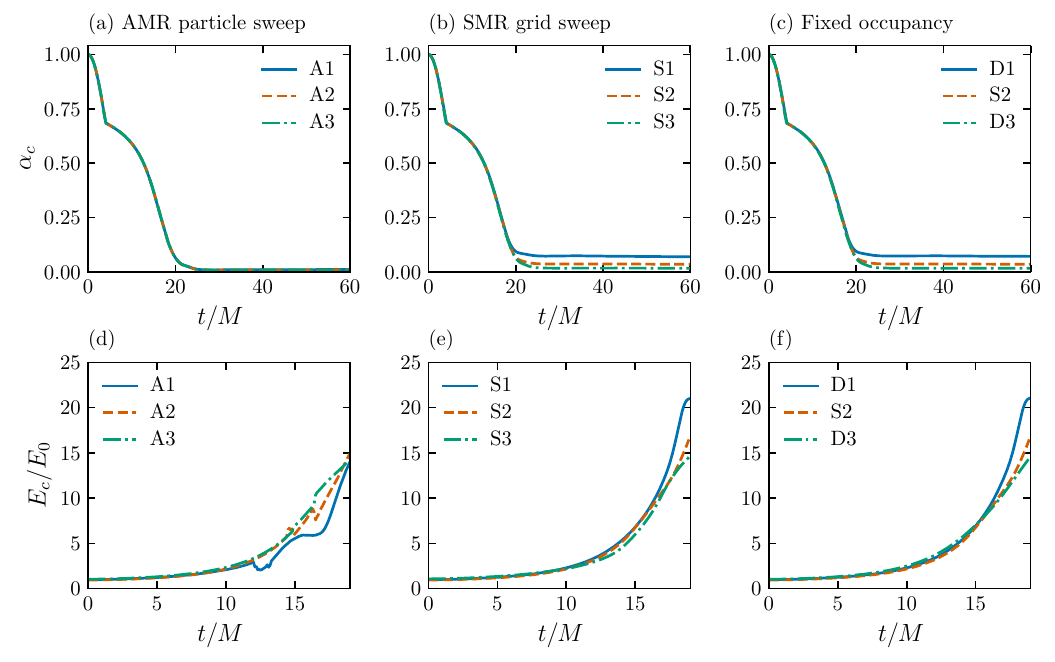}
\caption{Central evolution in the Oppenheimer--Snyder convergence tests. Panels (a)\hbox{--}(c) show the central lapse $\alpha_c$ and panels (d)\hbox{--}(f) the central Eulerian energy density $E_c/E_0$, both evaluated from the average of eight central cells. The columns compare the AMR particle-number sequence A1\hbox{--}A3, the static-refinement grid sequence S1\hbox{--}S3, and the fixed-occupancy sequence D1\hbox{--}S2\hbox{--}D3. Lapse time series extend to the termination time of the simulation, while density histories are restricted to times before the earliest particle excision in each sequence.}
\label{fig:os-history-new}
\end{figure*}

\begin{figure*}[t]
\centering
\includegraphics[width=\linewidth]{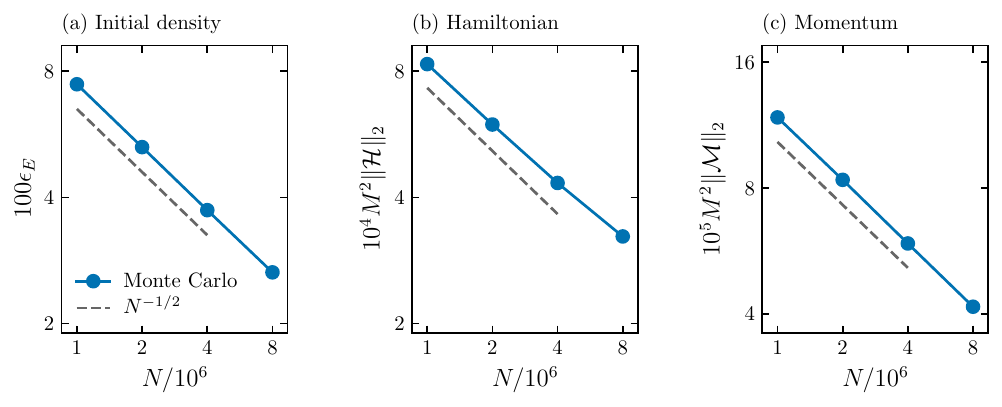}
\caption{Initial sampling error of the monoenergetic Shapiro-Teukolsky cluster with $y_c=0.819$ as a function of total particle number on a log-log scale. (a) Energy-density error as defined in Eq.~\eqref{eq:os-initial-error-new}, multiplied by a factor of $10^2$. (b,c) Hamiltonian and momentum volume-weighted norms averaged over the region with cluster matter, multiplied by $10^4$ and $10^5$ respectively. Dashed lines show $N^{-1/2}$ scaling in each panel.}
\label{fig:st-initial-new}
\end{figure*}

\begin{figure}[t]
\centering
\includegraphics[width=\linewidth]{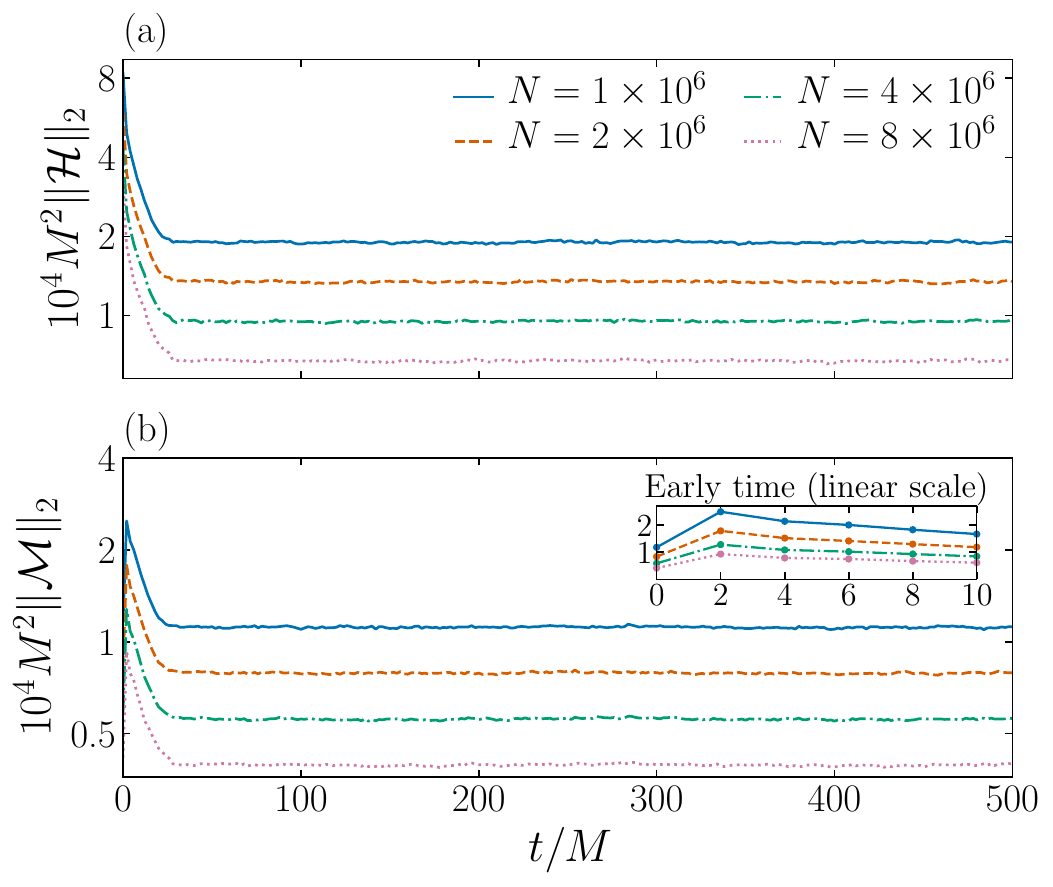}
\caption{Hamiltonian (top) and momentum (bottom) constraint evolution for the four particle counts. The volume-weighted average norm is calculated in the fixed central region $r<10M$, which covers the size of the full cluster. Vertical axes are logarithmic and include explicit multipliers $10^4$ and $10^5$ respectively. The inset in the bottom panel shows an early transient increase in the momentum constraint on a linear scale, followed by a drop on a timescale similar to that of the Hamiltonian constraint. These curves test constraint accuracy at a fixed mesh and do not assert convergence of the late-time lapse or shell oscillations.}
\label{fig:st-constraints-new}
\end{figure}

First, we quantify the initial sampling error in the deposited Eulerian energy density using a grid-volume-weighted estimator
\begin{equation}
\epsilon_E = \left[ \frac{\sum_{i\in\mathcal I}w_i[(E_i-E_0)/E_0]^2} {\sum_{i\in\mathcal I}w_i} \right]^{1/2},
\quad w_i=\psi_i^6\Delta V_i ,
\label{eq:os-initial-error-new}
\end{equation}
where $E_0$ is the analytical proper energy density and $\mathcal I$ is the set of cells within $r\leq0.7\bar r_0$ to exclude the sharp stellar surface. We document the values of $\epsilon_E$ of all eight runs in Table~\ref{tab:os-convergence-new}, and show the scaling law $\epsilon_E\propto [N(\Delta x/M)^3]^{-1/2}$ in the left panel of Fig.~\ref{fig:os-initial-new}. We find that for fixed grid resolution, increasing particle number decreases the sampling error according to $N^{-1/2}$, while for fixed particle number, higher grid resolution increases the sampling error inversely. The control parameter is particle number per cell, and the error is approximately constant across the D1\hbox{--}S2\hbox{--}D3 suite.

To compare the accuracy of the Monte Carlo sampler and the volume-weighted lattice sampler used in the fiducial test in Section~\ref{sec:os-collapse}, we show the Eulerian energy density error as a function of radius in the right panel of Fig.~\ref{fig:os-initial-new}. We use the same error estimator in Eq.~\eqref{eq:os-initial-error-new}, but averaged over cells belonging to certain radial shells. Both samplers use the same number of particles and the same grid resolution. The plot shows that the lattice sampler generates an initial error two orders of magnitude smaller than the Monte Carlo sampler in the stellar interior. Its advantage is much smaller in the surface region, where the discontinuity and truncation error dominate.

Next, we perform a convergence test on the evolution of the central lapse $\alpha_c$ and energy density $E_c$ normalized to the initial value $E_0$. Fig.~\ref{fig:os-history-new} shows the time evolution of these central quantities across different convergence suites. In the top row, we see from A1\hbox{--}A3 that particle number does not influence the evolution of the geometric field much, and with sufficient resolution from AMR, the central lapse already converges even with the minimum particle number. In the other two sequences, however, we find that the central lapse converges with increasing grid resolution.

In the bottom row, meanwhile, we find that both particle number and resolution affect the evolution of central density. Insufficient particle sampling introduces a large deficit from sampling noise in the density profile evolution in panel (d), while all sequences converge to the same analytical cycloid solution in Eq.~\eqref{eq:os-cycloid-time-density}.

Finally, we use the \texttt{BHaHAHA} library \cite{Etienne:2025auk} to calculate the irreducible mass $M_{\rm irr}$ of the horizon at $t=60M$, using the numerical spatial metric saved in that snapshot. {The quantity $\epsilon_M$ is the RMS relative horizon-mass error over the five seeds.} The AMR particle sequence gives $\epsilon_M=2.061\times10^{-3}$, $9.362\times10^{-4}$, and $7.347\times10^{-4}$ for A1, A2, and A3, respectively. Thus the final-mass RMS improves with increasing particle number
over the tested range, but this improvement is slower than the initial sampling-error reduction and is not assigned an $N^{-1/2}$ convergence law. We do not find a monotonic convergence trend of $M_{\rm irr}$ for the fixed-$N$ SMR sequence, and the constant-occupancy sequence gives a flat error across the sequence.\\

\section{Monte Carlo convergence of the Shapiro--Teukolsky cluster}
\label{sec:st-mc-convergence}

We test particle-number convergence using the unperturbed monoenergetic Shapiro--Teukolsky configuration with $y_c=0.819$, using $N=1$, 2, 4, and 8 million equal-rest-mass particles. All four runs use the same mesh structure and Monte Carlo equal-mass particle initial data sampler as discussed in Section~\ref{sec:os-collapse}. The particle rest mass is $M_0/N$, with $M_0\simeq1.03010648M$. Finite-particle sampling noise and density fluctuations therefore seed both Hamiltonian and momentum-constraint violations.

The mesh is fixed throughout this comparison: the domain is $[-128M,128M]^3$, with a $128^3$ root grid and two nested static refinement regions. The finest spacing is $\Delta x=M/2$ inside $[-32M,32M]^3$ uniformly. No coherent radial position perturbation, velocity perturbation, or density perturbation is applied. 

We compare the initial deposited energy density with the equilibrium source evaluated using the same error estimator as in Eq.~\eqref{eq:os-initial-error-new}. For this calculation, the set $\mathcal{I}$ covers the full cluster region, including the stellar surface. Meanwhile, for the constraints, we use the native Hamiltonian residual $\mathcal H$ and momentum residual $\mathcal M_i$. We present these constraint diagnostics as coordinate-volume-weighted RMS norms,
\begin{align}
\|\mathcal H\|_{2,\mathcal C}
&=\left(\frac{\sum_{\mathcal C}\mathcal H_i^2\Delta V_i}
                   {\sum_{\mathcal C}\Delta V_i}\right)^{1/2},
\label{eq:st-H-norm-new}\\
\|\mathcal M\|_{2,\mathcal C}
&=\left(\frac{\sum_{\mathcal C}
                 \gamma^{jk}\mathcal M_j\mathcal M_k\Delta V_i}
                   {\sum_{\mathcal C}\Delta V_i}\right)^{1/2},
\label{eq:st-M-norm-new}
\end{align}
where $\mathcal C$ is the set of cells involved in the average. Figure~\ref{fig:st-initial-new} shows that the initial density error, Hamiltonian norms, and momentum norms all decrease following the $N^{-1/2}$ scaling closely.

We also evolve all four configurations and follow their long-term constraint evolution. Figure~\ref{fig:st-constraints-new} shows both constraints through $t=500M$. The Hamiltonian residual decreases rapidly during the initial evolution and subsequently varies around a substantially lower level. The momentum residual initially rises from zero as the initial data settles, then decreases toward its longer-lived level. Increasing $N$ lowers both the central-region and full-domain curves over most of the evolution. We also measured the late-time asymptotic plateaus of the Hamiltonian and momentum constraints, which also scale as $N^{-1/2}$. This reduction establishes that increasing the MC sample improves constraint control at the tested fixed resolution.

\bibliography{references}

\end{document}